\documentclass[12pt,preprint]{aastex}

\shorttitle{Wide Field CCD photometry around nine open clusters}
\shortauthors{Sharma et al.}
\begin{document}

\title{Wide Field CCD photometry around nine open clusters }

\author{Saurabh Sharma\altaffilmark{1}, A. K. Pandey\altaffilmark{1}, K. Ogura\altaffilmark{2}, 
H. Mito\altaffilmark{3}, K. Tarusava\altaffilmark{3} and  R. Sagar\altaffilmark{1}}

\altaffiltext{1}{Aryabhatta Research Institute of Observational Sciences,
    Nainital, India - 263 129}
\altaffiltext{2}{Kokugakuin University, Higashi, Shibuya-ku, Tokyo 150-8440, Japan}
\altaffiltext{3}{Kiso Observatory, School of Science, University of Tokyo, Mitake-mura, Kiso-gun, Nagano 397-0101, Japan}

\date{Received ...../ Accepted .....}


\begin{abstract}
In this paper we study the evolution of core and corona of nine open clusters using 
the projected radial density profiles derived from homogeneous
CCD photometric data obtained through the 105-cm Kiso Schmidt telescope. The age and galactocentric distance of the
target clusters varies from 16 Myr to 2000 Myr and 9 kpc to 10.8 kpc respectively. Barring Be 62, which is
young open cluster, other clusters show a uniform reddening across the cluster region. The reddening in Be 62 varies 
from $E(B-V)_{min}$= 0.70 mag to $E(B-V)_{max}$= 1.00 mag. 
The corona of six of the clusters in the present sample is found to be elongated, 
however on the basis of the present sample it is not possible to establish any correlation between the age and shape of the
core. The elongated core in the case of young cluster Be 62 may reflect the initial conditions in the parental
molecular cloud. The other results of 
the present study are (i) Core radius `$r_c$' and corona size $`r_{cn}$'/cluster radius $`r_{cl}$' 
are linearly correlated. (ii) The $r_c/r_{cn}/r_{cl}$ are linearly correlated with the number of stars in that
region. (iii) In the age range 10-1000 Myr, the core and corona shrink with age. (iv) We find that in the galactocentric
 distance range 9 - 10 kpc, the core and corona/cluster extent of the clusters increase with the galactocentric distance.
\end{abstract}

\maketitle
\section{INTRODUCTION}

The study of galactic open clusters is of great interest in several 
astrophysical aspects. Young open clusters provide information about 
current star formation processes and are key objects for clarifying questions
of galactic structure, while observations of old and intermediate age open cluster
play an important role in studying the theories of stellar and galactic
 evolution. 

The nucleus and the corona (extended region of the star cluster)
are two main regions in open clusters (Kholopov 1969). The nucleus of a cluster
contains relatively bright and massive ($\ge$ 3 $M_\odot$) stars whereas
corona, which contains a large number of faint and low mass ($\le$ 1 $M_\odot$)
stars, has important bearing on studies related to the mass function (MF), 
the structure and the evolution of open clusters.
A detailed analysis of the structure of corona of open clusters is needed 
to understand the effects of external environments like the 
galactic tidal field and impulsive encounters with interstellar clouds etc. 
on dynamical evolution of open clusters (Pandey et. al. 1990).
Extensive studies of the coronal regions of clusters have not been carried
 out so far mainly because of non-availability of photometry in a large field
around open star clusters.

The $2K\times2K$ CCD mounted on Schmidt 
telescopes (Kiso, Japan), covering $\sim 50'\times 50'$ field
can be used to get photometry in a large field around open 
star clusters. The ability to obtain improved photometry of 
thousands of stars means that large-scale studies of open clusters can
be conducted to study the spatial structure and stability of galactic 
open clusters. With the addition of photometry of nearby field region, 
it is possible to construct luminosity 
function (LF)/mass function (MF), which are useful to understand 
the cluster-formation processes and the theory of star 
formation in open clusters (Miller \& Scalo 1979).

Considering the importance of low mass stars in the corona of star clusters, 
we have generated wide field photometric database around 9 open star clusters
with the aim to re-investigate the cluster's parameters e.g. reddening, distance,
age, their size and LF/MF using a homogeneous data base. The basic parameters of the cluster taken from 
WEBDA\footnote{http://obswww.unige.ch/webda} (Mermilliod 1995) are given in Table 1.

\section{OBSERVATIONS AND DATA REDUCTIONS}

 The broad band CCD photometric observations of clusters were carried out during 2001 November 19 to
  2001 November 25 using the 105 cm Schmidt telescope of the Kiso Observatory. The CCD
   camera used a SITe 2048$\times$2048 pixel$^2$ TK2048E chip having a pixel size 24 $\mu$m.
 At the Schmidt focus (f/3.1) each pixel of CCD corresponds to $1.^{\prime\prime}5$
   and the entire chip covers a field of $\sim 50'\times 50'$ on 
   sky. The read out noise and gain of the CCD are 23.2 $e^-$ and 3.4 $e^-$/ADU
   respectively. 

The observations of NGC 1528, NGC 2287, NGC 2301, NGC 2323,
NGC 2420, NGC 2437 and NGC 2548 were standardized by observing
standard stars in SA95 (Landolt 1992) having brightness
$12.2<V<15.6$ and color $0.45<(B-V)<1.51$ on four nights during 2001 November 22 to 2001 November 25.
To standardize the observation of Be 62 and NGC 1960 we generated secondary standard in
the field of these two clusters by observing stars in SA95 on 2001 November 24 and 2001 November 22 respectively.
A number of bias and dome flat frames were also taken during 
the observing runs. The log of observations is given in Table 2.
A number of short and deep exposure frames have been obtained for all
the clusters. 

The CCD data frames were reduced using computing facilities available at
Aryabhatta Research Institute of Observational Sciences (ARIES), Nainital.
Initial processing of the data frames was done
using the IRAF\footnote{IRAF is distributed by National Optical Astronomy
Observatories, USA} and ESO-MIDAS\footnote{ ESO-MIDAS is developed and 
maintained by the  European Southern Observatory.} data reduction packages. Photometry  of 
cleaned frames was carried out using DAOPHOT-II software (Stetson 1987).
The PSF was obtained for each frame using several uncontaminated
stars. The FWHM of the star images varied between $3^{\prime\prime}$
to $4^{\prime\prime}$ from night to night. 
When brighter stars were saturated on deep exposure frames, their 
magnitude has been taken from short exposure frames.
We used DAOGROW program for construction of an aperture growth curve required for 
determining the difference between aperture and profile fitting magnitude.
 Calibration of the instrumental magnitude to standard system was 
done using procedure outlined by Stetson (1992). The  photometric 
calibration equations used are as follow:

{\it
~~~~~~~~~~~~~~~~~~~~~~u = U + $c_1$ + $m_1$(U-B) + $k_u$X,
 
~~~~~~~~~~~~~~~~~~~~~~b = B + $c_2$ + $m_2$(B-V) + $k_b$X,
 
~~~~~~~~~~~~~~~~~~~~~~v = V + $c_3$ + $m_3$(V-I) ~+ $k_v$X,
 
~~~~~~~~~~~~~~~~~~~~~~r = R + $c_4$ + $m_4$(V-R) + $k_r$X,
 
~~~~~~~~~~~~~~~~~~~~~~i = ~I~ + $c_5$ + $m_5$(V-I) + $k_i$X, \\
}

 where {\it U,B,V,R {\rm and} I} are the standard magnitudes, {\it u,b,v,r {\rm and} i} 
are the instrumental magnitudes normalized for 1 second of exposure time
and {\it X} is the air mass; the $c_1,c_2,c_3,c_4,c_5$ and
$m_1,m_2,m_3,m_4,m_5$ are zero-point constants
and color-coefficients respectively; and $k_u,k_b$, $k_v$,$k_r$,$k_i$
are extinction coefficients in {\it U,B,V,R {\rm and} I} bands respectively.
The values of zero-point constants, color coefficients and extinction coefficients in various
bands on different nights are given in Table 3.
Figure 1 shows the standardization residual, $\Delta$, between standard and transformed V magnitude
and $(U-B),(B-V),(V-R)$ and $(V-I)$ colors of standard stars. The standard deviations in
$\Delta V, \Delta (U-B), \Delta (B-V), \Delta (V-R)$ and $\Delta (V-I)$
are 0.017, 0.060, 0.028, 0.023, 0.020 mag respectively. 
The typical DAOPHOT errors in magnitude and colors along with parameter $\chi$ and sharpness
as a function of $V$ magnitude in the case of clusters Be 62 and NGC 1960 are shown in Figure 2.
It can be seen that the errors become larger ($\ge$0.1 mag) for stars 
fainter than $V=20$ mag, therefore the measurements below this magnitude are not reliable.

\section{ COMPARISON WITH PREVIOUS STUDIES}

We have carried out a comparison of the present data with that
 available in the literature. The difference $\Delta$ (literature -
present data) as a function of $V$ magnitude is shown in Figure 3 and
 the statistical results are given in Table 4. The comparison 
indicates rather a large scatter in $\Delta$
for photographic data as compared to the $\Delta$ values for CCD photometry.
Comparison of the photometry in case of each cluster is discussed below. 

Be 62:  CCD $V$ magnitude and $(B-V)$ colors obtained by Phelps \& Janes (1994) are
in good agreement with those obtained in the present work. However the photoelectric
$V$ magnitudes in the range 13-14  obtained by Forbes (1981) are brighter by $\sim$0.13 mag than
that obtained in the present work. The $(U-B)$ colors obtained in the present work
for stars having $V\sim13-16$ mag are redder than those reported by Phelps \& Janes (1994). For stars
having $V>17$ mag, the $(U-B)$ colors obtained by Phelps \& Janes (1994) show a strong variation
with $V$ magnitude and becomes redder by $\sim$ 1.2 mag at $V\sim$18-19 mag. Our $(U-B)$ colors for star
$V$=13-14 mag are in agreement with the photoelectric observations by Forbes (1981).

NGC 1528: The $V$ magnitude and $(B-V)$ colors obtained in the present work are
in agreement with those reported by Hoag et al. (1961, hereafter H61).

NGC 1960: The comparison of present CCD photometry with the available
photoelectric (Johnson \& Morgan 1953) and photographic (Barkhatova et al. 1984)
 photometry indicates a fair agreement between the different measurements.
The $V$ magnitude and $(B-V)$ colors obtained by us for stars having $V\le15$ mag 
are in good agreement with those given by Sanner et al. (2000, CCD photometry),
whereas $V$ magnitude obtained by us for stars fainter than $V=15$ mag are systematically
fainter by $\sim$ 0.06 mag and $(B-V)$ color shows a trend with increasing $V$ 
magnitude in the sense that colors obtained by us become redder.

NGC 2287: Photoelectric $V$ magnitude by Eggen (1974) are in agreement with the present
observations whereas photoelectric $V$ magnitudes given by Harris et al. (1993),
Ianna et. al. (1987) and H61 are systematically fainter by 
$\sim$ 0.05 magnitude. The $(B-V)$ colors obtained in the present work are in agreement
with those reported by Harris et al. (1993), Eggen et al. (1974) \& Ianna et al. (1987).
The $(B-V)$ colors obtained by H61 are systematically blue by $\sim$
 0.06 mag than the $(B-V)$ colors reported in the present work.

NGC 2301: The $V$ magnitude and $(B-V)$ colors obtained in the present work are in
agreement with those reported in the literature.
The $(U-B)$ colors for stars having 
$V\le16$ mag by Kim et al. (2001) are in agreement with the present results, whereas stars having 
$V>16$ mag show a trend in $(U-B)$ color in the sense that present colors become
 bluer with increasing $V$ magnitude. The $(U-B)$ photographic colors reported
 by Mohan \& Sagar (1988) at $V\sim$10-13 mag are systematically redder by $\sim$
0.12 mag and become further red with increasing $V$ magnitude. Whereas, photographic
$(U-B)$ colors for stars having $V$ = 13-15 mag by H61 are systematically redder by $\sim$ 0.12 mag.

NGC 2323: The comparison indicates that the $V$ magnitude obtained by us are 
in general in agreement with those given in literature barring the magnitudes
given by H61. The V magnitudes reported by 
H61 are fainter by $\sim$ 0.06 mag. The $(B-V)$ colors also show a fair agreement
with the values reported  in the literature excepting 
$(B-V)$ colors for $V\ge13.0$ mag given by H61 which are bluer by $\sim$ 0.07 mag.
The $(U-B)$ colors given in the literature also show a fair agreement with those 
obtained in the present study.

NGC 2420: The comparison of present CCD observations indicates a good agreement 
with those  reported in the literature. 

NGC 2437: The comparison of the present CCD photometry with the available 
photographic, photoelectric and CCD photometry indicates a good agreement 
between various reported observations.

NGC 2548: The photoelectric $V$ magnitude by Pesch (1961) and Oja (1976) are 
systematically faint by about 0.05 magnitude.

\section{ ANALYSIS}

\subsection{ Radial stellar surface density and cluster radius }

As the area of present CCD observations is quite large, 
it can be used to study the radial extent, structure (core and corona region)
and evolutionary aspects of clusters.
Since, the projected distribution of stars in a cluster follows systematic
distribution from the center to the outer region,
the center of the cluster was estimated by convolving a Gaussian
 kernel with the stellar distribution and taking the point 
of maximum density as the center. This was performed for both the axis to
 get the center coordinates of the clusters.
To determine the radial surface density, the cluster was divided into
 a number of concentric rings. Projected radial stellar density in 
each concentric circle was obtained by dividing the number of 
stars in each annulus by its area, and the same are plotted in Figure 4
for various magnitude levels. The error bars derived assuming that the number
of stars in each annulus follows the Poisson statistics. 
The extent of the cluster `$r_{cl}$' is defined as the point where the radial
density becomes constant. The horizontal dashed line
in the plots indicates the density of contaminating field stars, which
is obtained from the region outside (about 1.5 cluster radius) the cluster area.

To check whether the density distribution shown in the left panel of Figure 4
is affected by contamination due to field stars, we selected a sample of stars
near a well defined main sequence (MS) in the color-magnitude diagram (CMD)
as mentioned by Pandey et al. (2001). The radial density distribution of the MS sample
is also shown in the right panel of Figure 4. Both the samples in general show almost similar radial density profiles.

The observed radial density profile of the clusters was parametrized following the 
approach by Kaluzny \& Udalski (1992) where projected radial density $\rho(r)$
is described as:
\begin{center}
$\rho (r) \propto {f_0\over \displaystyle{1+\left({r\over r_c}\right)^2}}$,
\end{center}
where the clusters core radius $`r_c$' is the radial distance at which
the value of $\rho (r)$ becomes half of the central density $f_0$.
The best fit obtained by $\chi^2$ minimization technique is shown in Figure 4.
Within uncertainties the King model (King 1962) reproduces well the radial density
profiles (RDPs) of the clusters studied in the present work.
Structural parameters obtained by fitting the King model surface density profile
to the observed radial density profile of main sequence stars having {$V<18$ mag} are given in Table 5.

Majority of the clusters indicate an increase in core radius when fainter cluster members 
included in the radial density profile, however two clusters namely 
NGC 1960 (log age $\sim$ 7.4) and NGC 2437 (log age $\sim$ 8.4)
yield a  smaller $r_c$ toward fainter magnitude. 
Whereas clusters NGC 2301 (log age $\sim$ 8.2) and NGC 2420 ( log age $\sim$ 9.3)
do not show any significant change in core radius as a function of brightness of the
cluster stars. The radial stellar surface density profiles displayed in Figure 4 indicate that
field star contamination is not significant up to $V\sim$18 mag, however it increases 
significantly at $V=20$ mag. We expect for the sample brighter than $V \sim18$ mag field star
contamination is not strong enough to smear the cluster properties and 
hence will have insignificant effect in the results derived below.
Poor radial surface density profile in the case of NGC 2287 is due to lack of clustering in  the object.

Recently various authors have used Two Micron All Sky Survey (2MASS)
data to study the radial structures of open clusters (Chen et al. 2004, hereafter C04;
Bonatto \& Bica 2005, hereafter BB05). We also used 2MASS data to obtain the radial 
density profile curve of the clusters studied in the present work and the same
are shown in the lowermost panels of Figure 4. The core radius `$r_c$' and extent of 
the cluster `$r_{cl}$' of the target open clusters estimated from the projected
radial density profile for main-sequence band stars are given in Table 5.
Table 5 indicates that the $r_c$ and $r_{cl}$  obtained using the 2MASS data are larger, in some cases, than
that obtained from optical data. The error in $r_c$ obtained from 2MASS data
is also relatively large in comparison to that for optical data.
Various recent estimates of radii of target clusters are given in Table 6, which indicates 
a fare agreement between present optical estimates and those by Nilakshi et al. (2002, hereafter N02).
To study various parameters, e.g. interstellar reddening, age, distances etc, 
in detail, we considered only those stars which are inside the cluster extent as determined from radial density profile.

\subsection {Isodensity contours}

It is well known that the internal interaction of two body relaxation due to encounters
among member stars and external tidal forces due to the galactic disk or giant molecular
clouds influence the morphology of the clusters.
To study the effects of external forces on the morphology of the
clusters, we obtained isodensity contours for the sample of MS stars ($V\le18$ mag) 
and the contours are shown in Figure 5.
The isodensity contours are least square fitted with ellipse to obtain the elongation
of the corona of the clusters. The elongation parameter `$e_p$' is defined as $e_p$ = b/a, where
a and b are semi-major and semi-minor axis of the ellipse. The parameter $e_p$ for each cluster
is given in Figure 5. $e_p$ = 1 indicates a spherical cluster. The fitting could not be done
in the case of Be 62 and NGC 1960. However a visual inspection of density contours of these
two clusters indicates an elongated corona of the clusters. C04 have also reported an elongated corona in the
case of NGC 1960. 

Figure 5 indicates that outer region (corona) of six clusters is found to be elongated.
The elongated morphology in the case of NGC 2287, 
NGC 2548 (Bergond et al. 2001, hereafter B01) and NGC 1893, Be 17, NGC 2420 (C04) have 
already been reported in the literature. Present results further support the elongated 
morphology of the coronal region of the clusters.

\subsection {Interstellar Extinction}

The extinction towards the clusters Be 62, NGC 1960, NGC 2301 and NGC 2323 was estimated using the $(U-B)/(B-V)$
two color diagram shown in Figure 6. We compared observed main-sequence (MS) 
in the cluster region with the intrinsic MS to 
estimate the value of reddening $E(B-V)$. In Figure 6, we show intrinsic MS for $Z=0.02$ by Schmidt-Kaler (1982)
shifted along the reddening vector having a normal slope of $E(U-B)/E(B-V)$ = 0.72. 
The interstellar extinction
in the case of other five clusters NGC 1528, NGC 2287, NGC 2420, NGC 2437 and NGC 2548 was estimated using the $V/(B-V)$ or
$V/(V-I)$ color-magnitude diagrams (CMDs, cf. $\S$ 4.4). The estimated values of $E(B-V)$ are given in Table 7
which are in good agreement with those given in the literature. Barring the cluster Be 62, all other
clusters show a uniform reddening across the cluster region. The reddening in Be 62 varies from $E(B-V)_{min}=0.70$ mag
to $E(B-V)_{max}=1.00$ mag. The reddening for individual stars ($V_{error}\le 0.1$ mag) having spectral
type earlier than $A0$ has also been derived using the Q-method (Johnson \& Morgan 1953).

To study the nature of the extinction law in the cluster region, we used two-color diagrams (TCDs)
as described by Pandey et al. (2001, 2003). 
The TCDs of the form of ($\lambda-V$) vs. ($B-V$), where ($\lambda-V$) is one of the 
wavelength of the broadband filters ($R,I,J,H,K,L$) provide an effective method for 
separating the influence of the normal extinction produced by the diffuse interstellar medium
from that of the abnormal extinction arising within regions having a peculiar distribution of dust sizes
(cf. Chini \& Wargan 1990, Pandey et al. 2000).
The near infrared (NIR) data have been taken from the 2MASS.
For illustration, the TCDs for the clusters Be 62, NGC 2420 and NGC 1960 are 
shown in Figure 7 and slopes of the distribution, $m_{cluster}$,
in the case of these three clusters are given in Table 8.

The ${E(\lambda-V)}\over {E(B-V)}$ values in the cluster region are estimated using the 
following approximate relation:
\begin{center}
${{E(\lambda-V)}\over {E(B-V)}} ={ { m_{cluster}\over m_{normal}} \times [{{{E(\lambda-V)}\over {E(B-V)}}]_{normal}}}$
\end{center}
as described by Pandey et al. (2003), where $m_{cluster}$ and $m_{normal}$ are the slope of the 
distribution in the cluster region and slope of the theoretical
MS. The slopes of the theoretical MS, $m_{normal}$, obtained from the stellar models by Bertelli et al. (1994)
are also given in Table 8. The obtained slope $m_{cluster}$ indicates
that barring Be 62 and NGC 2420 all other clusters show a normal reddening law towards the cluster region.
The color excess ratio $E(\lambda-V)\over E(B-V)$ for Be 62 and NGC 2420
are smaller than the normal ones, which indicates that there may be 
anomalous reddening law towards these cluster
regions. Since Be 62 shows a differential extinction indicating that the stars are still embedded in the parent
molecular gas and dust, an anomalous reddening law is expected in the cluster region. The value of $R$ 
towards the cluster Be 62 region estimated to be $2.95\pm0.12$. However it is quite surprising
in the case of NGC 2420 which is an old cluster having a uniform reddening across the cluster. 
Keeping the large error in mind we used a normal value of $R$ for all the cluster regions in the ensuing discussions.

\subsection{Color magnitude diagrams, distance and age of the clusters}

Color magnitude diagrams (CMDs) for stars lying within cluster region as mentioned in Table 5 are
shown in Figure 8. The CMDs show a well defined MS.
%
Barring Be 62, other clusters manifest a uniform reddening in the cluster region and the error
in magnitude estimation is $\sim$ 0.05 mag for stars having $V\le$18.0 mag. Therefore we can conclude that the 
presence of probable binaries and field stars should be the main cause for broadening of the MS in these clusters.
In the case of Be 62 variable reddening in the cluster region along with presence of probable
binaries and field stars should be the cause of broad MS.
It is difficult to separate field stars from the cluster stars, however we can reduce the contamination
due to field stars if we restrict the sample to the central portion of the cluster only.

We supplemented present CCD data for some cluster with the photoelectric photometry for bright stars available in the
literature as these bright stars were saturated even on present short exposure frames. The 
distance and age of the cluster were obtained by visual fitting of isochrones by Bertelli et al. (1994)
for $Z=0.02$ to the blue envelope of the observed MS except in the case of NGC 2420 where we used 
isochrones for $Z=0.008$ as Lee et al. (2002) have reported $Z=0.009$ for the cluster NGC 2420. 
The fitted isochrones are shown in Figure 8.
Since extinction is uniform in clusters under study except Be 62,
we used mean value of $E(B-V)$ and following relations,
$E(U-B)/E(B-V)$=0.72, $A_V=3.1\times E(B-V)$, $E(V-R)=0.60\times E(B-V)$, $E(V-I)=1.25\times E(B-V)$
to fit the theoretical isochrones to the observations. In the case of Be 62 
individual reddening for stars earlier than
$A0$ was obtained using the $Q$ method and the same was applied to get intrinsic magnitude and colors,
whereas for other stars mean reddening of the nearby stars was applied to get the intrinsic magnitude and colors.
The unreddened CMDs along with the fitted isochrones in the case of Be 62 are shown in Figure 9. The age and distance 
obtained for the target clusters are given in Table 8.
The $K/(J-K)$ CMDs for the clusters obtained from 2MASS data are shown in Figure 10. The theoretical 
isochrones by Bertelli et al. (1994) using the parameters obtained from the optical data (cf. Table 7) 
are also plotted in Figure 10 which nicely follow the observations.
Comparison of previous estimates available in the literature with the present estimates is
given in Table 9.

\section {RESULTS}

We have determined basic parameters of 9 open clusters by analyzing the 
color-color diagrams and CMDs of the clusters.
We assumed solar abundance for all the clusters except NGC 2420 where metallicity is reported to be $Z = 0.009$ by
Lee et al. (2002). The discussion on individual cluster follows, where we have compared the radii of the cluster obtained in the present study with those reported by N02, BB05, C04 and Kharchenko et al. (2005, hereafter K05).
The morphological parameters of open clusters derived by N02 and K05 are based on optical observations whereas
C04 and BB05 have used 2MASS data. In the subsequent sections we use the core radius 
 `$r_c$' and cluster extent `$r_{cl}$' obtained from the present
optical photometric and 2MASS data.

\subsection {Be 62}

The reddening $E(B-V)$, distance and age for the cluster has been estimated as $0.86$ mag, $2.05\pm0.24$
kpc and 10 Myr by Forbes (1981) using the photoelectric photometry, whereas Phelps \& Janes (1994) using the 
CCD observations reported $E(B-V)$=0.82 mag, distance=2.7 kpc and an age of $\sim$ 10 Myr for this cluster.
 
In Figure 8, $V/(B-V)$, $V/(V-R)$, $V/(V-I)$ CMDs for stars within  the cluster region i.e. $r_{cl}=10.'0$ (6.8 pc)
are presented.  The CMDs of the Be 62 shows a broad MS which should be due to variable reddening in the cluster region. 
To get unreddened CMDs, the reddening $E(B-V)$ for stars having spectral type $A0$ or 
earlier has been calculated using the $Q$ method and these stars
were unreddened individually. For remaining stars the average reddening of the nearby stars was applied. The
unreddened CMDs shown in Figure 9. To calculate ${\it V_O}$, the value of $R=3.1$ is assumed. 
The distance and age of the cluster comes out to be 2.32 kpc and 16 Myr (log age = 7.2). 
The distance obtained in the present work is in between the values 2.05 kpc and 
2.70 kpc reported by Forbes (1981) and Phelps \& Janes (1994) respectively.

The core of the cluster using the optical MS data is estimated to be $\sim2.'2\pm0.'3$ ($1.5\pm0.2$ pc). 
The 2MASS data  gives the value for core as $\sim2.'5\pm1.'0$ ($1.7\pm0.7$ pc).
The optical and 2MASS data yields the extents of cluster as $\sim10'$ (6.8 pc) and $\sim12'$ (8.1 pc) respectively.
The core and the coronal region of the cluster are found to be elongated.

\subsection {NGC 1528}

The radial density profile of the cluster yields its extent $r_{cl}\sim15'$ (4.8 pc)
with a core radius $r_c\sim8.'3\pm1.'5$ ($2.6\pm0.5$ pc) which is in agreement with the
value ($8.'4$) by K05, however the $r_{cl}$ obtained in the 
present work ($\sim15'$) is significantly smaller than the value ($26'.4$) by K05.
The RDP obtained using the 2MASS data is quite noisy and it yields significantly
different value of $r_c$ ($\sim18.'5\pm7.'4$; $5.9\pm2.4$ pc).

 The CMDs of the cluster within $r_{cl}\sim15'$ show a well defined but broad MS.
Since the cluster has uniform reddening, the presence of binary stars
should be a probable reason for the broadening of the MS. The CMDs show the turn-off
of the MS which can be fitted nicely with a isochrones of log age = 8.6 (400 Myr) and $Z=0.02$. 
The distance to the cluster comes out to be 1.09 kpc which is in agreement with those reported in the literature.
Isodensity curves shown in Figure 5 indicate that both the core and the corona have elongated morphology.

\subsection {NGC 1960 (M36)}

This cluster situated in Auriga constellation is reported to have diameter of $10^\prime$ 
(Lynga \& Palous 1987). Sanner et al. (2000)
reported a distance modulus $(m-M)_o = 10.6\pm0.2$ mag ($1318\pm120$ pc), 
reddening $E(B-V)$ = $ 0.25\pm0.02$ mag, age = $16^{+10}_{-5}$ Myr and metallicity $Z = 0.02$ for the cluster.

The CMDs of the NGC 1960 show a well defined narrow MS. By fitting the isochrones for 
log age = 7.4 (25 Myr) 
to the stellar distribution in the CMDs using $E(B-V)$ = 0.22 mag, we obtained a distance modulus $(m-M)_V$ = 11.3 mag
corresponding to a true distance modulus $(m-M)_o$ = 10.62 mag, 
or a distance of 1.33 kpc. The estimated age, distance and 
reddening in the cluster region are in good agreement with 
the values obtained in earlier studies (e.g. Sanner et al. 2000).

The core radius ($r_c\sim3.'2\pm0.'5$; $1.2\pm0.1$ pc) and cluster extent ($r_{cl}\sim14'$; 5.4 pc) obtained in 
the present study are in good agreement with the values obtained by N02 ($r_c\sim3.'2$, $r_{cl}\sim15.'4$).
The core radius ($r_c\sim5.'4$) obtained by K05 
is larger than the present value. The cluster extent ($r_{cl}\sim14'$) obtained in the 
present work is in agreement with the value ($\sim16.'2$)
by K05 whereas it is larger than that ($\sim7.'6$) reported by C04.
From 2MASS data, we obtained $r_c \sim 3.'8\pm0.'6$ ($1.5\pm0.2$ pc)
and $r_{cl} \sim 21'$ (8.1 pc). The core of the cluster is rather spherical 
symmetric whereas the outer region of the cluster clearly shows the effect of external forces.

\subsection {NGC 2287 (M41)}

Despite being a bright cluster, no CCD photometry is available for this cluster. This cluster has been studied
for proper motion (Ianna et al. 1987), radial velocities (Amieux 1988) and spectroscopic properties 
(Harris et al. 1993). Battinelli et al. (1994) and B01 have reported 
$R_{lim}$ = 4.0 pc ($19'$) and $R_{tidal}$ = 4.1 pc ($20.'3$) respectively for the cluster.
K05 have reported $r_c\sim9.'6$ and $r_{cl} \sim 30'$. BB05 have estimated
$R_{core}$ = 1.1 pc ($4.'7$) and $R_{lim}$ = 7.0 pc ($30.'1$). In present study we found that optical 
data for brighter limiting visual magnitude 
($V\sim14$ mag) and fainter limiting magnitude ($V\le18$ mag) were yielding quite different values for the morphology of the
cluster. For brighter sample we obtained $r_c\sim5'.5\pm0.'9$ ($1.1\pm0.2$ pc), $r_{cl}\sim17'$ (3.5 pc) 
whereas for sample having V$\le$18 mag these
values are estimated as $\sim 1.'4\pm0.'3$ ($0.3\pm0.2$ pc) and $\sim 12'$ (2.5 pc) respectively.
The radial density profile for 2MASS data is quite noisy with $r_c\sim12.'7\pm3.'8$ ($2.6\pm0.8$ pc) and
$r_{cl}\sim 16'$ (3.3 pc). For further study of the cluster, we assumed
$r_{cl}\sim12'$ (2.5 pc). The morphology of the cluster indicates that both the
central region and outer region are elongated.
The center coordinate for the cluster reported by B01 are shifted towards 
$15'$ SW of ``classical'' coordinates given in WEBDA. However, BB05 find the same coordinates for the center
as given in WEBDA, whereas in the present work we find a shifted center for the cluster 
towards north of the center as given in WEBDA.

The CMDs of the clusters show a very well defined MS along with red giant clump. The distance 
and age for the cluster estimated to be 0.71 kpc and 250 Myr (log age = 8.4) 
which are in agreement with recent findings.

\subsection {NGC 2301}

A comparison of the theoretical isochrones with the observations yields
a distance modulus $(m-M)_V$ = 9.8 mag, which corresponds to a true distance modulus $(m-M)_0$ = 9.71 mag
or a distance of 0.87 kpc and an age of 160 Myr (log age = 8.2) for the cluster. 
The age, distance and reddening towards the cluster region obtained in the present work 
are in good agreement with those reported in the literature. 

The estimated core radius ($r_c\sim1.'9\pm0.'3$; $0.5\pm0.1$ pc) is in good agreement with the value 
($r_c\sim1.'9$; 0.5 pc) by N02 whereas it is significantly smaller than $r_c\sim4.'8$ by K05. 
The extent of cluster ($\sim9'$; 2.3 pc) obtained in the present work is smaller
than that reported by N02 ($12'$; 3.0 pc) and K05 ($\sim15'$). Here
it is worthwhile to mention that the core radius ($r_c\sim4.'5\pm1.'0$; $1.1\pm0.2$ pc) 
and cluster extent ($r_{cl}\sim20'$; 5.1 pc) obtained from the
2MASS data are almost double to that values obtained from the optical observations. The core and 
coronal region of the cluster NGC 2301 are found to be highly elongated indicating
a strong effect of external forces on the cluster.

\subsection {NGC 2323 (M50)}

The distribution of MS stars having V$\le$18 mag yields $r_c \sim 6'.5\pm1'.5$ ($1.8\pm0.4$ pc)
and $r_{cl} \sim 17'$ (4.7 pc).
The core radius obtained in the present study is larger than that obtained by N02 ($2.'6$)
but in agreement with that obtained by K05 ($6'$).
The extent of the cluster is in agreement with that reported by N02 ($16.'7$), however it is smaller than that reported by 
K05 ($\sim22'.2$). The 2MASS data yield $r_c\sim6.'7\pm1.'3$ ($1.9\pm0.4$ pc) and
$r_{cl}\sim22'$ (6.1 pc).
The radial density profile derived from optical and 2MASS data thus gives comparable core radius.
The core and the coronal region of the cluster are found to be spherical.

The cluster NGC 2323 shows a well defined MS towards the brighter end ($V\le15$ mag). 
The effect of photometric error becomes significant towards the fainter end. A comparison of the observations with the 
theoretical isochrones yields a distance of 0.95 kpc and age of $\sim$ 100 Myr (log age = 8.0) to the cluster.

\subsection{ NGC 2420}

The $BV$ and $VI$ CCD photometry for the cluster NGC 2420 has been carried out by 
Anthony-Twarog et al. (1990) and Lee et al. (2002)
respectively. Leonard (1988), using the star counts on Palomar Sky Survey, reported a quite large
extent of $\sim20'$ for the cluster. In the present work we find that optical data 
and 2MASS data indicate a radial extent of the cluster is $\sim10'$ (7.2 pc) and $\sim9'$ (6.5 pc)
respectively which is in fair agreement with that reported by C04 ($\sim11.'6$) but smaller 
than that reported by N02 ($\sim13.'2$). 
The core radius is estimated to be $1.'4\pm0.'1$ ($1.0\pm0.1$ pc) which is in agreement
with that obtained by N02 ($1.'5$). 
2MASS data also yield nearly same core radius $\sim1.'3\pm0.'2$ ($0.9\pm0.1$ pc). 
The isodensity curves indicate a rather spherical core and coronal regions of the cluster. 
C04 have obtained a spherical core with an elongated coronal region of the cluster.

The CMDs within the cluster region displayed in Figure 8 show a well defined MS. 
As can be seen in the CMDs, the contamination
due to field stars is almost negligible. The CMDs manifest a sequence of evolved stars which is nicely reproduced by
the theoretical isochrones for $Z=0.008$ and log age = 9.3 (2 Gyr).

The distance to the cluster is estimated to be 2.48 kpc, which is in good agreement with the values reported by
Anthony-Twarog et al. (1990), Lee at al. (2002) and West (1967)
whereas distance reported by McClure et al. (1974) is 1.9 kpc.
Because of its age and metallicity, NGC 2420 is an important object to test 
theoretical stellar evolutionary models.

\subsection {NGC 2437 (M46)}

The star count peak obtained in the present work is shifted by $\sim3'$ East and $\sim1'$ south to the center
coordinates given in WEBDA. 
The radial density profile obtained for optical as well as 2MASS data shows a well
defined distribution of stellar density around center of the cluster. However the extent of the cluster obtained for two distributions  differs. The $r_{cl}$ estimated to be $\sim20'$ (8.8 pc) and $25'$ (11.0 pc) for the optical
and 2MASS data while the corresponding values of $r_{c}$ are $6.'8\pm1.'0$ ($3.0\pm0.4$ pc) 
and $9.'6\pm0.'6$ ($4.2\pm0.3$ pc) respectively. 
The values of $r_{c}$ reported by K05 and N02 are $7.'2$ and $5.'2$ respectively.
The corresponding $r_{cl}$ values are $22.'8$ and $26.'6$ respectively.
The core of the cluster is elongated whereas the corona of the cluster
is found to have spherical symmetry.

The $V/(B-V)$ and $V/(V-I)$ CMDs for the cluster indicate a broad and well defined MS. 
The effect of field star contamination and errors is clearly visible towards fainter end ($V>15$ mag).
A few stars on the red giant clump region can also be noticed. Using $E(B-V)=0.10$ mag
we obtained a distance of $\sim$ 1.51 kpc and an age $\sim250$ Myr (log age = 8.4) for the cluster. Since the cluster is of intermediate age and no indication of presence of parental molecular gas and dust, the broad
MS may be probably due to presence of binary stars.

\subsection {NGC 2548 (M48)}

The point of maximum stellar density i.e. adopted center of the cluster is found towards south ($\sim1.'6$) of
the center given in WEBDA. The center reported by BB05 is in excellent agreement with that obtained 
in the present work. However, B01 obtained the center of this cluster further south of ($\sim1.'4$)
of the coordinate obtained in the present work. The extent of 
the cluster is reported to be 4.8 pc ($21.'5$), 8.8 pc ($37.'8$)
by B01 and BB05 respectively. K05 have reported a significantly larger value 
($43.'8$) for the cluster extent. The core radius $r_c$ is reported as $3.'9\pm0.'9$
($0.9\pm0.2$ pc) by BB05
whereas K05 reported a significantly larger value for the core as $r_c\sim15'$. In the present work
we found $r_c\sim 1.'5\pm0.'2$ ($0.3\pm0.1$ pc) and $\sim2.'4\pm1.'5$ ($0.5\pm0.3$ pc) for optical 
and 2MASS data respectively,
whereas the extent of the cluster for both of the samples is found to be $\sim8'$ (1.8 pc). 
The elongated morphology of the coronal region of the cluster manifests
that cluster structure has been affected by a significant amount of external forces.

The cluster shows a well defined narrow MS in the $V/(V-I)$ CMD with a few evolved stars on the giant branch.
The shape of the turn-off point is that of typical intermediate-age open cluster and can be nicely explained
by an isochrones having $Z=0.02$ and age  $\sim400$ Myr (log age = 8.6). The distance to the cluster is estimated to be 0.77 kpc.  

\section {DISCUSSION AND SUMMARY}

\subsection { The evolution of core and corona of clusters}

The isodensity curves shown in Figure 5 indicate  that the coronal regions of the six clusters show an
elongated morphology. The youngest cluster of the present sample, Be 62 (log age = 7.2) has an elongated
core, whereas NGC 1960 (log age = 7.4) has somewhat spherical core. 
On the other hand, the core of the cluster NGC 2548 (log age = 8.6) 
shows an elongated morphology whereas the core
 of the NGC 2420, oldest cluster in the present sample (log age = 9.3) shows rather a spherical symmetry.
C04 have also reported a spherical core in the case of NGC 2420. On the basis of present small 
homogeneous data sample it is not possible to establish any correlation between age and shape of the core.
However C04 found that the core tends to circularize as a star cluster ages. It seems that initial 
morphology of the open cluster at the time of formation of the cluster is governed by the initial conditions
in the parent molecular clouds (cf. C04), because observed shapes of molecular cloud cores also
show highly elongated morphology (Curry 2002; Gutermuth et al. 2005). 
C04 suggested that the later evolution of the cluster may be governed by both
 the internal gravitational interaction and external tidal forces.

Danilov \& Seleznev (1994) showed that the parameter $\zeta$ (=$R_1\over R_2$) and
$\mu$ (=$N_1 \over N_2$) follow the relationship 
$\zeta \propto \mu^{0.35}$, where $R_1$ and $R_2$ are the radii of the cluster core and halo,
and $N_1$ and $N_2$ are numbers of star in the core and halo. This relationship may be 
caused by an approximate equality between the rates of transfer of star from the core to corona
and vice-versa, as a result of stellar encounters (Danilov 1997). 
In order to study the core-corona structure, we selected a mass limited sample,
since evolution of clusters, their stability
and their parameters depend on the mass of the cluster (cf. Pandey et al. 1991; BB05).
We estimated various parameters for the target clusters for mass limited sample having $M_V\le 7$ mag 
and these are given in Table 10. In the case of Be 62, it is not possible to estimate the parameters for
$M_V\le 7$ mag ($V\le 21$ mag), we used parameters obtained for $V\le 20$ mag ($M_V\le 6$ mag) for further analysis.
Data for a few cluster studied earlier by us have been taken from literature
and relevant parameters for the clusters are given in Table 11. 
In Figure 11 we plot ($r_c\over r_{cn}$) vs. ($n_c\over n_{cn}$) and ($r_c\over r_{cl}$) vs. ($n_c\over n_{cl}$)
diagrams for the target clusters where $r_c$, $r_{cn}$ (= $r_{cl}-r_c$) and $r_{cl}$ represents 
core radius, size of corona and cluster extent respectively and $n_c$, $n_{cn}$ and $n_{cl}$ 
represents number of stars in the core, corona and cluster respectively. 
Figure 11a indicates that clusters under study except NGC 1528 follow the relation 
($r_c\over r_{cn}$) $\propto$ $({n_c\over n_{cn}})^{0.35}$ as suggested 
by Danilov \& Seleznev (1994). The anomalous location of NGC 1528 in Figure 11
is probably due to large error in estimating the core radius
because of noisy radial density profile. If we take the core radius obtained from all star radial density profile
as shown in Figure 4, the cluster follows the expected relation as shown in Figure 11.
Danilov \& Seleznev (1994) have carried out simulation of isolated cluster dynamics and found that
the parameters $\zeta$ \& $\mu$ obtained from simulations also follow the relation
$\zeta \propto \mu^{0.35}$. They concluded that the evolution of the core and corona
of the clusters are mainly controlled by internal relaxation processes.

In Figure 12 we plot various structural parameters of the clusters, given in Tables 10 and 11, 
which indicates that $r_c$ has a good correlation
with $r_{cn}$ and $r_{cl}$. Similarly $r_c$, $r_{cn}$ and $r_{cl}$ indicate a linear correlation with the
 number of stars in the core, corona and the entire cluster in the sense larger core, corona can accommodate
larger number of stars. The central density $f_0$ decreases exponentially
with increase in core radius.
In Figure 13, various structural parameters from Tables 10 and 11 are plotted
as a function of age which indicate that in the age range 10-1000 Myr, 
the core and extent of the cluster seems to shrink with age. In the case of less massive
clusters, the work of BB05 (their Figure $7g$) also indicates a similar correlation
between age and core radius for the cluster having age $\le1000$ Myr. Because of the above correlation between
age and core/corona radius, the projected surface density in core ($\rho_c$) and corona ($\rho_{cn}$) is higher for
intermediate/old age (100-1000 Myr) clusters as compared to younger clusters. The same trend can be 
noticed in Figure {7\it i} and Figure {7\it l} of BB05 for less massive clusters.
The concentration parameter $r_{cl}/r_c$ is plotted as a function of age in Figure 13
which indicates no trend with age. N02 have also concluded the same.

Figure 13 also indicates that the core and corona of two oldest open clusters in the 
present sample (age $> 10^9$ yr), namely NGC 2420 ($r_c$ = 1.0 pc, $r_{cl}$ = 10 pc)
and Be 20 ($r_c$ = 1.6 pc, $r_{cl}$ = 6.6 pc), are found to be relatively larger
than the expected value of $r_c$ ($\sim$ 0.5 pc) and $r_{cl}$ ($\sim$ 2 pc) at age
$\sim10^9$ yr. These clusters are located at large distances (NGC 2420; $z$ = 844 pc \& 
Be 20; $z$ = -2700 pc) from the galactic plane. The large corona of NGC 2420 and Be 20
probably indicates that these clusters may be in process of disintegration as suggested by C04.

Dependence of cluster sizes on the galactocentric distances $R_G$ is displayed in Figure 14.
A vertical dashed line is drawn to delineate the sample of clusters having $R_G<10$ kpc
and $R_G>10$ kpc. The figure indicates
that in the range $9<R_G<10$ kpc, the core, corona and total cluster size increase with the 
galactocentric distance. The study of N02 also indicates a similar correlation between cluster size and $R_G$
in the range $R_G>9$ kpc.

\subsection{Summary}

This paper analyzes homogeneous wide field CCD data of nine open cluster (ages in the range 16 Myr to
2 Gyr) taken from Kiso Schmidt telescope. Based on the well defined $(U-B)/(B-V)$ two color diagram and CMDs,
we estimated the reddening towards the cluster region, the distance and age of the target clusters. The 
structural parameter for target clusters were obtained using the projected radial density profile.
Most of the clusters show an increase in core radius when faint cluster members included in the sample.
The radial structure of all clusters can be well explained by King's empirical model (1962). 
The target clusters sample indicate an elongated core for young (Be 62, log age = 7.2) as well as for intermediate
age open clusters (NGC 2548, log age = 8.6). C04 mentioned that the
clusters have tendency towards spherical shaping away from the disk, in particular in the case of old systems
due to internal dynamical relaxation. However we find that relatively young cluster NGC 1960 (log age = 7.4, $z$ = 24 pc)
has a rather spherical core.

Present sample of the clusters follows the relation ${r_c\over r_{cn} }\propto{ ({n_c\over n_{cn}})^{0.35}} $
as suggested by Danilov \& Seleznev (1994). It is found that the core radius and corona size/cluster
radius are linearly correlated. It is also found that in the age range 10-1000 Myr, the core
and corona of the cluster shrink with age. In the galactocentric distance range 9-10 kpc, the core
and corona/cluster extent of the clusters increase with galactocentric distance.
 
\acknowledgments
We would like to thank the anonymous referee for his constructive comments.
This work is partly supported by the Department of Science and Technology (DST) India
and Japan Society for the Promotion of Science (JSPS) Japan. AKP is thankful to the staff
of KISO observatory for their help during his stay there.

\begin{figure*}[h]
\hbox{
\includegraphics[height=9cm,width=9cm]{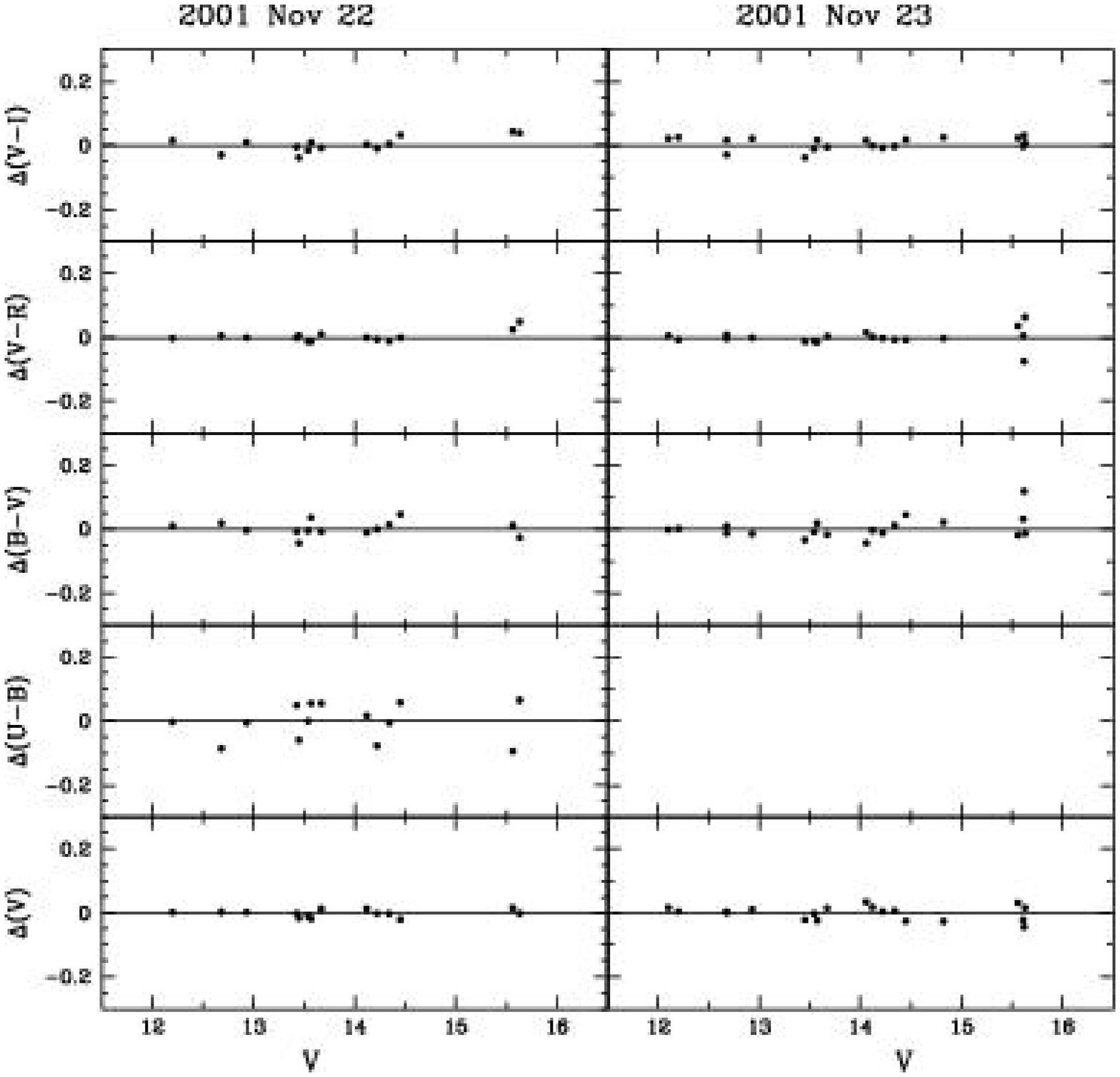}
\includegraphics[height=9cm,width=9cm]{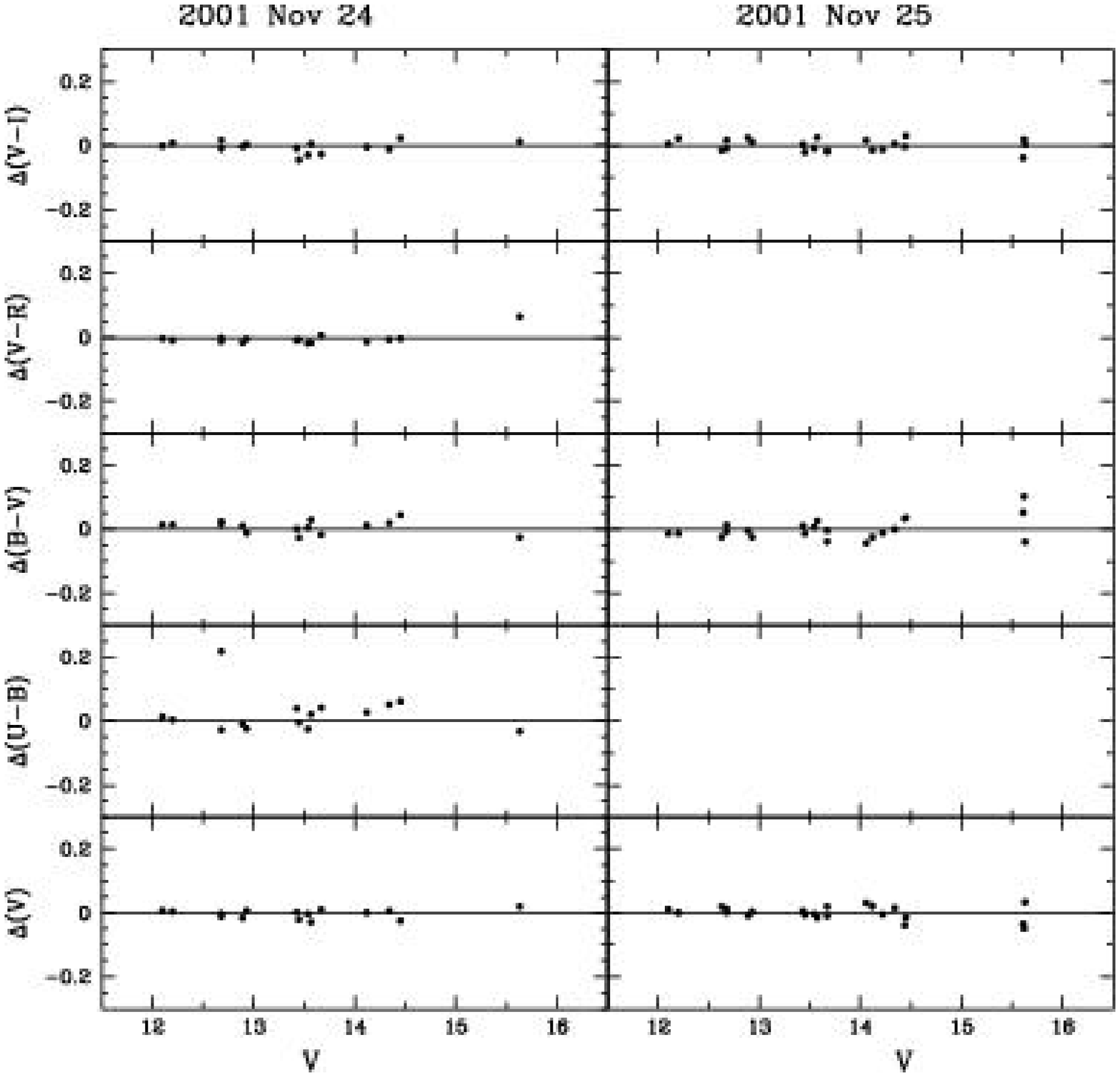}
}
\caption{ Residuals between standard and transformed magnitude and colors of standard 
stars plotted against standard magnitude.}

\end{figure*}

\begin{figure*}[h]
\includegraphics[height=14cm,width=18cm]{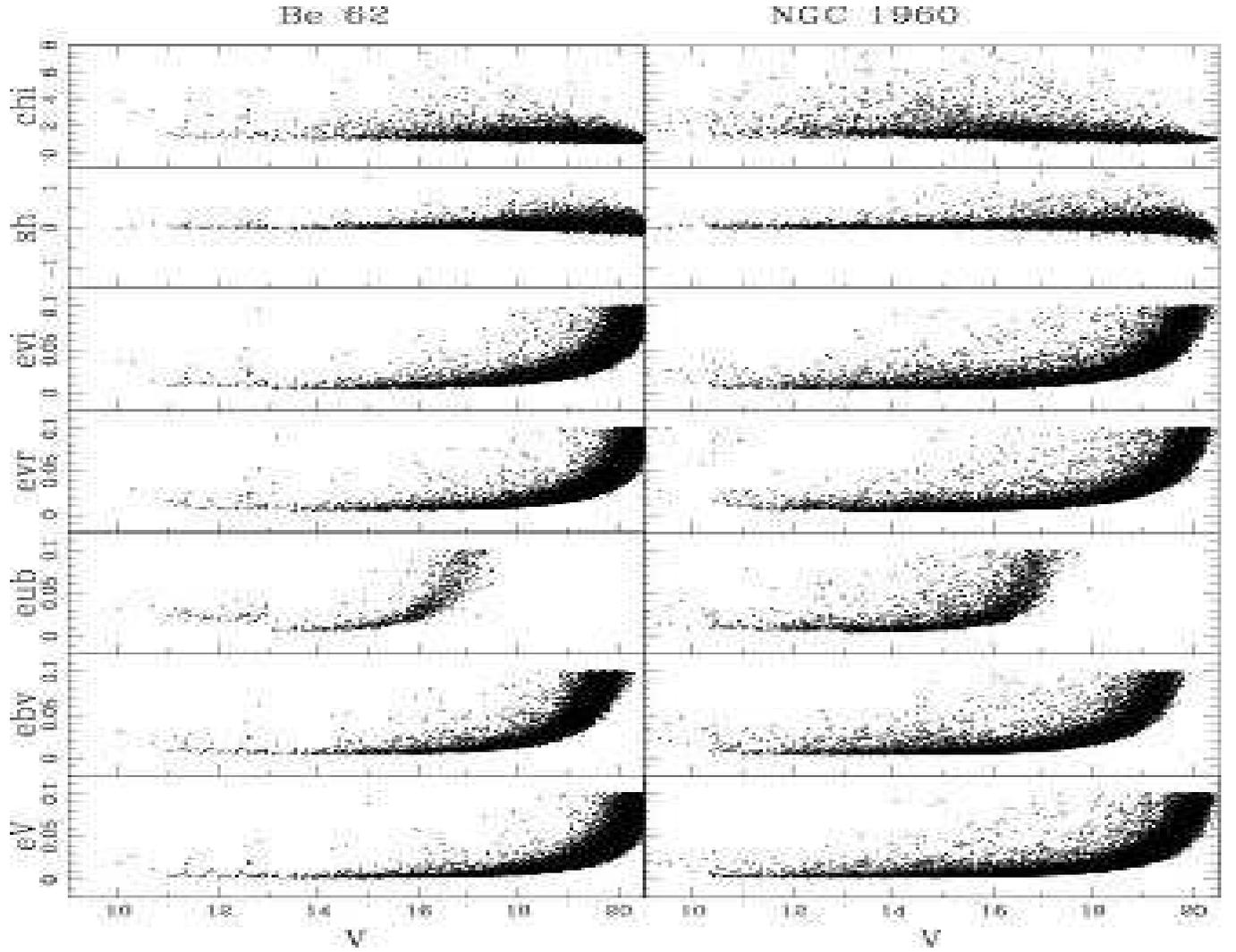}
\caption{ The DAOPHOT errors, image parameter $\chi$ and sharpness as a function of $V$ 
magnitude for the measurements in the case of Be 62 and NGC 1960.}
\end{figure*}

\begin{figure*}[h]
\includegraphics[height=14cm,width=18cm]{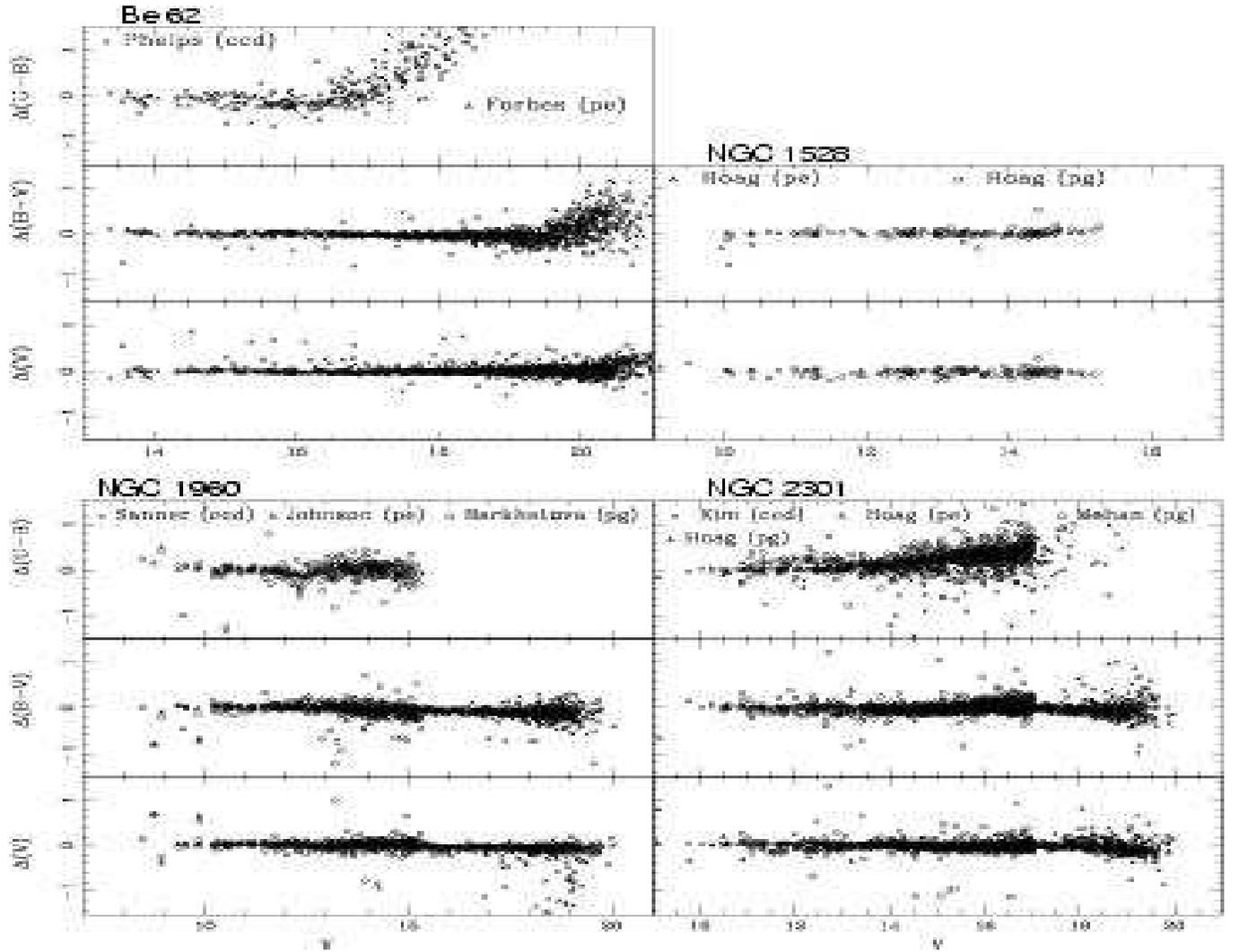}
\caption{Comparison of the present CCD photometry with the data available in the literature. The difference $\Delta$
(literature - present data) as a function of $V$ mag for all the target clusters is shown in figure.
In the figure pe, pg, ccd indicate for photoelectric, photographic and CCD photometry respectively.}
\end{figure*}

\setcounter{figure}{2}

\begin{figure*}[h]
\includegraphics[height=14cm,width=18cm]{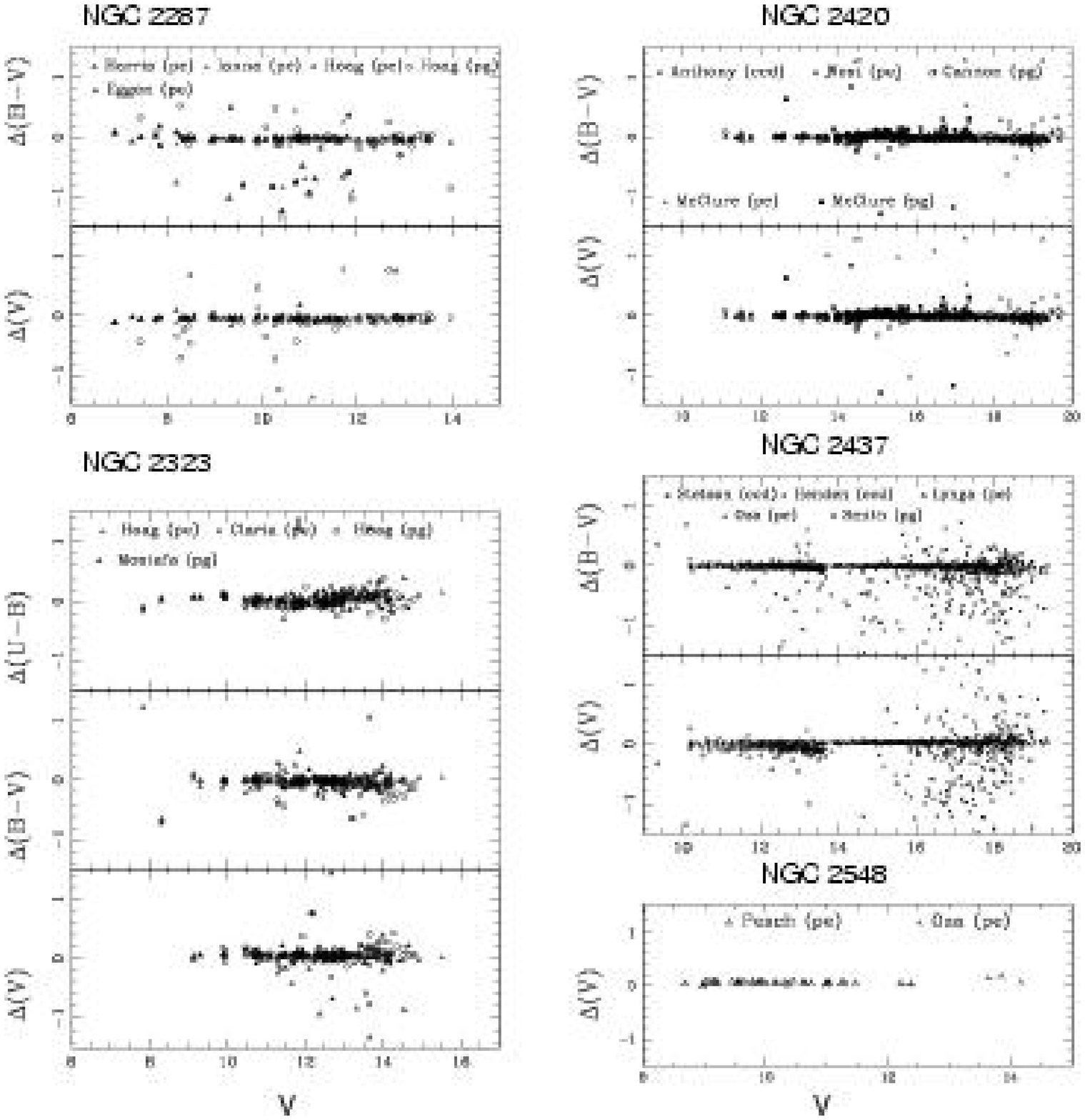}
\caption{Continued}
\end{figure*}

\begin{figure*}[h]
\hbox{
\includegraphics[height=7cm,width=9cm]{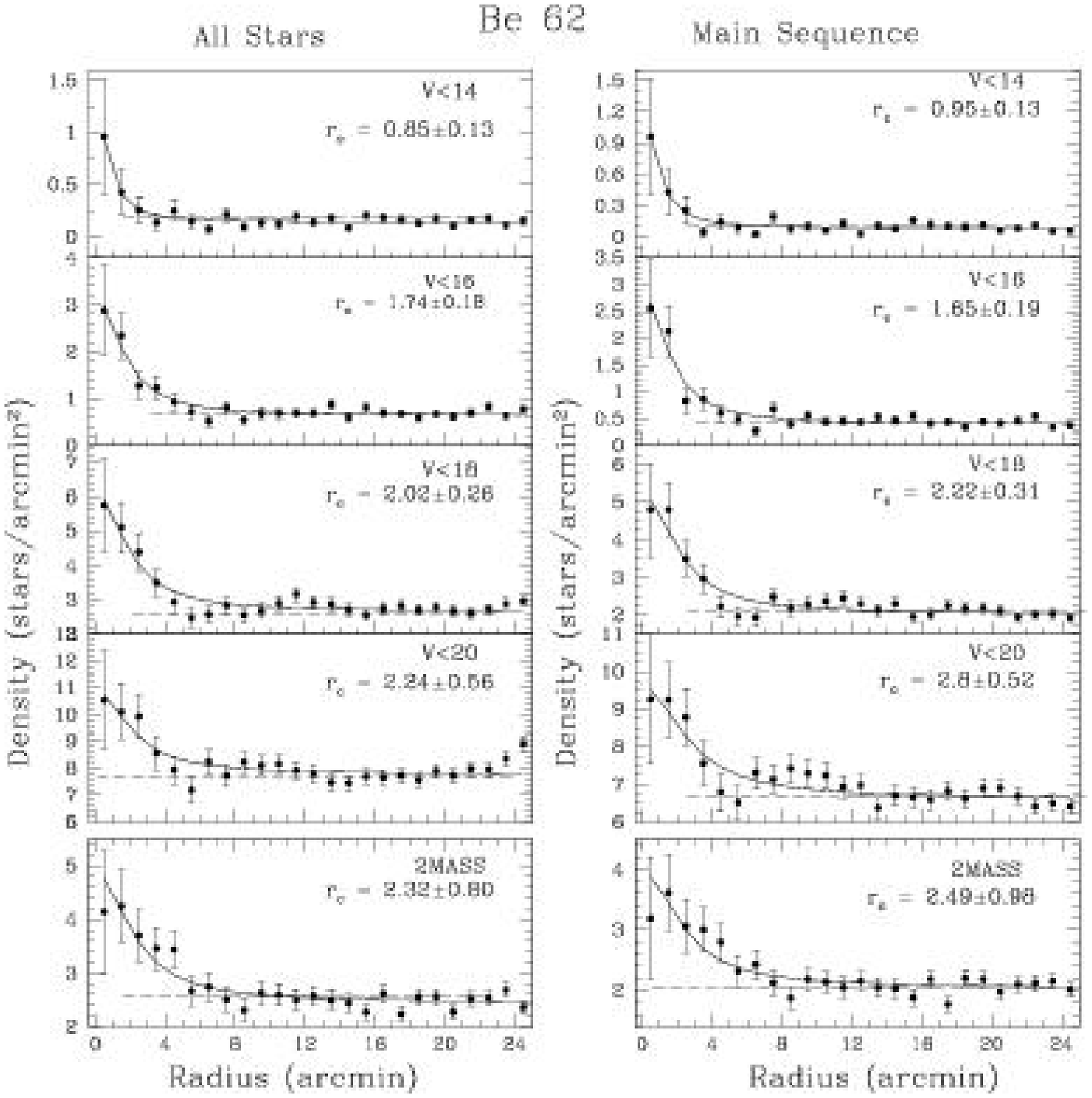}
\includegraphics[height=7cm,width=9cm]{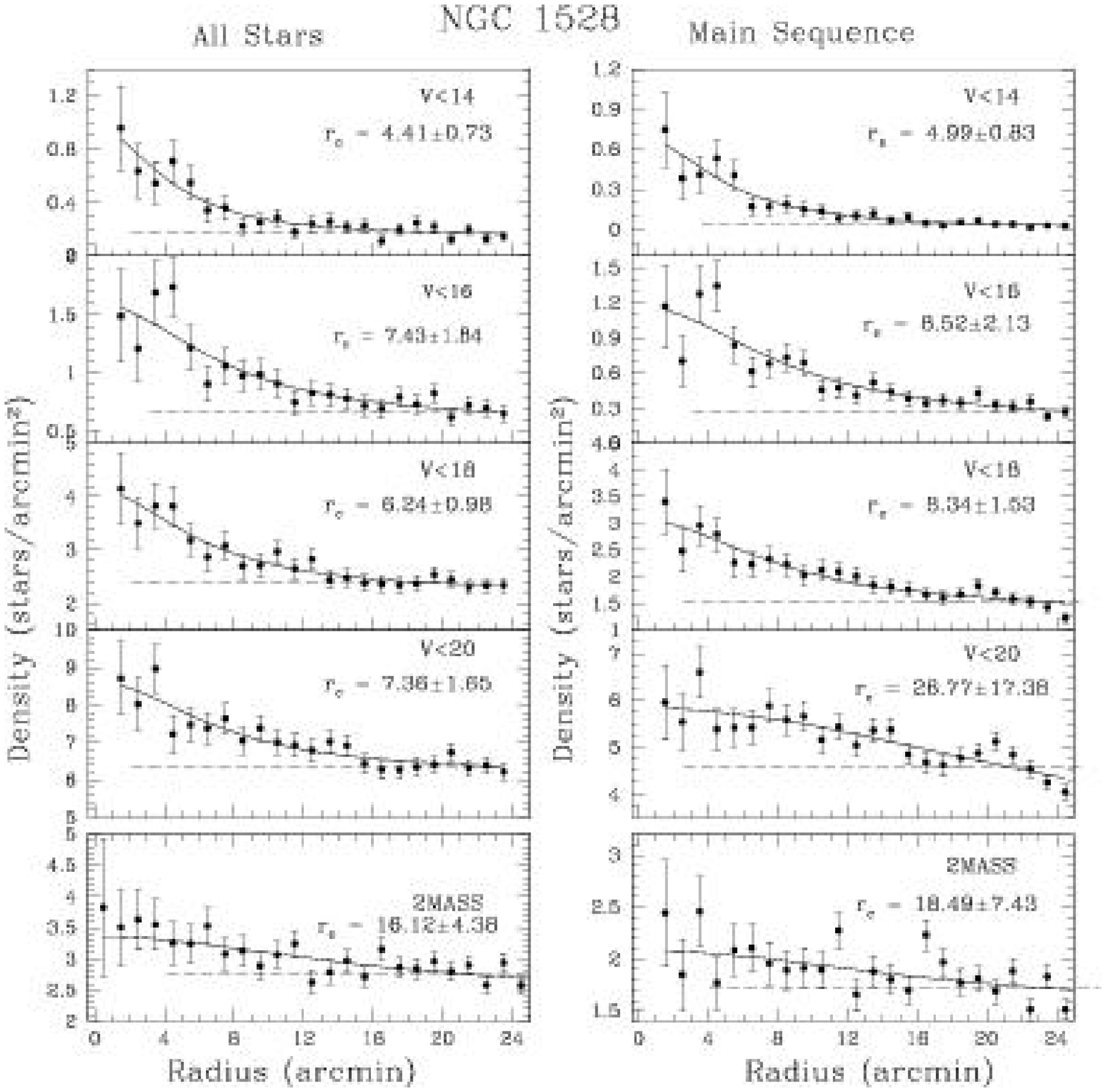}
}
\hbox{
\includegraphics[height=7cm,width=9cm]{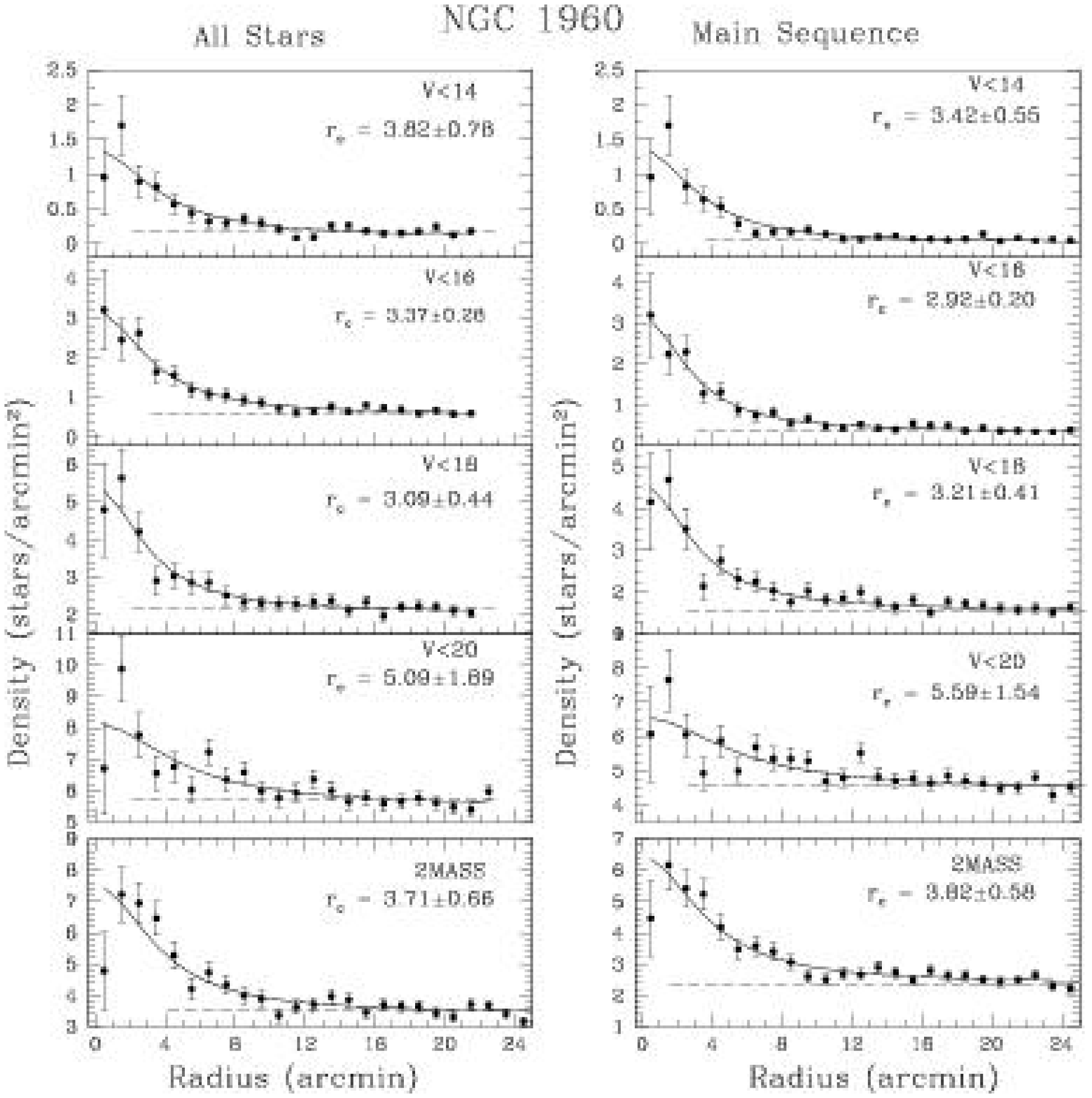}
\includegraphics[height=7cm,width=9cm]{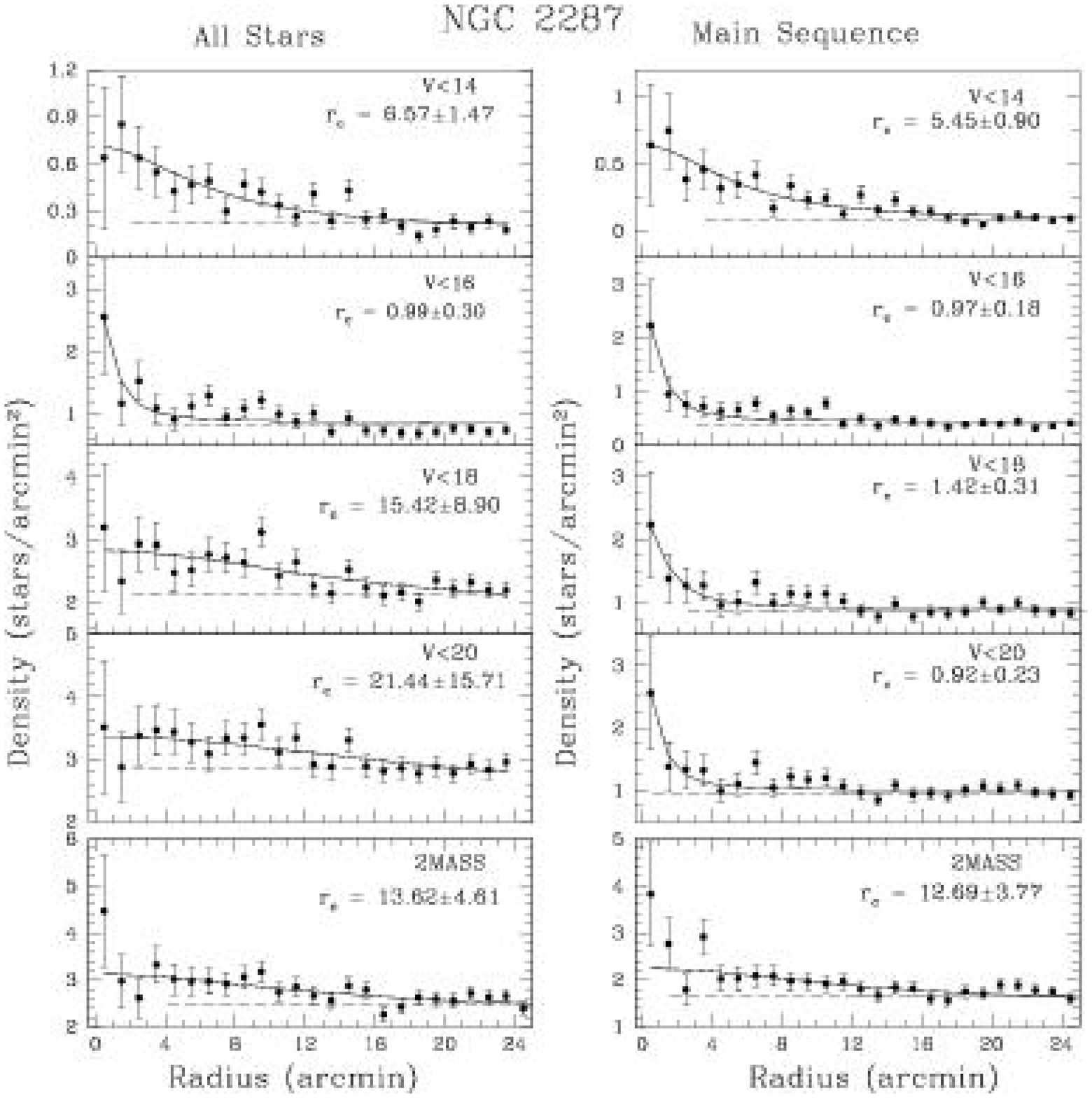}
}
\caption{Radial density profile of clusters Be 62, NGC 1528, NGC 1960 and NGC 2287 for different magnitude levels using present optical and 2MASS data. Solid curve shows a least square fit of the King (1962) profile to the observed data points (see text). The error bars represent $1\over {\sqrt{(N)}}$ errors. The dashed line indicates the density of field stars.}
\end{figure*}

\setcounter{figure}{3}

\begin{figure*}[h]
\hbox{
\includegraphics[height=7cm,width=9cm]{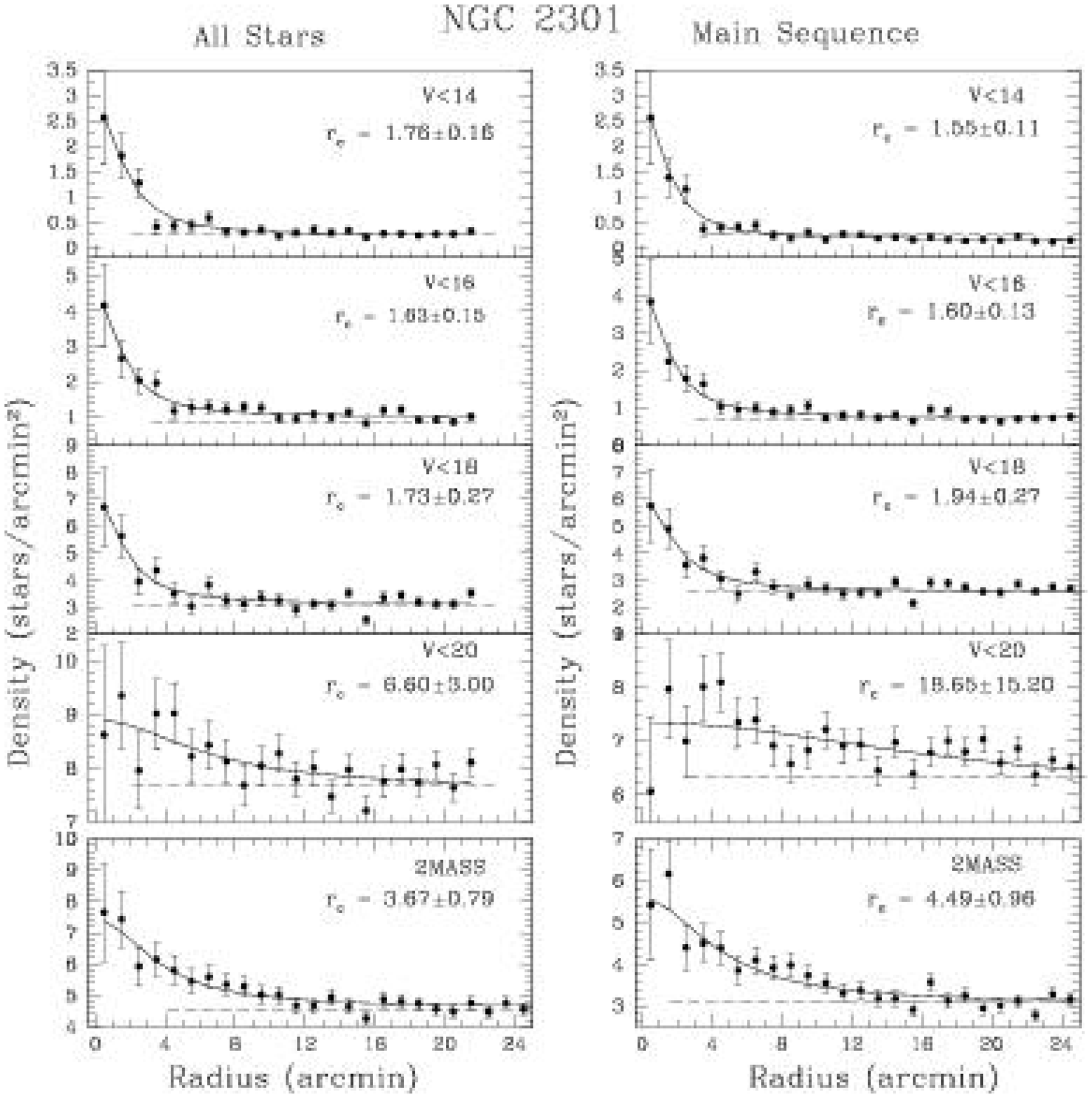}
\includegraphics[height=7cm,width=9cm]{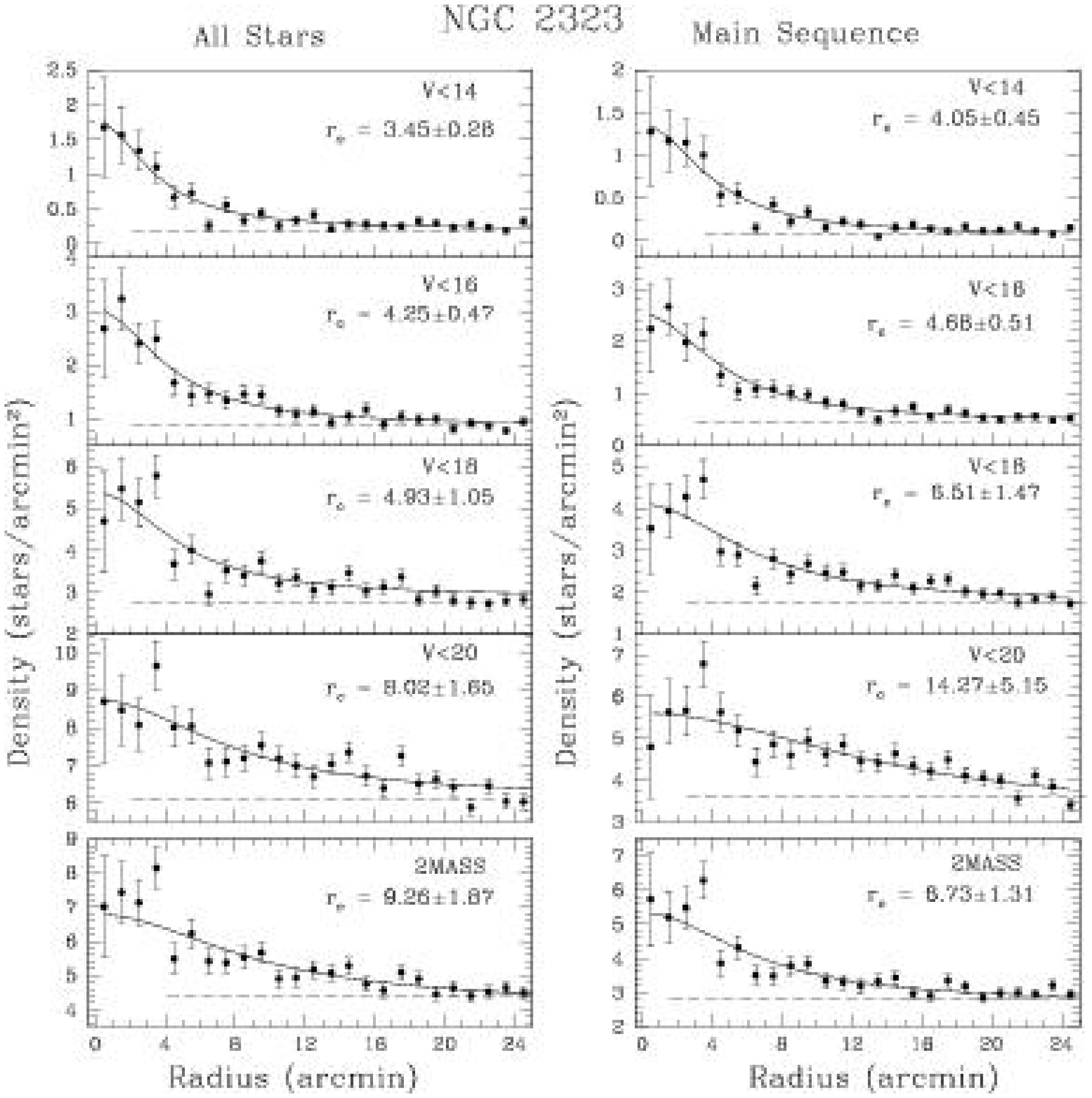}
}
\hbox{
\includegraphics[height=7cm,width=9cm]{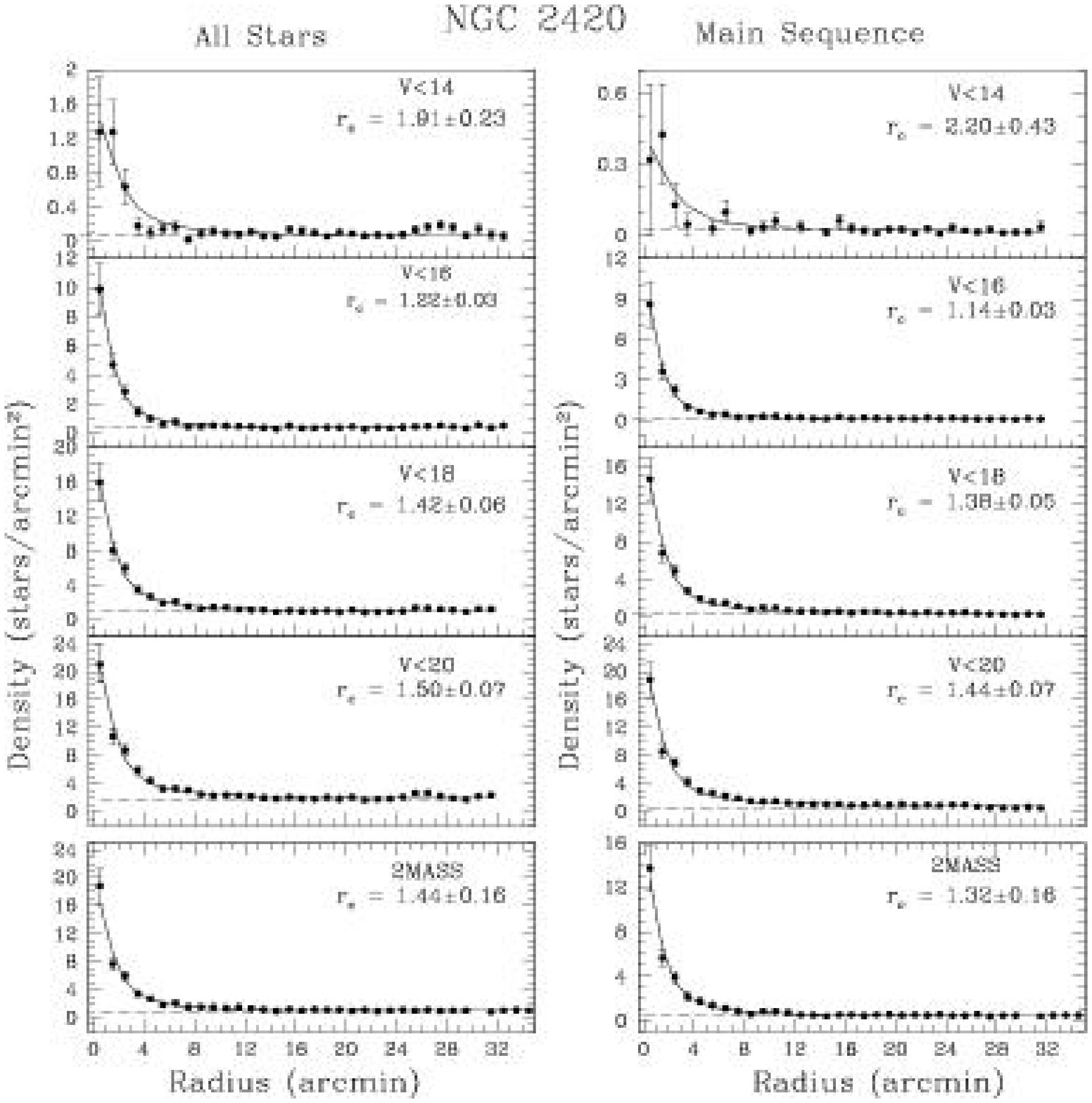}
\includegraphics[height=7cm,width=9cm]{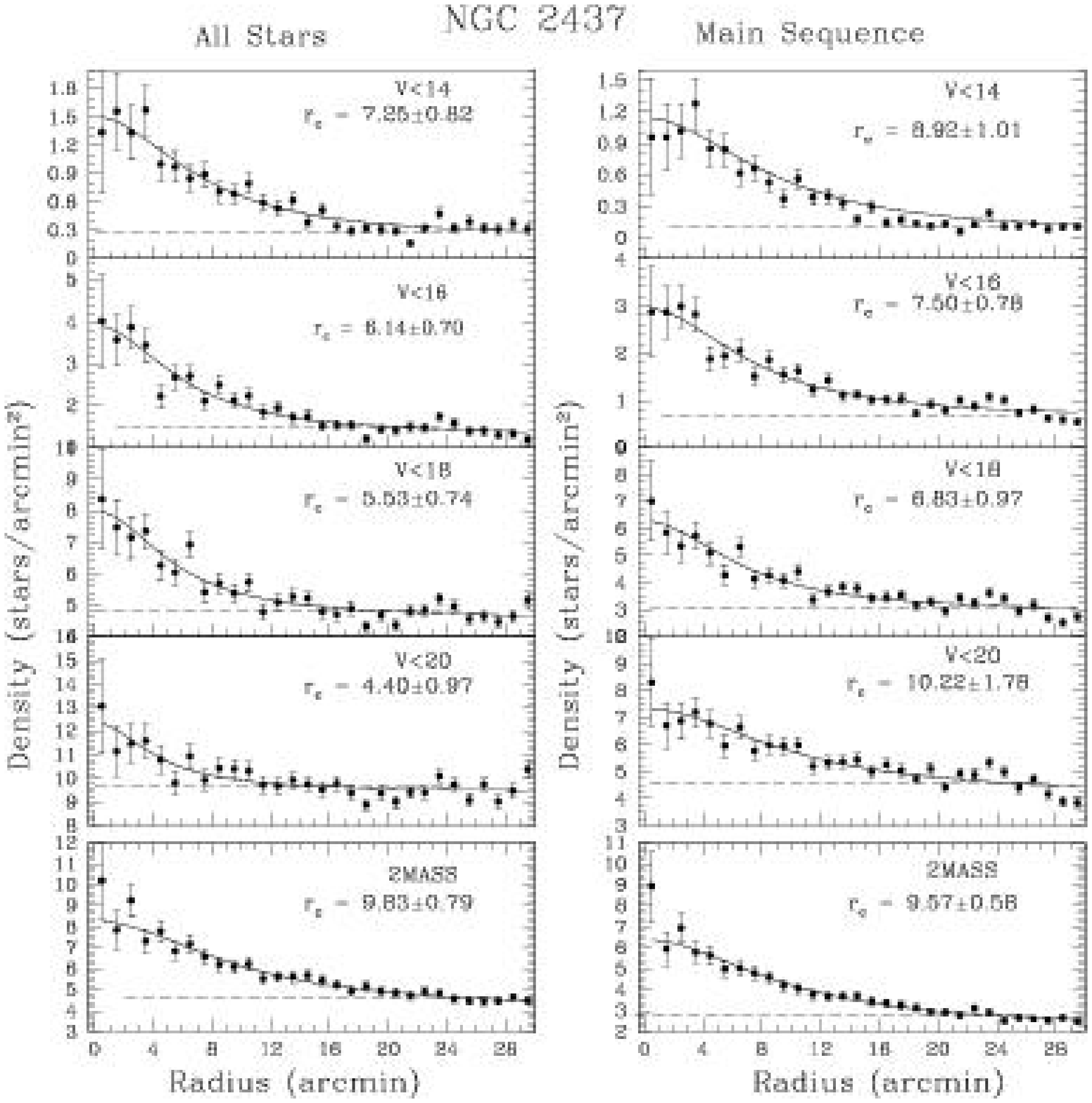}
}
\includegraphics[height=7cm,width=9cm]{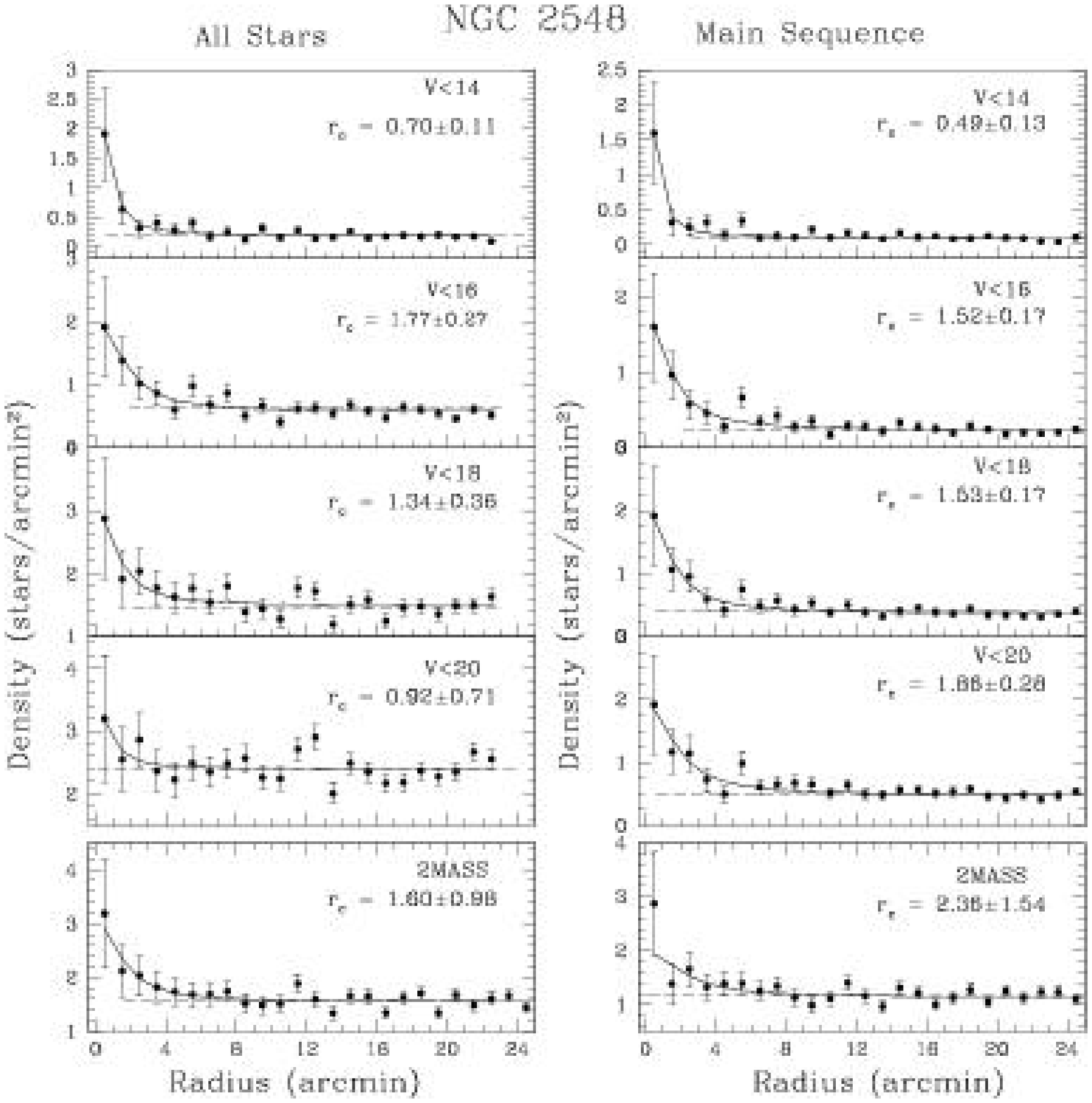}
\caption{Continued for clusters NGC 2301, NGC 2323, NGC 2420, NGC 2437 and NGC 2548}
\end{figure*}

\clearpage

\begin{figure*}[h]
\hbox{
\includegraphics[height=5.5cm,width=5.5cm]{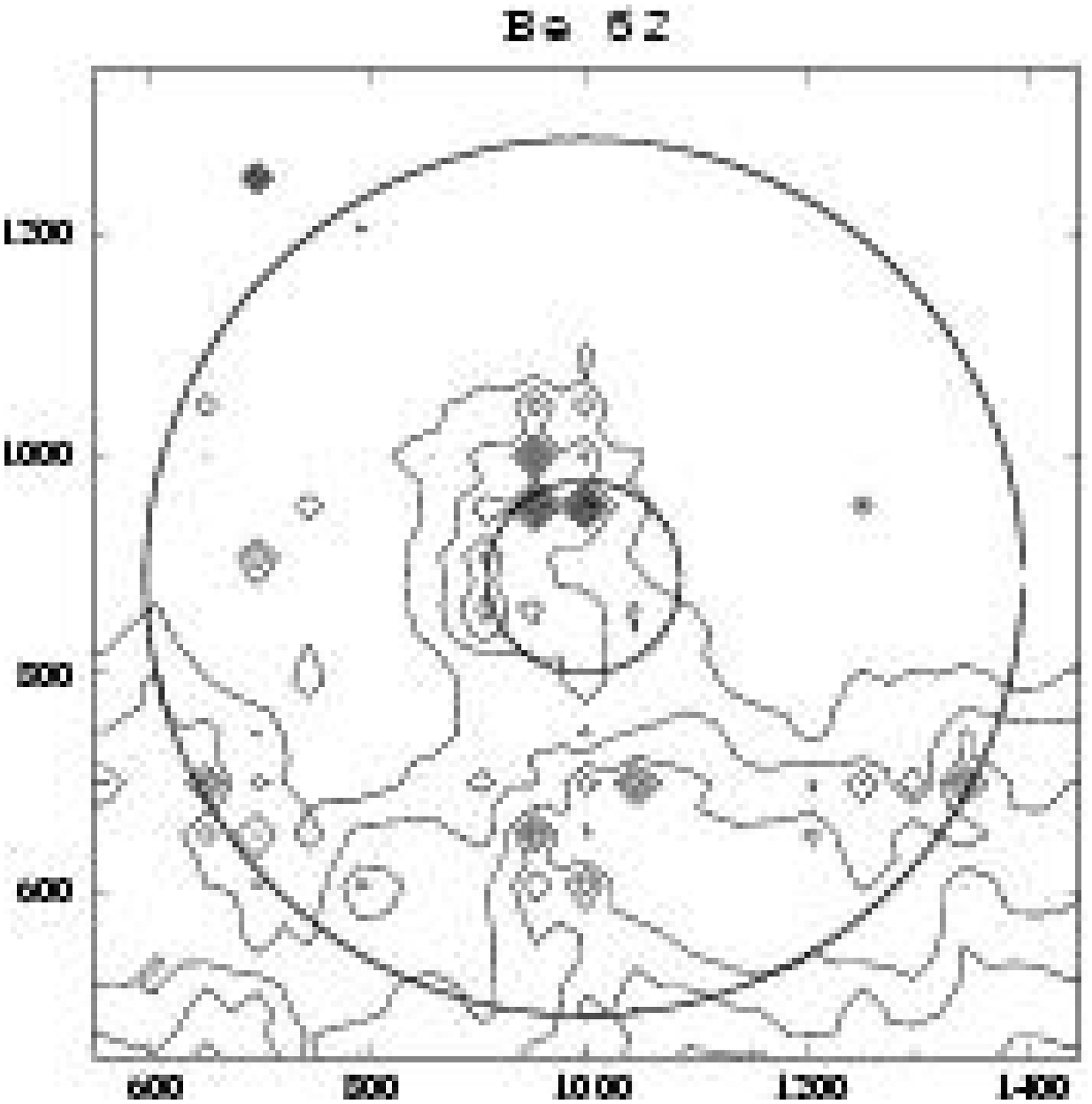}
\includegraphics[height=5.5cm,width=5.5cm]{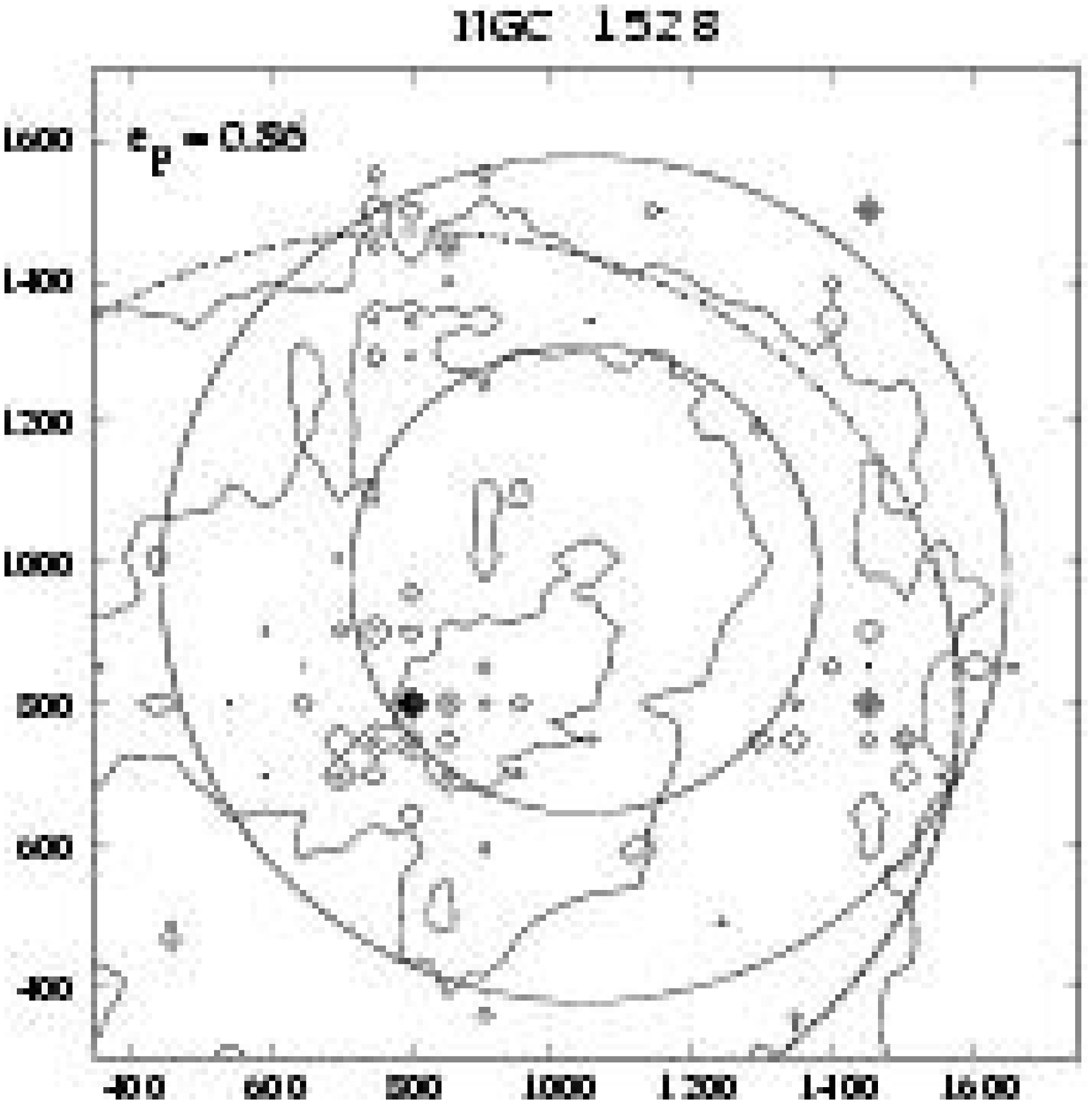}
\includegraphics[height=5.5cm,width=5.5cm]{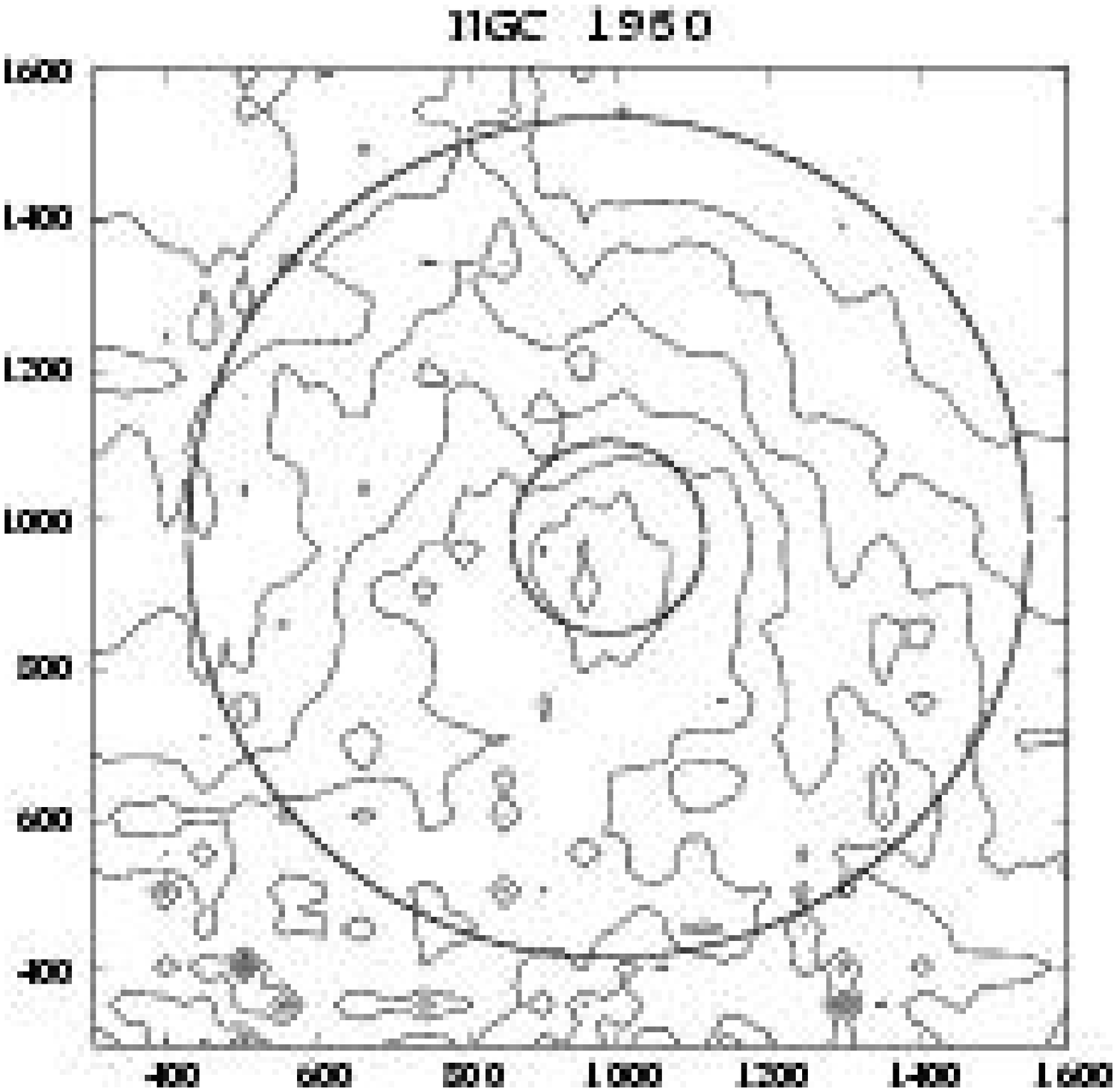}
}
\vspace{0.4cm}
\hbox{
\includegraphics[height=5.5cm,width=5.5cm]{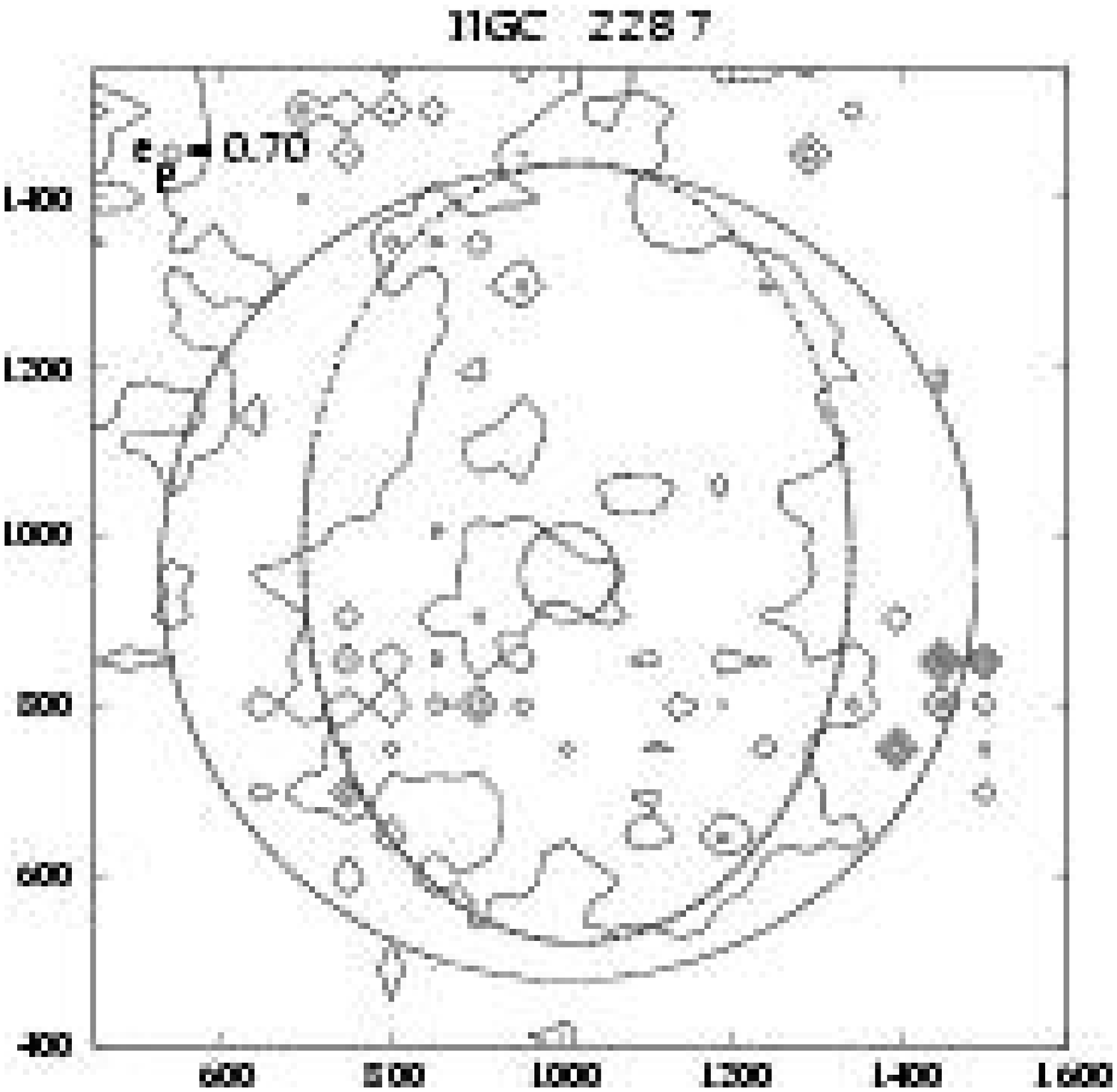}
\includegraphics[height=5.5cm,width=5.5cm]{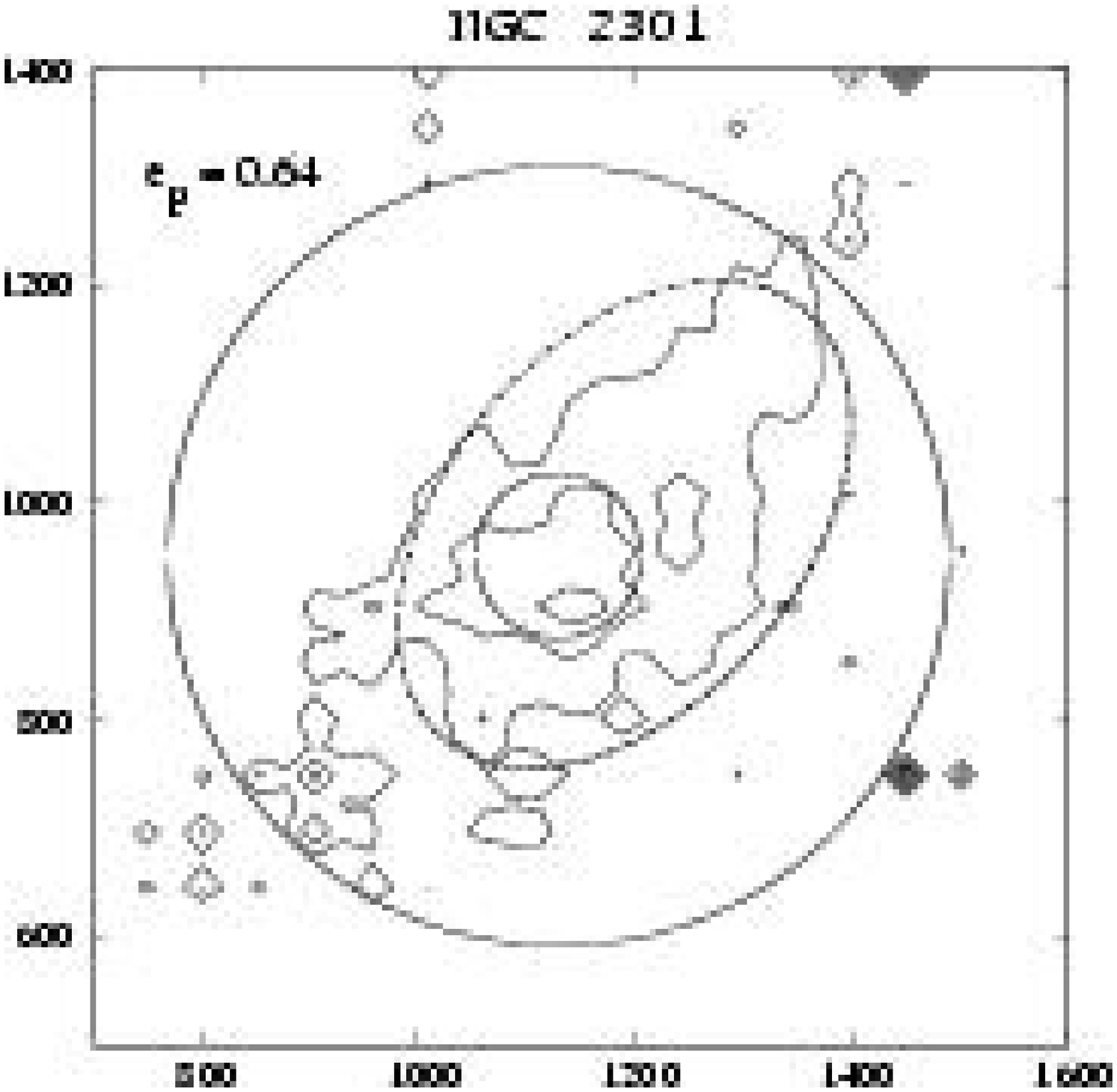}
\includegraphics[height=5.5cm,width=5.5cm]{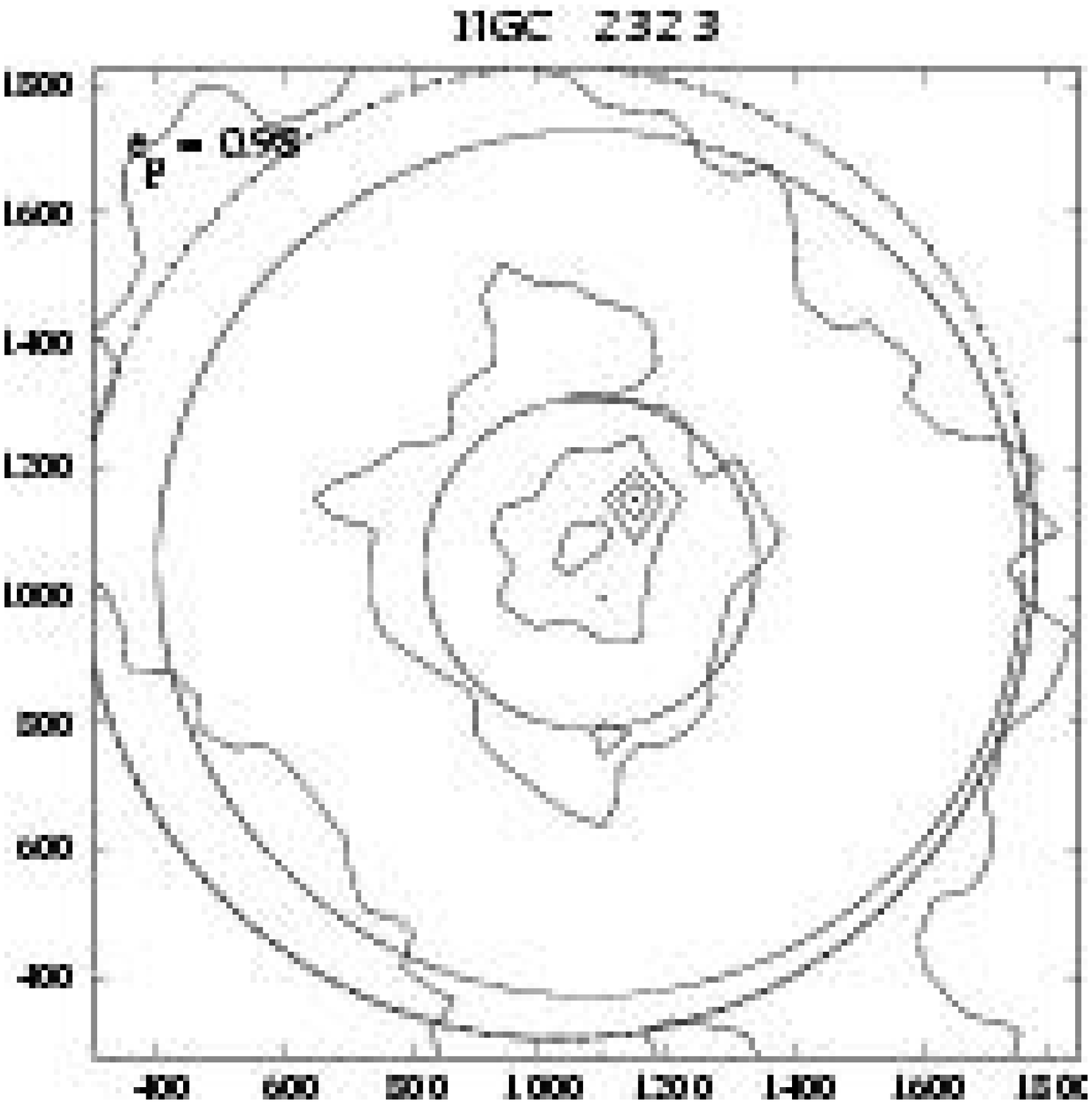}
}
\vspace{0.4cm}
\hbox{
\includegraphics[height=5.5cm,width=5.5cm]{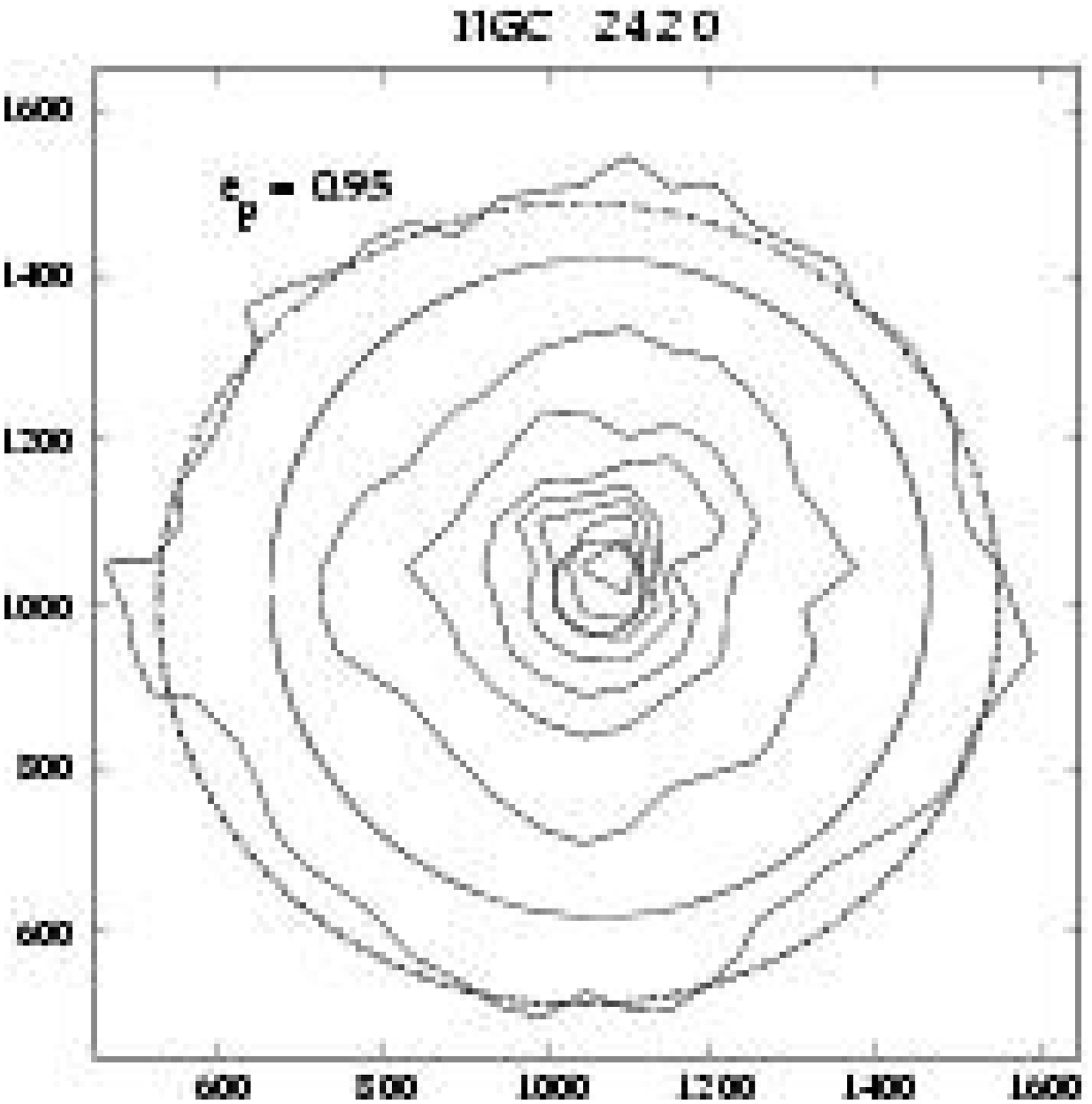}
\includegraphics[height=5.5cm,width=5.5cm]{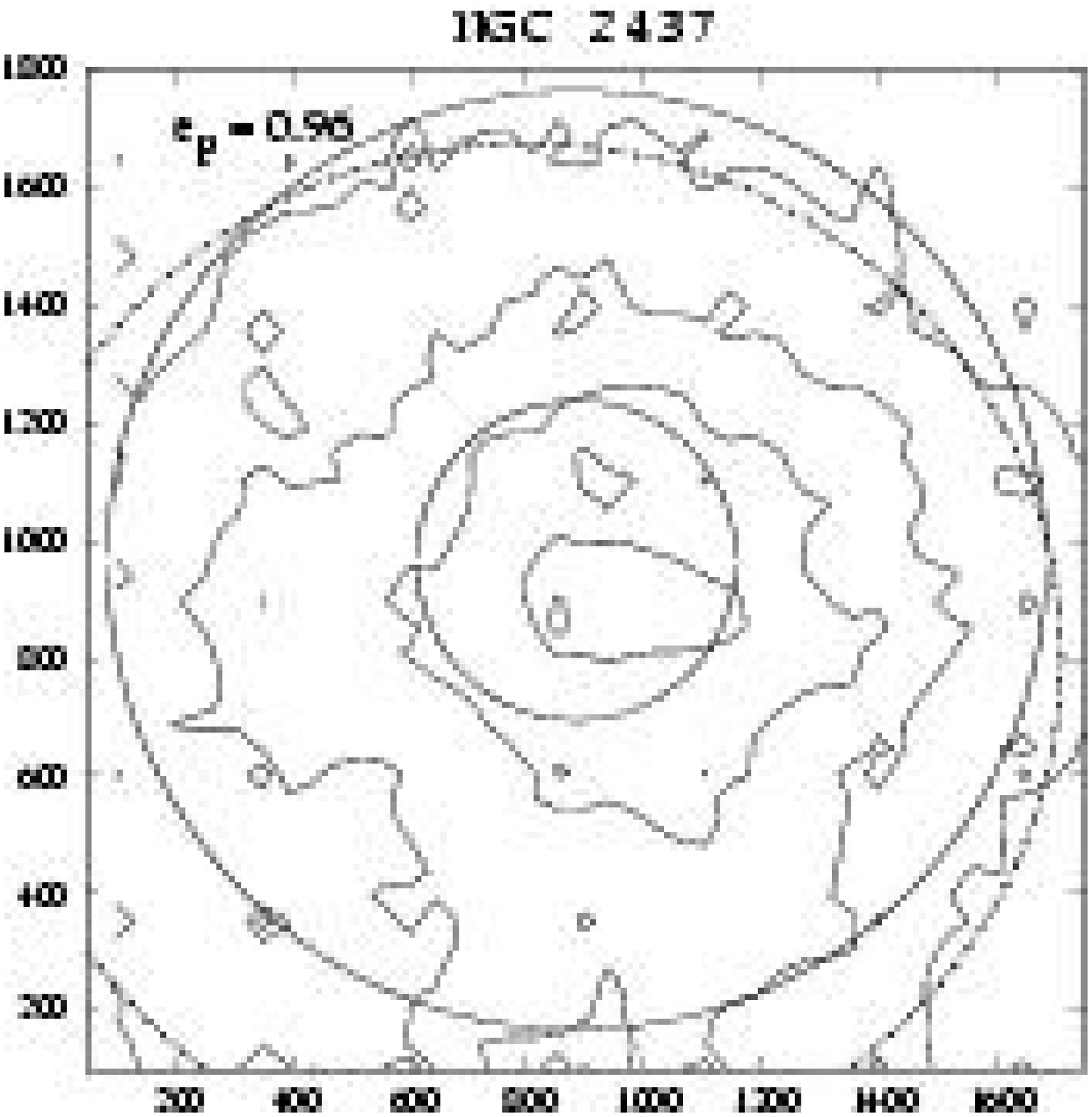}
\includegraphics[height=5.5cm,width=5.5cm]{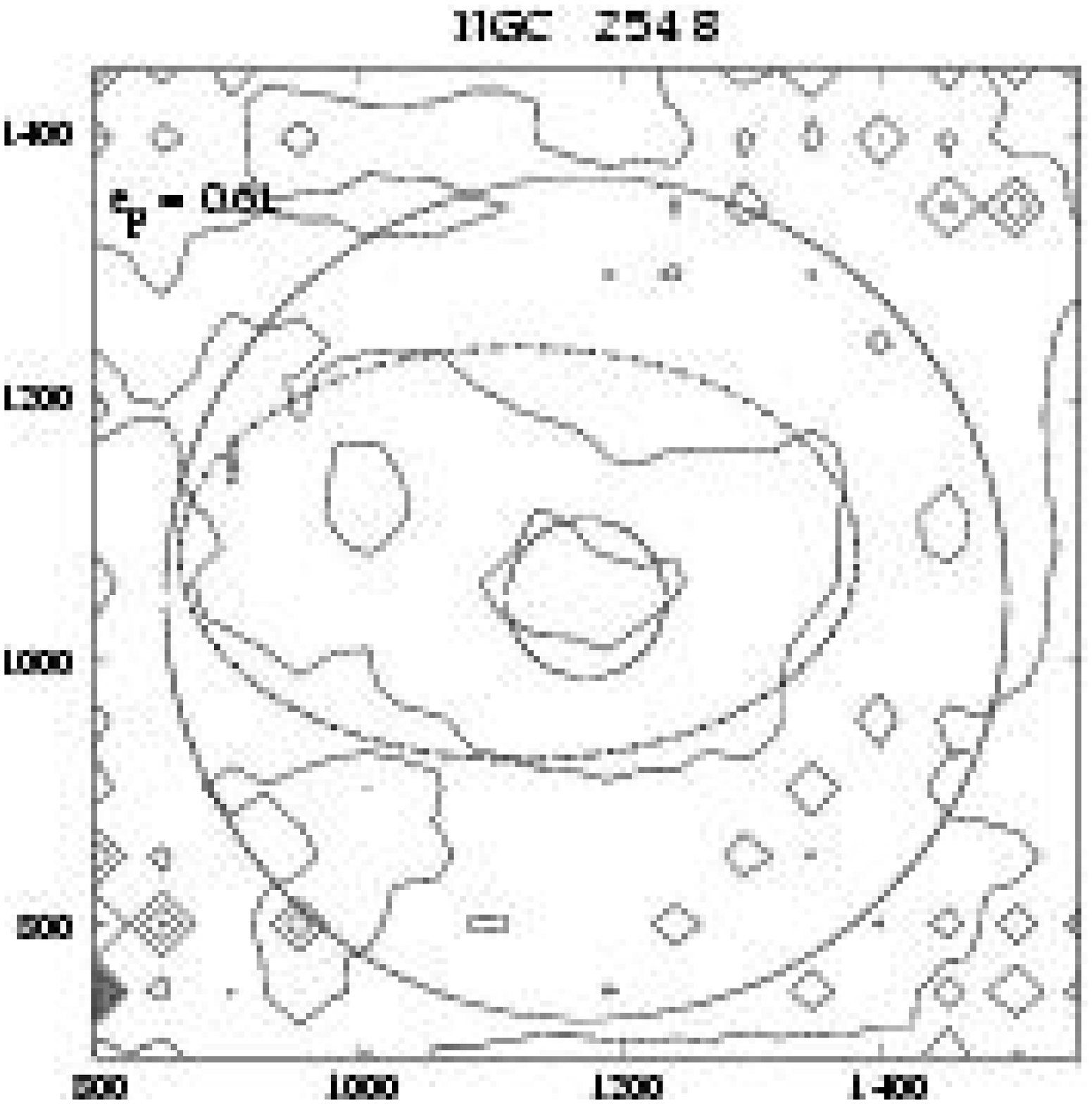}
}
\caption{Isodensity contours drawn for the target clusters. The X and Y axis are in pixels.
The inner and outer circles represent the core and cluster radius as obtained in the present study.
Dashed curve shows the least square fitted curve to the outer region of the clusters.
$e_p$ is the elongation parameter (see text).}
\label{figure:5}
\end{figure*}

\clearpage

\begin{figure*}
\includegraphics[height=18cm,width=18cm]{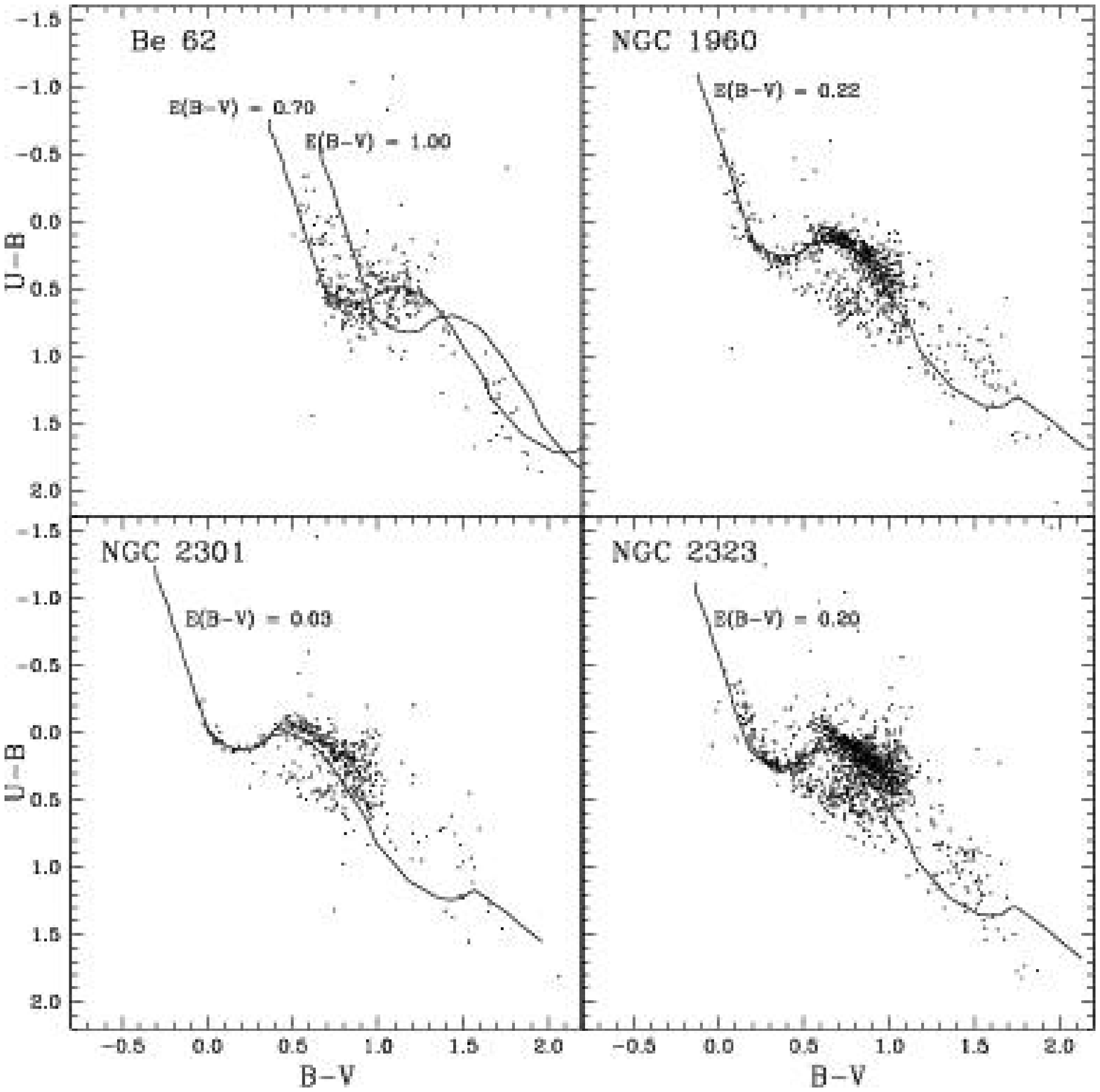}
\caption{The $(U-B)/(B-V)$ color-color diagram for the stars within the region 
$ r \le r_{cl}$ of clusters Be 62, NGC 1960, NGC 2301 
and NGC 2323. The continuous curve represents intrinsic MS for $Z=0.02$ by 
Schmidt-Kaler (1982) shifted along the reddening vector of 0.72.}
\end{figure*}

\begin{figure*}
\includegraphics[height=18cm,width=14cm,angle=-90]{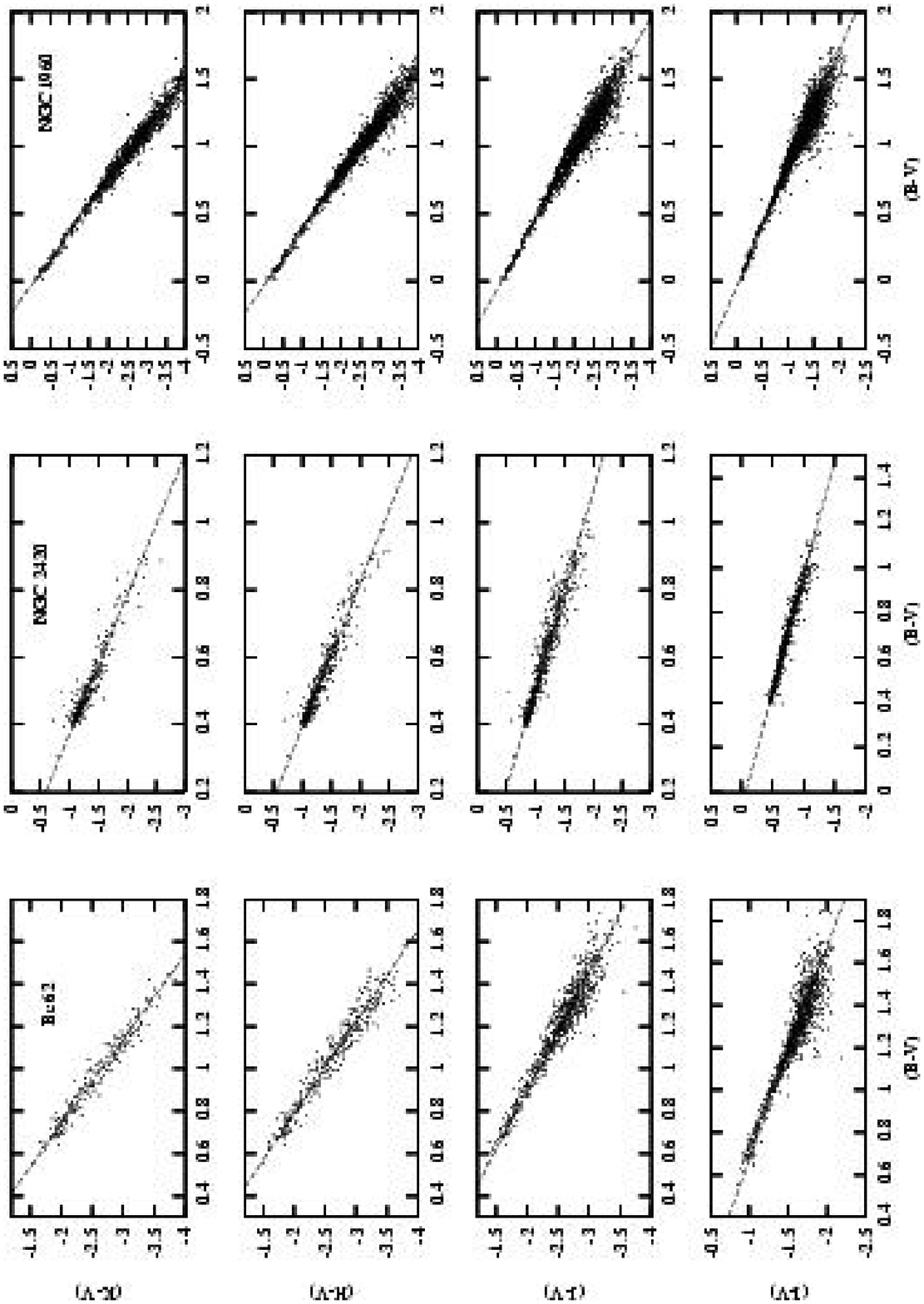}
\caption{$(I-V), (J-V), (H-V), (K-V)$ vs.  $(B-V)$ two color diagrams, within the
cluster region ($r\le r_{cl}$), for the clusters Be 62, NGC 2420 and NGC 1960.
Straight line shows a least square fit to the data. }
\end{figure*}

\begin{figure*}
\includegraphics[height=18cm,width=18cm]{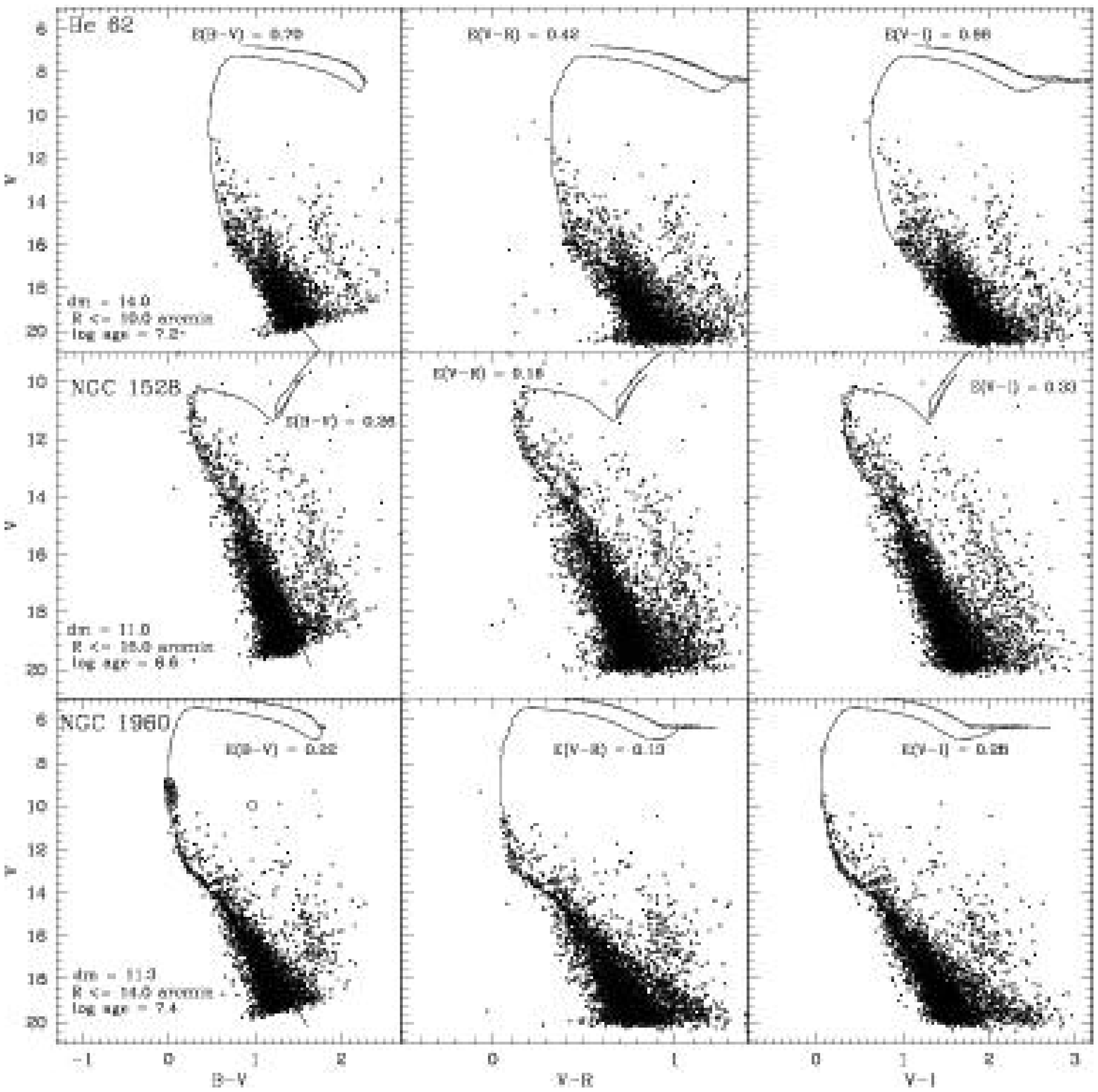}
\caption{The $V$ vs. $(B-V), (V-R), (V-I)$ CMDs for clusters Be 62, NGC 1528 and NGC 1960. The isochrones by Bertelli et al. (1994) for solar metallicity and indicated logarithmic age are also shown. The open circle represents photoelectric data taken from literature.}
\end{figure*}

\setcounter{figure}{7}

\begin{figure*}
\includegraphics[height=18cm,width=18cm]{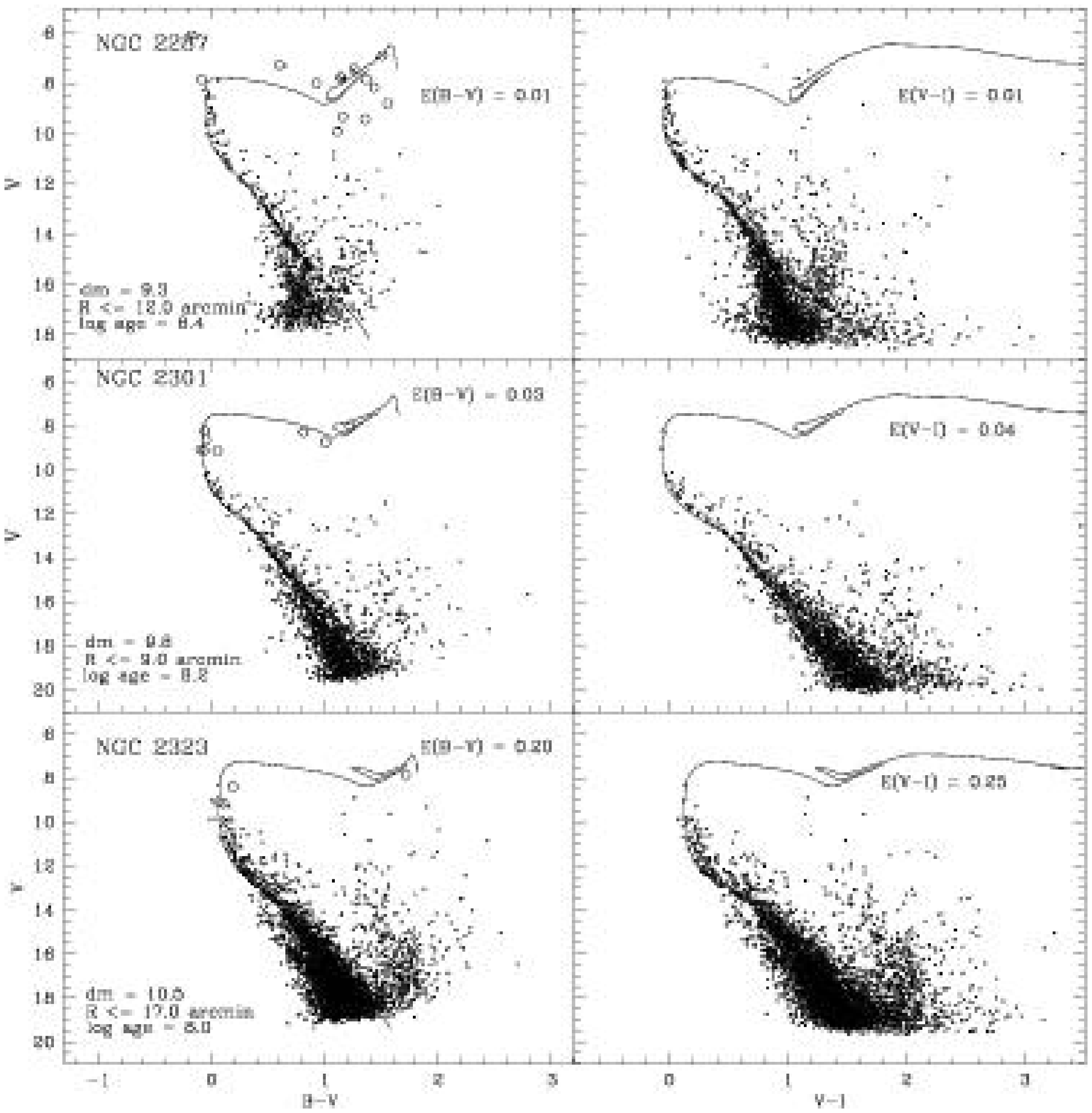}
\caption{Continued - The $V$ vs. $(B-V), (V-I)$ CMDs for clusters NGC 2323, NGC 2301 and NGC 2287.}
\end{figure*}

\setcounter{figure}{7}

\begin{figure*}
\includegraphics[height=18cm,width=18cm]{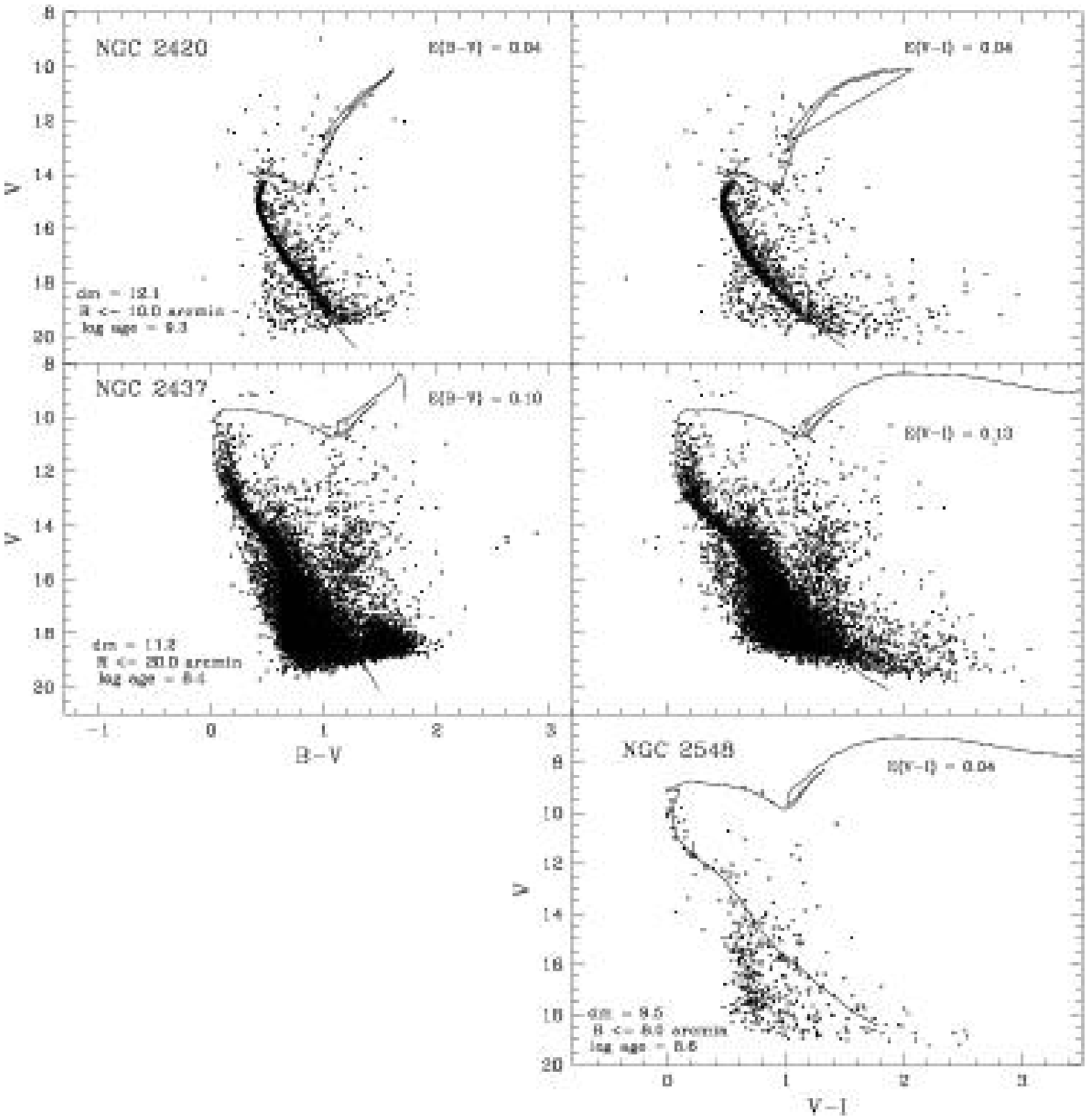}
\caption{Continued - The $V$ vs. $(B-V)$, $(V-I)$ CMDs for clusters NGC 2420, NGC 2437 and NGC 2548.}
\end{figure*}

\begin{figure*}
\includegraphics[height=8cm,width=18cm]{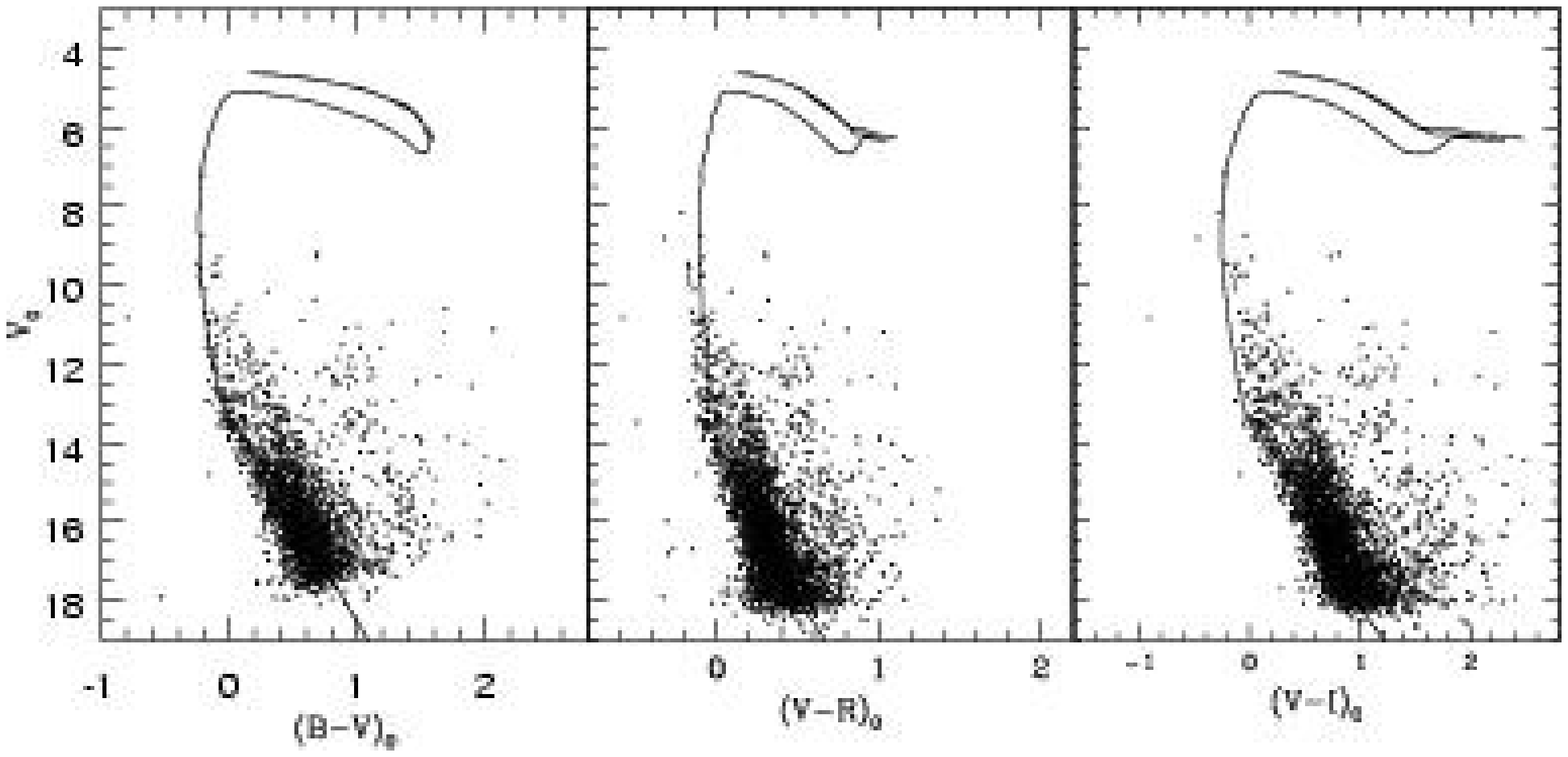}
\caption{The unreddened CMDs for the cluster Be 62. The isochrone by Bertelli et al. (1994) for solar metallicity
 and log age = 7.2 is also shown. }
\end{figure*}

\begin{figure*}
\includegraphics[height=18cm,width=18cm]{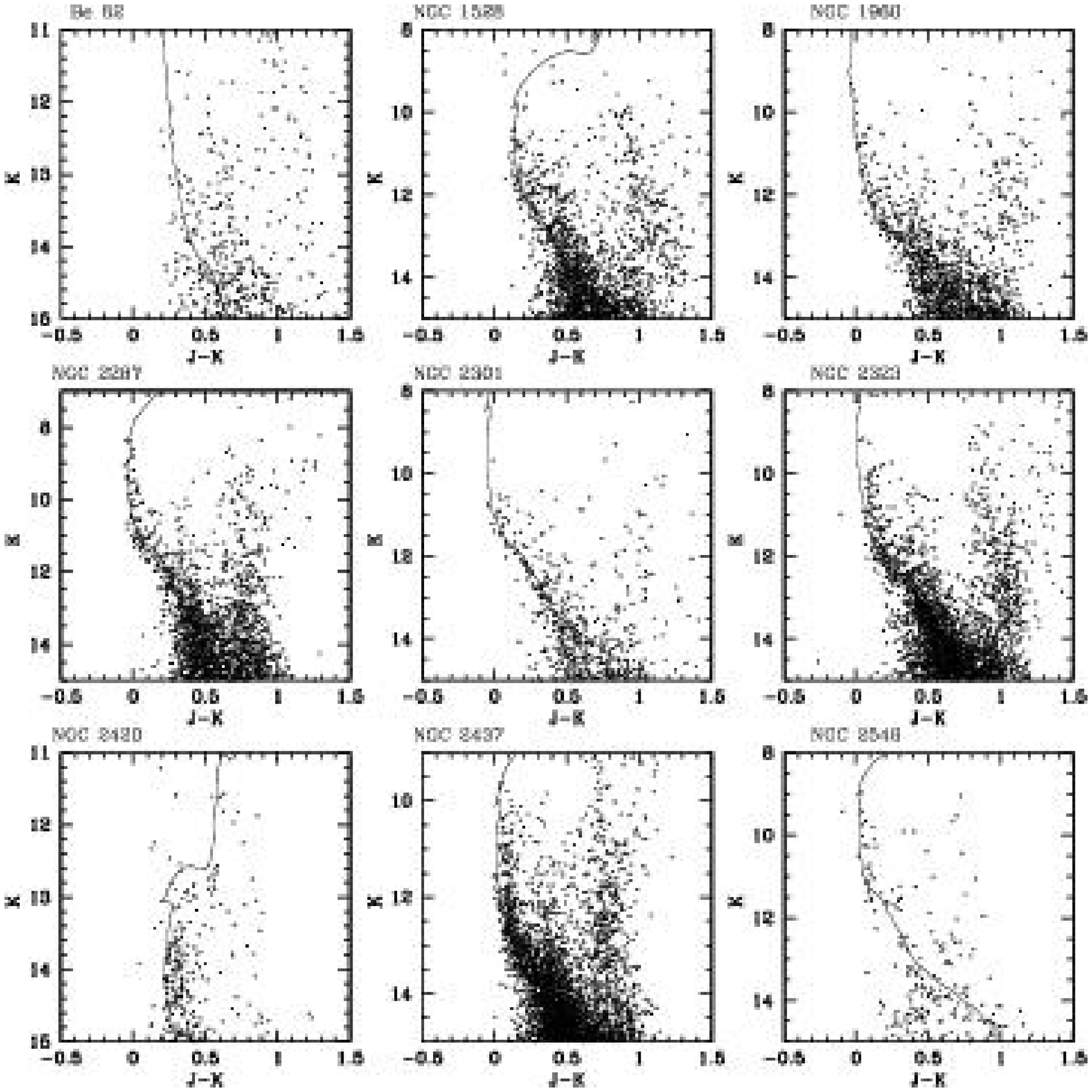}
\caption{$K$ vs. $(J-K)$ CMDs for target clusters obtained from 2MASS data. The isochrones by Bertelli et al. (1994) 
for the age obtained in the present work (cf. Table 8) are also shown. The values of $E(B-V)$
and distance modulus for the clusters has been adopted from Table 8.}
\end{figure*}

\begin{figure*}
\hbox{
\includegraphics[height=9cm,width=9cm]{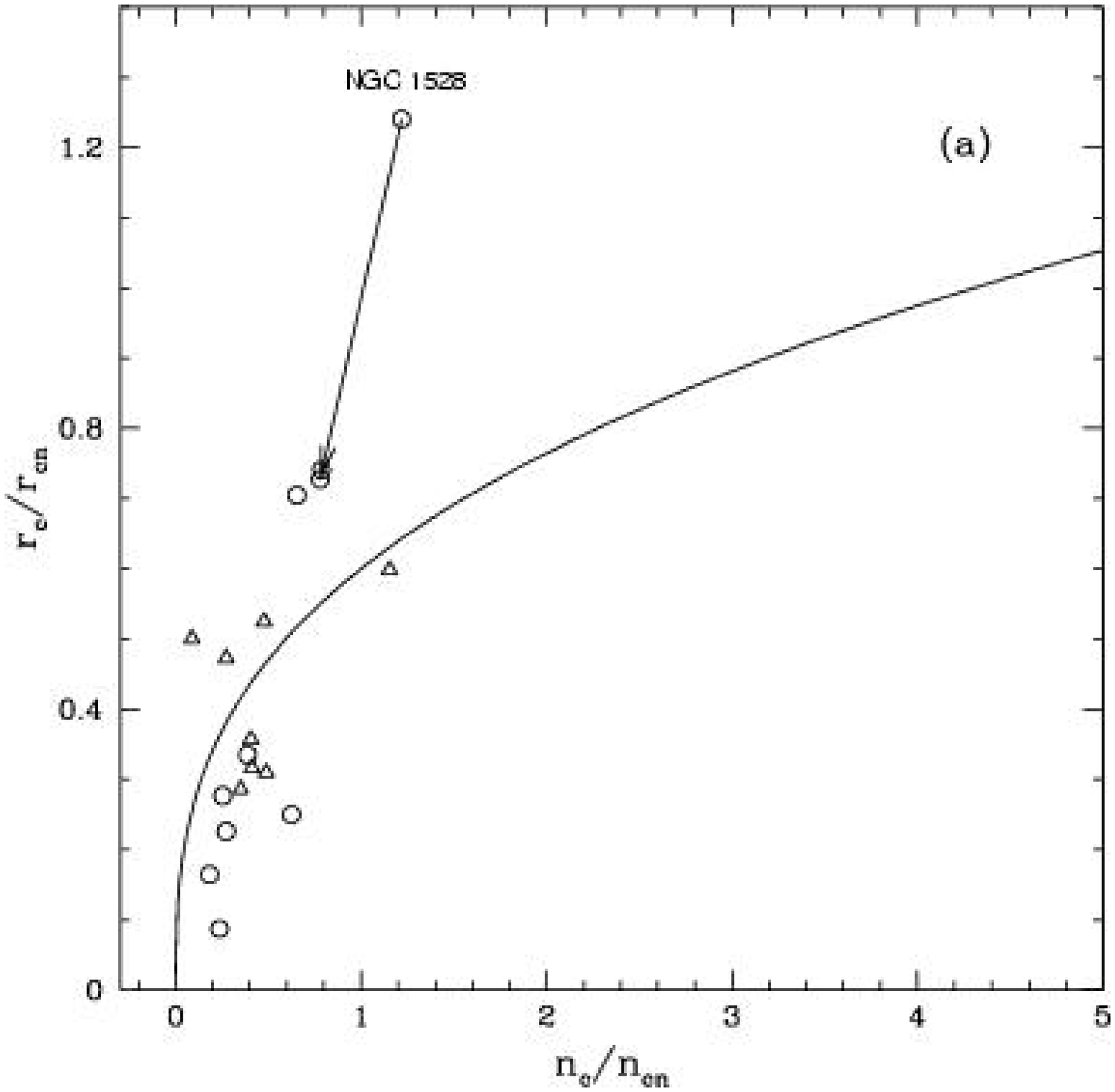}
\includegraphics[height=9cm,width=9cm]{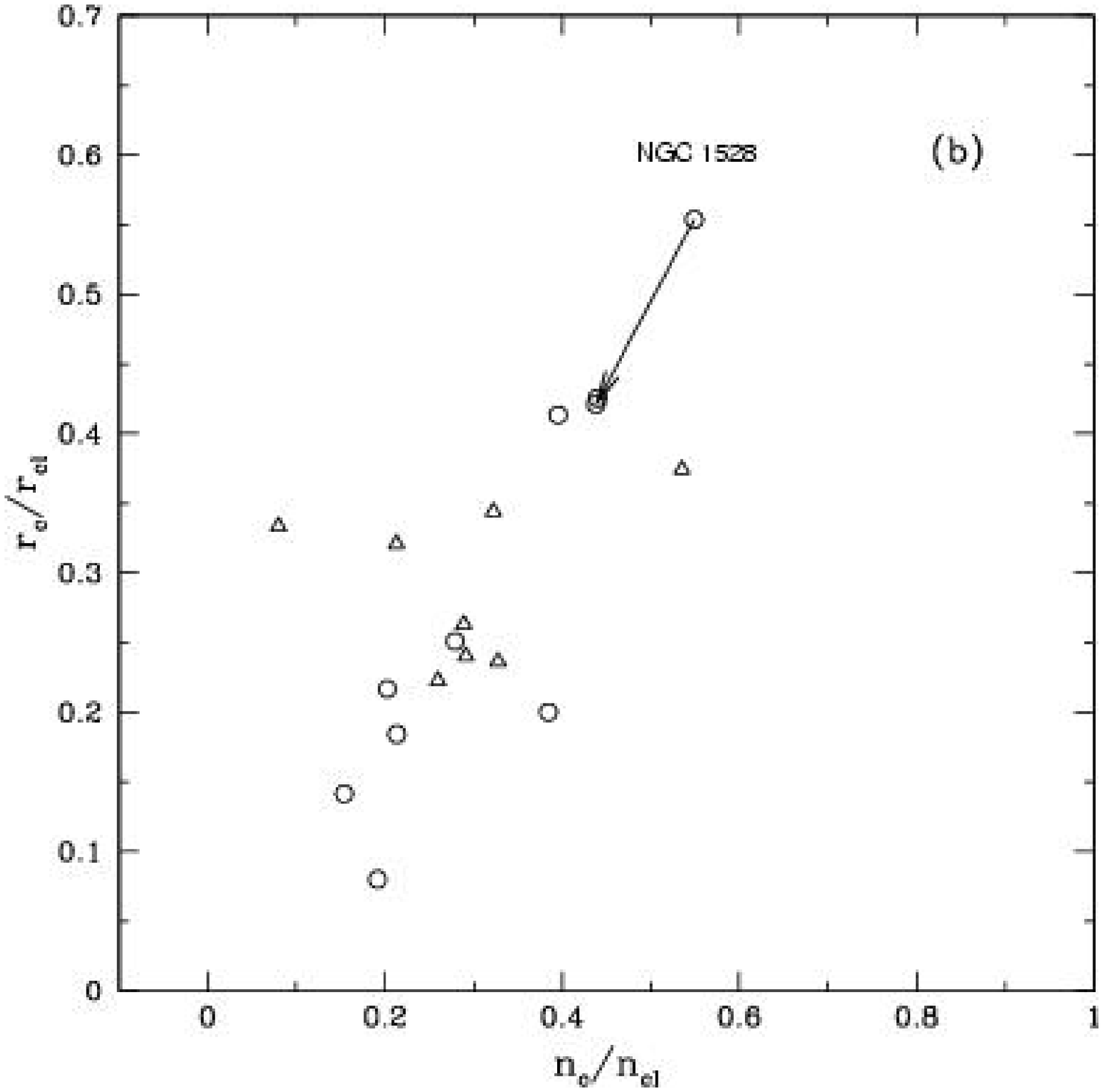}
}
\caption{(a) The $r_c/r_{cn}$ and $n_c/n_{cn}$ diagram and (b) the  $r_c/r_{cl}$ and $n_c/n_{cl}$
diagram for the clusters. Circles and triangle represent data taken from present study
and  previous studies respectively.
The $r_c$, $r_{cn}$ (=$r_{cl}-r_c$) and $r_{cl}$ represents core radius, corona size and cluster extent
respectively and $n_c$, $n_{cn}$ and $n_{cl}$ represents number of stars in the core, corona and
total number of stars in the cluster respectively. The continuous curve represents relation 
$({r_c\over r_{cn}}) \propto({n_c\over n_{cn}})^{0.35}$ (cf. $\S$ 6). 
Arrow indicates the revised location of NGC 1528
(see text).}
\end{figure*}

\begin{figure*}
\includegraphics[height=14cm,width=18cm]{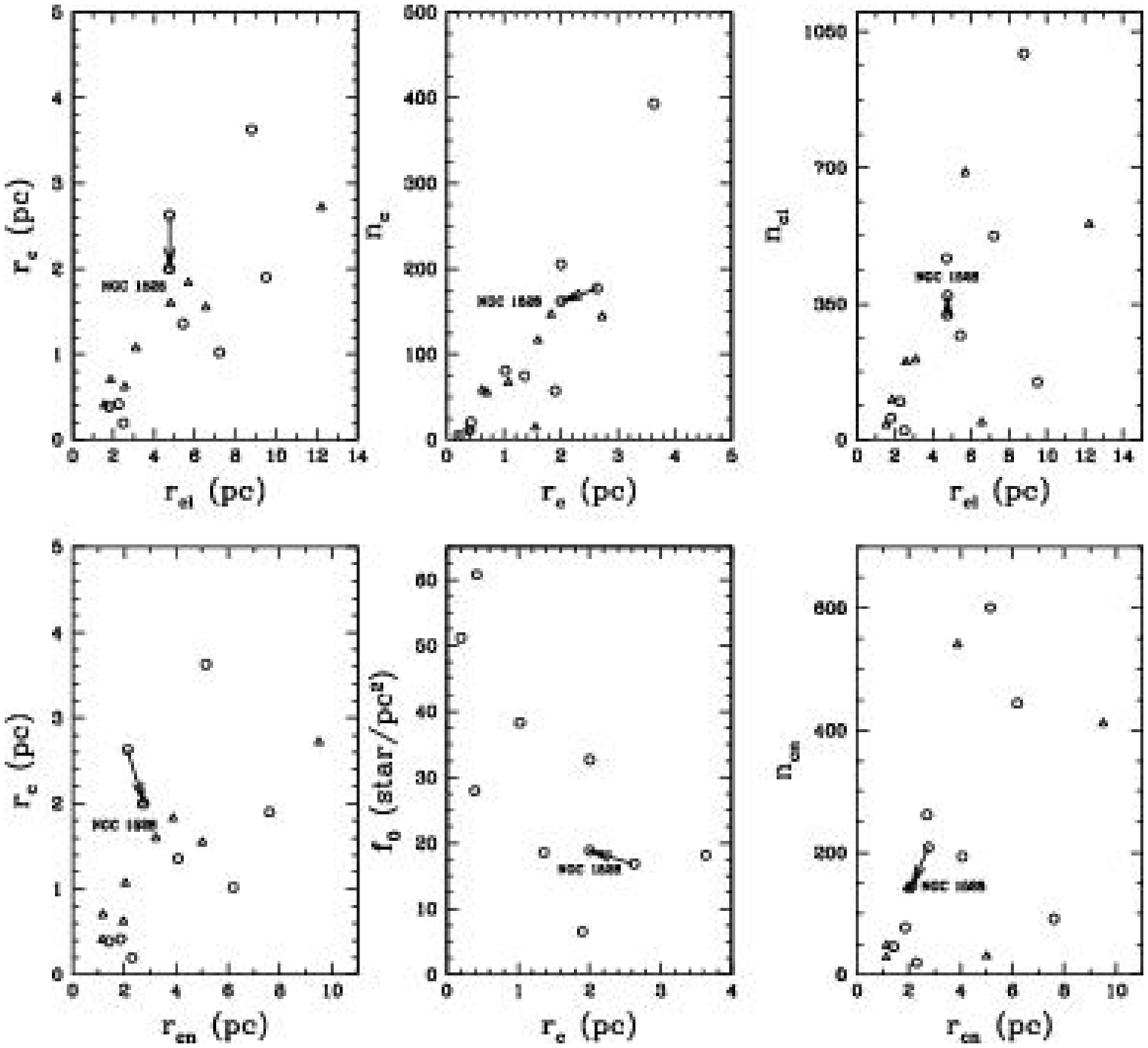}
\caption{Correlation between various structural parameters of the clusters. Symbols are same as in Figure 11. }
\end{figure*}

\begin{figure*}
\includegraphics[height=14cm,width=18cm]{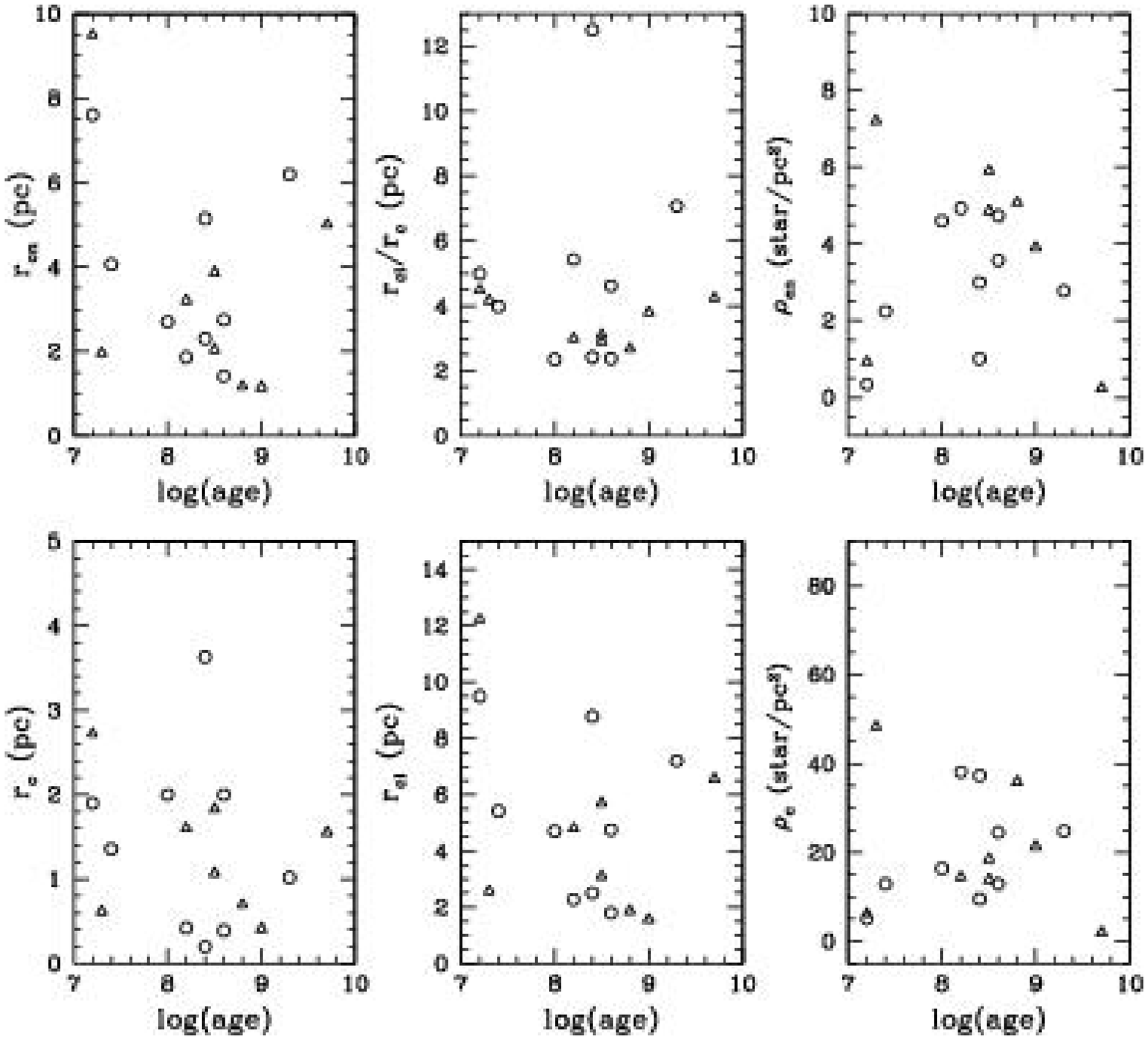}
\caption{Various structural parameters of clusters as a function of age. Symbols are same as in Figure 11. }
\end{figure*}

\begin{figure*}
\includegraphics[height=7cm,width=18cm]{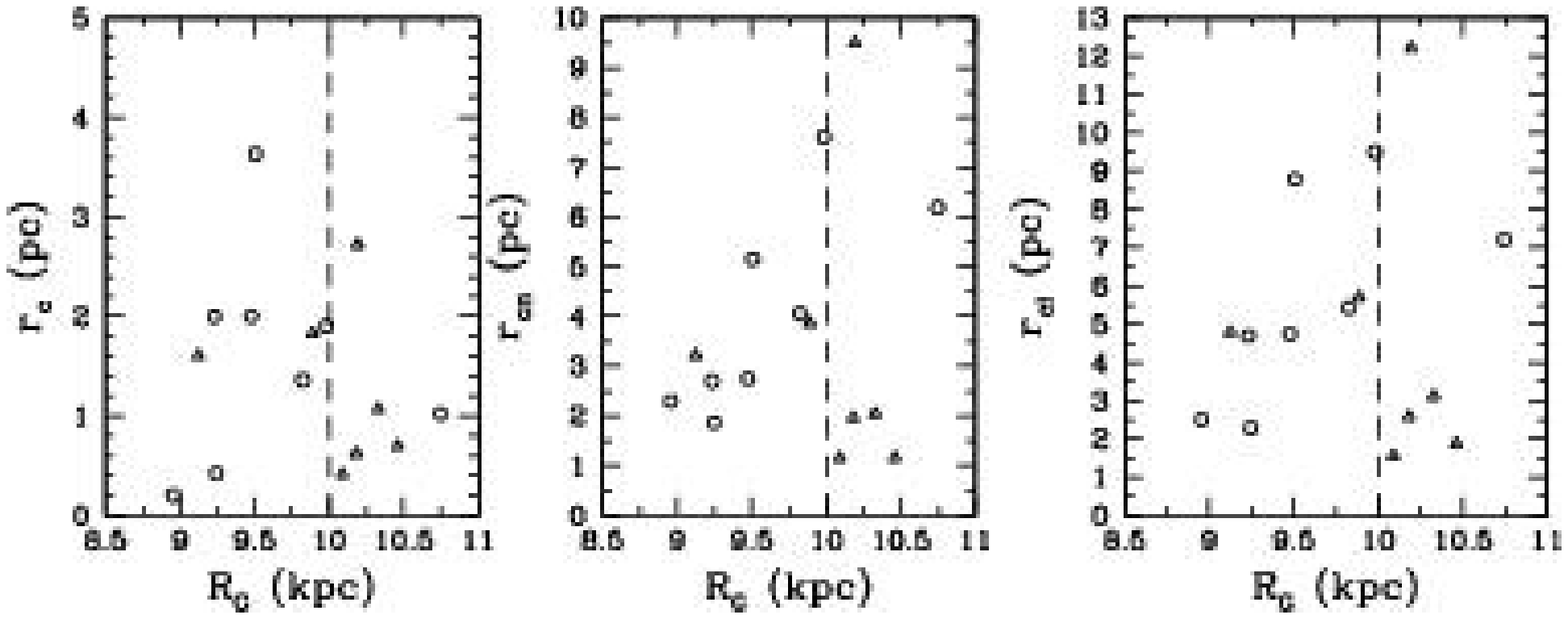}
\caption{Dependence of cluster sizes on galactocentric distance $`R_G$'. Symbols are same as in Figure 11.
The $r_c$, $r_{cn}$ and $r_{cl}$ of the clusters located in $9<R_G<10$ kpc show an increasing trend
with increasing $R_G$. The vertical dashed line is drawn to delineate the sample of clusters having
$R_G<10$ kpc and $R_G>10$ kpc.}
\end{figure*}

\clearpage

\begin{deluxetable}{ r r r r r r r r r r r }
\rotate
\tablewidth{0pt}
\tablehead{
\colhead {Cluster}&\colhead { Cluster no.}&\colhead {Trumpler class} &\colhead { $\alpha_{2000}$ }  &\colhead {$\delta_{2000}$}  &\colhead {$l$}& \colhead {$b$}  & \colhead {Log age}  &\colhead {Distance}  & \colhead {$E(B-V)$}\\
       &\colhead {(IAU)}&       &\colhead {(h:m:s) }       &\colhead { (d:m:s)} & \colhead {(degree)}&\colhead{(degree)}&\colhead { (yr) } &\colhead { (pc) }  & \colhead { (mag)}
}
\tabletypesize{\footnotesize}
\tablecaption{Cluster parameters from WEBDA \label{Table: 1}}
\startdata
    
Be 62    &  C0057+636 &III2m&$01:01:00$&$  63:57:00$& $ 123.98$&$  1.10$&7.2 &1837  &  0.85\\
NGC 1528 &  C0411+511 & II2m&$04:15:23$&$  51:12:54$& $ 152.06$&$  0.26$&8.6 & 776  &  0.26\\
NGC 1960 &  C0532+341 &  I3r&$05:36:18$&$  34:08:24$& $ 174.52$&$  1.07$&7.5 &1318  &  0.22\\
NGC 2287 &  C0644-206 &  I3r&$06:46:01$&$ -20:45:24$& $ 231.02$&$-10.44$&8.4 & 693  &  0.03\\
NGC 2301 &  C0649+005 &  I3r&$06:51:45$&$  00:27:36$& $ 212.56$&$  0.28$&8.2 & 872  &  0.03\\
NGC 2323 &  C0700-082 & II3r&$07:02:42$&$ -08:23:00$& $ 221.67$&$ -1.33$&8.1 & 929  &  0.21\\
NGC 2420 &  C0735+216 &  I1r&$07:38:23$&$  21:34:24$& $ 198.11$&$ 19.63$&9.1 &3085  &  0.03\\
NGC 2437 &  C0739-147 & II2r&$07:41:46$&$ -14:48:36$& $ 231.86$&$  4.06$&8.4 &1375  &  0.15\\
NGC 2548 &  C0811-056 &  I3r&$08:13:43$&$ -05:45:00$& $ 227.87$&$ 15.39$&8.6 & 769  &  0.03\\
\enddata
\end{deluxetable}

\begin{deluxetable}{ c c c c }
\tablewidth{0pt}
\tablehead{
\colhead { Field } &\colhead {Filter}&\colhead {Exposure (seconds)$\times$ No. of frames} & \colhead {Date}
}
\tabletypesize{\scriptsize}
\tablecaption{Log of Observations\label{Table: 2}}
\startdata
Be 62     & $U$      &  $ 180\times11, 60\times3               $&   2001 Nov. 20 \\
          & $B$      &  $  90\times9,  30\times2,  20\times5   $&   2001 Nov. 20 \\
          & $V$      &  $  90\times9,  20\times6               $&   2001 Nov. 20 \\
          & $R$      &  $ 60\times12,  10\times8               $&   2001 Nov. 21 \\
          & $I$      &  $  60\times9,  20\times1,  10\times6   $&   2001 Nov. 21 \\
\hline                                                          
NGC 1528  & $B$      &  $  60\times7, 10\times7                $&  2001 Nov. 23  \\
          & $V$      &  $  60\times7, 10\times7                $&  2001 Nov. 23  \\
          & $R$      &  $  60\times7, 10\times7,  5\times9     $&  2001 Nov. 23  \\
          & $I$      &  $  60\times7, 10\times7,  5\times9     $&  2001 Nov. 23  \\
\hline                                                          
NGC 1960  & $U$      &  $ 180\times8, 30\times6                $&  2001 Nov. 21  \\
          & $B$      &  $  60\times6, 20\times6,  10\times3    $&  2001 Nov. 21  \\
          & $V$      &  $  60\times6, 30\times3,  10\times2    $&  2001 Nov. 21  \\
          & $R$      &  $  30\times9, 10\times6                $&  2001 Nov. 21  \\
          & $I$      &  $  30\times9, 10\times6,  5\times2     $&  2001 Nov. 21  \\
\hline                                                          
NGC 2287  & $B$      &  $  60\times4, 10\times4,  5\times3     $&   2001 Nov. 25 \\
          & $V$      &  $  60\times4, 10\times4,  5\times3     $&   2001 Nov. 25 \\
          & $I$      &  $  60\times1, 30\times3, 10\times4, 5\times5$ &   2001 Nov. 25 \\
\hline                                                          
NGC 2301  & $U$      &  $ 180\times8, 60\times8                $&   2001 Nov. 22 \\
          & $B$      &  $  60\times8, 20\times1, 10\times7     $&   2001 Nov. 22 \\
          & $V$      &  $  60\times9, 10\times7                $&   2001 Nov. 22 \\
          & $I$      &  $  60\times8, 10\times7                $&   2001 Nov. 22 \\
\hline                                                          
NGC 2323  & $U$      &  $ 180\times9, 60\times9                $&   2001 Nov. 19 \\
          & $U$      &  $ 180\times3                           $&   2001 Nov. 24 \\
          & $B$      &  $  60\times4, 10\times12               $&   2001 Nov. 19 \\
          & $B$      &  $  60\times2                           $&   2001 Nov. 24 \\
          & $V$      &  $  60\times3, 10\times9                $&   2001 Nov. 20 \\
          & $V$      &  $  60\times2                           $&   2001 Nov. 24 \\
          & $I$      &  $  30\times6, 10\times9                $&   2001 Nov. 20 \\
          & $I$      &  $  10\times1                           $&   2001 Nov. 24 \\
\hline                                                          
NGC 2420  & $B$      &  $  60\times2, 10\times1                $&   2001 Nov. 24 \\
          & $V$      &  $  60\times9, 20\times6,  10\times3    $&   2001 Nov. 24 \\
          & $I$      &  $  60\times9, 10\times9                $&   2001 Nov. 24 \\
\hline                                                          
NGC 2437  & $B$      &  $  60\times9, 20\times6                $&   2001 Nov. 24 \\
          & $B$      &  $  60\times1, 20\times1                $&   2001 Nov. 25 \\
          & $V$      &  $  60\times9, 20\times6, 10\times3     $&   2001 Nov. 24 \\
          & $V$      &  $  60\times1, 20\times1                $&   2001 Nov. 25 \\
          & $I$      &  $  60\times6, 30\times3, 10\times3     $&   2001 Nov. 25 \\
\hline                                                          
NGC 2548  & $V$      &  $  60\times6, 10\times6                $&   2001 Nov. 25 \\
          & $I$      &  $  60\times4, 10\times4, 5\times2      $&   2001 Nov. 25 \\
\enddata
\end{deluxetable}

\clearpage

\begin{deluxetable}{r r r r r }
\tablewidth{0pt}
\tablehead{
\colhead {Parameters} &\multicolumn{4}{c}{Date}
}
\tabletypesize{\small}
\tablecaption{The zero-point constants, color coefficients and extinction coefficients on different nights\label{Table: 3}}
\startdata
       &  2001 Nov. 22        &   2001 Nov. 23  &2001 Nov. 24&2001 Nov. 25\\
Zero-point constant\\
$c_1$& $ 6.305\pm0.005      $&$                     $&$ 6.189\pm0.011    $&$                  $\\
$c_2$& $ 2.553\pm0.005      $&$ 2.562\pm0.003       $&$ 2.488\pm0.003    $&$ 2.920\pm0.006    $\\
$c_3$& $ 2.882\pm0.005      $&$ 2.893\pm0.004       $&$ 2.841\pm0.004    $&$ 3.198\pm0.008    $\\
$c_4$& $ 2.564\pm0.006      $&$ 2.567\pm0.004       $&$ 2.531\pm0.005    $&$                  $\\
$c_5$& $ 3.545\pm0.007      $&$ 3.532\pm0.007       $&$ 3.491\pm0.005    $&$ 3.762\pm0.010    $\\

Color coefficient\\
$m_1$&  $ -0.045\pm0.008    $&$                     $&$-0.014\pm0.010    $&$                  $\\
$m_2$&  $ -0.116\pm0.004    $&  $ -0.125\pm0.003    $&$-0.112\pm0.003    $&$-0.136\pm0.006    $\\
$m_3$&~~~~$  0.068\pm0.005  $&~~~ $  0.068\pm0.004  $&~~~$0.070\pm0.003  $&~~~$0.067\pm0.006  $\\
$m_4$&~~~~$  0.031\pm0.009  $&~~~ $  0.040\pm0.007  $&~~~$0.047\pm0.007  $&$                  $\\
$m_5$&  $ -0.076\pm0.006    $&  $ -0.076\pm0.006    $&$-0.061\pm0.004    $&$-0.083\pm0.008    $\\

Extinction coefficient\\
$k_u$& $    0.661\pm0.028   $&$                     $&$    0.683\pm0.028 $&$                  $\\
$k_b$& $    0.261\pm0.012   $&$    0.352\pm0.007    $&$    0.295\pm0.007 $&$    0.211\pm0.012 $\\
$k_v$& $    0.176\pm0.010   $&$    0.218\pm0.012    $&$    0.183\pm0.008 $&$    0.133\pm0.015 $\\
$k_r$& $    0.150\pm0.009   $&$    0.181\pm0.012    $&$    0.128\pm0.010 $&$                  $\\
$k_i$& $    0.047\pm0.009   $&$    0.140\pm0.021    $&$    0.096\pm0.010 $&$    0.037\pm0.020 $\\

\enddata 
\end{deluxetable}

\clearpage

\begin{deluxetable}{ l r r r r r r }
\tablewidth{0pt}
\tablehead{
\colhead{V range}&\colhead{$\Delta(V)$} && \colhead{$\Delta(B-V)$}&&\colhead{$\Delta(U-B)$}&\\
&\colhead{($Mean\pm \sigma$)}&\colhead{(N)} &\colhead{($Mean\pm \sigma$)}&\colhead{(N)}&\colhead{( $Mean\pm \sigma$)}&\colhead{(N)}
}
\tabletypesize{\scriptsize}
\tablecaption{\label{Table:4}Comparison of the present photometry with the available photometry in the literature. The difference $\Delta$ (literature-present data) is in magnitude. Mean and $\sigma$ are based on N stars in a V magnitude bin.}

\startdata
Be 62&&&&&&\\    
Phelps \& Janes (1994, ccd) &&&&&&\\
 13-14&$  0.010\pm0.033$& 9  &$  0.022\pm0.023$&9  &$ -0.112\pm0.123$&9\\
 14-15&$  0.021\pm0.066$&30   &$ -0.021\pm0.068$&30 &$ -0.073\pm0.106$&27\\
 15-16&$  0.002\pm0.053$&65   &$ -0.005\pm0.057$&65 &$ -0.129\pm0.126$&57\\
 16-17&$ -0.002\pm0.063$&92   &$ -0.017\pm0.052$&92 &$ -0.023\pm0.223$&74\\
 17-18&$ -0.002\pm0.057$&165  &$ -0.045\pm0.060$&165&$  0.394\pm0.321$&64\\
 18-19&$  0.007\pm0.070$&239  &$ -0.064\pm0.087$&239&$  1.227\pm0.420$&30\\
 19-20&$  0.010\pm0.089$&328  &$ -0.022\pm0.193$&315&$  1.480\pm0.000$&1\\
 20-21&$  0.043\pm0.149$&285  &$  0.267\pm0.318$&199& $ $ 	     &$ $\\  
Forbes (1981, pe)&&&&&&\\
  13-14&$ -0.129\pm0.024$&4    &$  0.062\pm0.030$&4  &$ -0.014\pm0.122$&4\\
\hline
NGC 1528 &&&&&&\\
Hoag et al. (1961, pe) &&&&&&\\
          $<$11 &$ 0.024\pm0.031$& 6    &$ 0.001\pm0.014$&5   &&\\
          11-12 &$ 0.011\pm0.008$&4   & $ -0.006\pm0.006$&4   &&\\
          12-13 &$-0.001\pm0.015$&5   & $ -0.006\pm0.008$&5   &&\\
          13-14 &$ 0.014\pm0.017$&5   & $ -0.006\pm0.011$&5   &&\\
          14-15 &$-0.011\pm0.010$&5   & $  0.013\pm0.037$&5   &&\\
Hoag et al. (1961, pg) &&&&&&\\                                 
         $<$11 &$-0.011\pm0.115$&7     &$ -0.067\pm0.102$&6   &&\\
          11-12 &$-0.079\pm0.066$&12   &$  0.021\pm0.048$&12  &&\\
          12-13 &$-0.034\pm0.066$&23   &$  0.010\pm0.043$&23  &&\\
          13-14 &$-0.024\pm0.066$&30   &$ -0.005\pm0.056$&30  &&\\
          14-15 &$-0.023\pm0.080$&38   &$  0.028\pm0.099$&38  &&\\
          15-16 &$-0.064\pm0.023$&3    &$  0.046\pm0.022$&3   &&\\
\hline                  
NGC 1960&&&&&&\\
Sanner et al. (2000, ccd)&&&&&&\\
$<$10&$   0.008\pm0.028$&3   &$  -0.033\pm0.058$&3  &&\\
10-11&$  -0.009\pm0.016$&10  &$   0.035\pm0.022$&10 &&\\
11-12&$  -0.005\pm0.015$&15  &$   0.019\pm0.013$&15 &&\\
12-13&$  -0.006\pm0.016$&25  &$  -0.007\pm0.033$&25 &&\\
13-14&$  -0.001\pm0.050$&33  &$  -0.002\pm0.034$&33 &&\\
14-15&$  -0.001\pm0.092$&24  &$  -0.014\pm0.053$&24 &&\\
15-16&$  -0.040\pm0.054$&91  &$  -0.060\pm0.053$&91 &&\\
16-17&$  -0.046\pm0.048$&99  &$  -0.079\pm0.044$&99 &&\\
17-18&$  -0.057\pm0.065$&148 &$  -0.105\pm0.077$&148&&\\
18-19&$  -0.058\pm0.068$&268 &$  -0.123\pm0.127$&265&&\\
19-20&$  -0.080\pm0.103$&75  &$  -0.132\pm0.178$&62 &&\\
Johnson \& Morgan (1953, pe)&&&&&&\\
$<$10 &$  -0.014\pm0.067$&6   &$  -0.024\pm0.053  $&5   &$ 0.169\pm0.214$&5\\
10-11 &$   0.019\pm0.009$&11  &$   0.009\pm0.011  $&11  &$-0.005\pm0.043$&10\\
11-12 &$   0.016\pm0.016$&16  &$  -0.003\pm0.013  $&16  &$-0.012\pm0.048$&15\\
12-13 &$   0.003\pm0.025$&4   &$   0.001\pm0.008  $&4   &$-0.221\pm0.228$&4\\
Barkhatova et al. (1984, pg)&&&&&&\\
$<10$&$  -0.032\pm0.129$&5   &$  -0.076\pm0.124$&4   &$ 0.171\pm0.164$&4 \\
10-11 &$   0.018\pm0.054$&18  &$  -0.010\pm0.085$&18  &$-0.010\pm0.078$&16\\
11-12 &$   0.027\pm0.065$&27  &$   0.026\pm0.078$&27  &$-0.027\pm0.099$&27\\
12-13 &$  -0.008\pm0.074$&52  &$   0.046\pm0.080$&51  &$-0.088\pm0.141$&50\\
13-14 &$   0.002\pm0.080$&99  &$  -0.005\pm0.134$&98  &$ 0.033\pm0.159$&86\\
14-15 &$   0.023\pm0.080$&118 &$  -0.058\pm0.114$&117 &$ 0.003\pm0.131$&90\\
15-16 &$   0.043\pm0.073$&34  &$  -0.148\pm0.102$&33  &$-0.075\pm0.191$&17\\
\hline
NGC 2287 &&&&&&\\ 
 Harris et al. (1993, pe)&&&&&&\\ 
 $<$ 8&$ -0.061\pm0.032$&5    &$ -0.032\pm0.084$&5 &&\\
 8-9  &$ -0.054\pm0.009$&7    &$ -0.031\pm0.014$&7 &&\\
 9-10 &$ -0.058\pm0.048$&10   &$ -0.032\pm0.006$&9 &&\\
 10-11&$ -0.047\pm0.059$&16   &$ -0.026\pm0.029$&16&&\\
 11-12&$ -0.056\pm0.017$&10   &$ -0.029\pm0.019$&10&&\\
 12-13&$ -0.055\pm0.015$&7    &$ -0.040\pm0.022$&7 &&\\
 13-14&$ -0.058\pm0.010$&3    &$ -0.045\pm0.019$&3 &&\\
Eggen (1974, pe)&&&&&&\\                                
$<$8  &$ -0.065\pm0.041$&4    &$ -0.008\pm0.092$&4 &&\\
 8-9  &$ -0.014\pm0.045$&8    &$ -0.010\pm0.043$&8 &&\\
 9-10 &$ -0.025\pm0.009$&3    &$ -0.043\pm0.003$&3 &&\\
 10-11&$ -0.024\pm0.017$&5    &$ -0.021\pm0.012$&4 &&\\
 11-12&$ -0.049\pm0.028$&2    &$ -0.030\pm0.037$&2 &&\\
 12-13&$ -0.022\pm0.010$&2    &$ -0.050\pm0.003$&2 &&\\
 13-14&$ -0.049\pm0.000$&1    &$ -0.042\pm0.000$&1 &&\\
Ianna et al. (1987, pe)&&&&&&\\                          
$<$8  &$ -0.061\pm0.030$&5    &$ -0.032\pm0.073$&5 &&\\
 8-9  &$ -0.057\pm0.022$&9    &$ -0.038\pm0.035$&9 &&\\
 9-10 &$ -0.025\pm0.064$&9    &$ -0.031\pm0.012$&9 &&\\
 10-11&$ -0.042\pm0.055$&24   &$ -0.025\pm0.020$&24&&\\
 11-12&$ -0.059\pm0.025$&16   &$ -0.024\pm0.018$&16&&\\
Hoag et al. (1961, pe)&&&&&&\\                            
 $<$8 &$ -0.076\pm0.000$&1    &$  0.075\pm0.000$&1 &&\\
  8-9 &$ -0.136\pm0.179$&3    &$ -0.024\pm0.090$&3&&\\
  9-10&$ -0.045\pm0.015$&2    &$ -0.031\pm0.040$&2&&\\
 10-11&$ -0.035\pm0.017$&6    &$ -0.056\pm0.040$&6&&\\
 12-13&$ -0.032\pm0.020$&3    &$ -0.060\pm0.029$&3&&\\
 13-14&$ -0.071\pm0.054$&2    &$ -0.039\pm0.086$&2&&\\
 Hoag et al. (1961, pg)&&&&&&\\                           
 8-9  &$ -0.120\pm0.015$&3    &$ -0.058\pm0.063$&3&&\\
 9-10 &$ -0.097\pm0.022$&4    &$ -0.050\pm0.023$&3&&\\
 10-11&$ -0.111\pm0.106$&9    &$ -0.029\pm0.087$&9&&\\
 11-12&$ -0.104\pm0.048$&10   &$ -0.037\pm0.036$&10&&\\
 12-13&$ -0.055\pm0.043$&17   &$ -0.075\pm0.054$&17&&\\
 13-14&$ -0.038\pm0.038$&9    &$ -0.012\pm0.038$&9&&\\
\hline                  
NGC 2301&&&&&&\\
Kim et al. (2001, ccd)&&&&&&\\
10-11 &$  0.021\pm0.016  $&11   &$  0.014\pm0.016 $&11  &$ 0.002\pm0.022$&11\\
11-12 &$  0.043\pm0.054  $&18   &$  0.002\pm0.019 $&18  &$-0.030\pm0.191$&18\\
12-13 &$  0.026\pm0.057  $&27   &$ -0.002\pm0.035 $&27  &$ 0.005\pm0.069$&23\\
13-14 &$  0.020\pm0.021  $&21   &$  0.013\pm0.020 $&21  &$-0.027\pm0.085$&21\\
14-15 &$  0.014\pm0.080  $&55   &$  0.003\pm0.055 $&55  &$-0.047\pm0.113$&42\\
15-16 &$  0.015\pm0.049  $&67   &$ -0.001\pm0.049 $&67  &$-0.010\pm0.186$&61\\
16-17 &$  0.016\pm0.083  $&98   &$  0.002\pm0.077 $&98  &$ 0.096\pm0.244$&73\\
17-18 &$  0.001\pm0.068  $&157  &$ -0.031\pm0.069 $&152 &$ 0.374\pm0.337$&33\\
18-19 &$ -0.028\pm0.106  $&255  &$ -0.062\pm0.116 $&247 &$ 0.539\pm0.652$&9\\
19-20 &$ -0.082\pm0.130  $&128  &$ -0.053\pm0.208 $&105 &&\\
 Hoag et al. (1961, pe)   &&&&&&\\
10-11&$  0.052\pm0.031   $&3    &$ -0.011\pm0.022  $&3  &$ -0.027\pm0.015  $&2\\
11-12&$  0.038\pm0.011   $&8    &$  0.006\pm0.013  $&8  &$  0.001\pm0.058  $&8\\
12-13&$  0.031\pm0.010   $&7    &$  0.017\pm0.020  $&7  &$  0.011\pm0.047  $&7\\
13-14&$  0.004\pm0.029   $&4    &$  0.026\pm0.018  $&4  &$  0.090\pm0.015  $&4\\
 Mohan \& Sagar (1988, pg)&&&&&&\\
11-12&$ -0.043\pm0.088  $&26   &$ -0.050\pm0.099  $&26 &$  0.125\pm0.146  $&23\\
12-13&$ -0.026\pm0.086  $&44   &$ -0.043\pm0.122  $&44 &$  0.132\pm0.145  $&37\\
13-14&$  0.005\pm0.062  $&69   &$ -0.075\pm0.086  $&68 &$  0.106\pm0.108  $&66\\
14-15&$ -0.011\pm0.057  $&153  &$ -0.045\pm0.073  $&153&$  0.209\pm0.191  $&140\\
15-16&$ -0.038\pm0.055  $&220  &$  0.011\pm0.119  $&220&$  0.297\pm0.129  $&196\\
16-17&$ -0.018\pm0.068  $&335  &$  0.047\pm0.108  $&335&$  0.397\pm0.187  $&247\\
 Hoag et al. (1961, pg)   &&&&&&\\
10-11&$ -0.019\pm0.211  $&4    &$  0.045\pm0.075  $&4  &$  0.003\pm0.118  $&4\\
11-12&$  0.054\pm0.000  $&1    &$ -0.010\pm0.000  $&1  &$ -0.023\pm0.000  $&1\\
12-13&$ -0.029\pm0.117  $&11   &$  0.058\pm0.068  $&11 &$  0.032\pm0.170  $&10\\
13-14&$ -0.044\pm0.098  $&13   &$  0.069\pm0.100  $&13 &$  0.097\pm0.088  $&12\\
14-15&$  0.019\pm0.050  $&8    &$  0.038\pm0.038  $&8  &$  0.166\pm0.076  $&6\\
\hline                  
NGC 2323 &&&&&&\\
 Hoag et al. (1961, pe) &&&&&&\\
$<$10  & $0.058\pm0.029$  &6 & $ -0.007\pm0.054$  &6&$ 0.078\pm0.036 $  &6\\
 11-12 & $0.058\pm0.000$  &1 & $ -0.006\pm0.000$  &1&$  0.056\pm0.000 $  &1\\
 12-13 & $0.052\pm0.019$  &5 & $ -0.024\pm0.031$  &5&$ 0.018\pm0.062$  &5\\
 13-14 & $0.106\pm0.074$  &2 & $ -0.081\pm0.066$  &2&$ 0.061\pm0.007$  &2\\
 14-15 & $0.053\pm0.039$  &3 & $ -0.070\pm0.015$  &3&$ 0.102\pm0.118$  &2\\
 15-16 & $0.017\pm0.000$  &1 & $  0.051\pm0.000$  &1&$ 0.157\pm0.000$  &1\\
Claria et al. (1988, pe)&&&&&&\\
 $<$10 & $0.038\pm0.033$  &6 & $  0.009\pm0.049$  &6 &$ 0.074\pm0.023$&6\\
 10-11 & $0.046\pm0.010$  &12& $ -0.007\pm0.015$  &12&$ 0.045\pm0.023$&12\\
 11-12 & $0.051\pm0.017$  &23& $ -0.013\pm0.018$  &23&$ 0.000\pm0.040$&22\\
 12-13 & $0.023\pm0.067$  &33& $ -0.002\pm0.022$  &33&$ 0.012\pm0.047$&33\\
 13-14 & $0.011\pm0.037$  &10& $  0.016\pm0.025$  &10&$ 0.025\pm0.072$&9\\
 14-15 &$-0.007\pm0.037$  &2 & $ -0.001\pm0.042$  &2 &$ 0.062\pm0.059$&2\\
 Hoag et al. (1961, pg) &&&&&&\\
 $<$11 & $0.092\pm0.063$  &11 & $  0.001\pm 0.06$  &11&$-0.030\pm0.055$  &11\\
 11-12 & $0.049\pm0.094$  &24 & $ -0.018\pm0.078$  &22&$-0.041\pm0.050$  &22\\
 12-13 & $0.052\pm0.044$  &36 & $ -0.030\pm0.050$  &35&$-0.001\pm0.096$  &35\\
 13-14 & $0.111\pm0.087$  &28 & $ -0.072\pm0.075$  &28&$-0.105\pm0.054$  &25\\
 14-15 & $0.104\pm0.074$  &13 & $ -0.143\pm0.111$  &13&$ 0.028\pm0.195$  &10\\
Mostafa et al. (1983, pg)&&&&&\\ 
 $<$10 &$ 0.079\pm0.095$  &5 & $ -0.045\pm0.069$  &5 &$ 0.094\pm0.037$  &5\\
 10-11 &$ 0.058\pm0.075$  &9 & $ -0.048\pm0.097$  &9 &$-0.004\pm0.032$  &9\\
 11-12 &$ 0.027\pm0.117$  &17& $  0.006\pm0.123$  &17&$ 0.065\pm0.089$  &13\\
 12-13 &$ 0.036\pm0.075$  &29& $ -0.072\pm0.103$  &28&$ 0.049\pm0.090$  &28\\
 13-14 &$ 0.035\pm0.113$  &22& $ -0.027\pm0.120$  &22&$-0.019\pm0.015$  &21\\
 14-15 &$ 0.002\pm0.082$  &12& $ -0.022\pm0.078$  &12&$ 0.021\pm0.099$  &11\\
\hline
NGC 2420 &&&&&&\\
Anthonny-Twarog et al. (1990, ccd)&&&&&&\\ 
 11-12&$  0.024\pm0.011  $&2    &$  0.023\pm0.014  $&2   &&\\
 12-13&$  0.011\pm0.015  $&8    &$ -0.003\pm0.021  $&8   &&\\
 13-14&$  0.014\pm0.027  $&11   &$ -0.013\pm0.016  $&11  &&\\
 14-15&$  0.002\pm0.017  $&55   &$ -0.016\pm0.011  $&55  &&\\
 15-16&$ -0.006\pm0.014  $&64   &$ -0.013\pm0.015  $&64  &&\\
 16-17&$ -0.006\pm0.040  $&78   &$ -0.017\pm0.023  $&78  &&\\
 17-18&$ -0.013\pm0.040  $&80   &$ -0.017\pm0.026  $&79  &&\\
 18-19&$ -0.027\pm0.076  $&110  &$ -0.002\pm0.075  $&110 &&\\
 19-20&$ -0.006\pm0.084  $&48   &$  0.056\pm0.014  $&48  &&\\
West (1967, pe)&&&&&&\\ 
 11-12&$  0.013\pm0.037   $&5&$  0.010\pm0.011   $&5&&\\
 12-13&$ -0.004\pm0.012   $&8&$  0.021\pm0.073   $&8&&\\
 13-14&$ -0.010\pm0.005   $&3&$ -0.004\pm0.012   $&3&&\\
 14-15&$ -0.005\pm0.008   $&7&$ -0.007\pm0.034   $&7&&\\
 15-16&$ -0.014\pm0.012   $&6&$ -0.015\pm0.022   $&6&&\\
 17-18&$  0.002\pm0.017   $&2&$ -0.028\pm0.036   $&2&&\\
Mc Clure et al. (1974, pe)&&&&&&\\ 
 11-12&$  0.009\pm0.019   $&5&$  0.011\pm0.011  $&5&&\\
 12-13&$  0.012\pm0.011   $&7&$  0.011\pm0.012  $&7&&\\
 13-14&$  0.006\pm0.006   $&4&$ -0.006\pm0.013  $&4&&\\
 14-15&$  0.015\pm0.009   $&5&$ -0.010\pm0.012  $&5&&\\
 15-16&$  0.015\pm0.000   $&1&$  0.023\pm0.000  $&1&&\\
Mc Clure et al. (1974, pg)&&&&&&\\ 
 12-13&$  0.024\pm0.021  $&2&$   0.005\pm0.042 $ &2 &&\\
 13-14&$  0.027\pm0.043  $&10&$  0.004\pm0.018 $ &10&&\\
 14-15&$  0.027\pm0.031  $&50&$ -0.008\pm0.019 $ &50&&\\
 15-16&$  0.002\pm0.041  $&62&$ -0.012\pm0.027 $ &62&&\\
 16-17&$  0.001\pm0.058  $&70&$ -0.003\pm0.040 $ &70&&\\
 17-18&$  0.015\pm0.068  $&36&$ -0.017\pm0.041 $ &35&&\\
Cannon \& Lloyd (1970, pg) &&&&&&\\
11-12&$  -0.015\pm0.059  $&5&$   0.055\pm0.075  $& 5&&\\
12-13&$   0.002\pm0.016  $&11&$  0.021\pm0.024  $&11&&\\
13-14&$  -0.031\pm0.036  $&21&$  0.031\pm0.050  $&21&&\\
14-15&$  -0.056\pm0.308  $&80&$  0.004\pm0.070  $&80&&\\
15-16&$   0.010\pm0.181  $&59&$ -0.049\pm0.044  $&59&&\\
\hline           
NGC 2437&&&&&&\\ 
Stetson (2000, ccd)&&&&&&\\
11-12 &$  0.029\pm0.002  $&3  &$  0.018\pm0.005  $&3 &&\\
12-13 &$  0.028\pm0.014  $&16 &$  0.017\pm0.013  $&16&&\\
13-14 &$  0.029\pm0.015  $&14 &$  0.009\pm0.018  $&14&&\\
14-15 &$  0.029\pm0.014  $&15 &$  0.002\pm0.012  $&15&&\\
15-16 &$  0.030\pm0.013  $&23 &$  0.001\pm0.015  $&23&&\\
16-17 &$  0.030\pm0.021  $&15 &$  0.004\pm0.020  $&15&&\\
17-18 &$  0.019\pm0.018  $&20 &$  0.008\pm0.031  $&20&&\\
18-19 &$  0.049\pm0.076  $&14 &$  0.001\pm0.090  $&14&&\\
19-20 &$  0.046\pm0.030  $&6  &$ -0.053\pm0.045  $&6 &&\\
 Henden (2003, ccd)&&&&&&\\
11-12 &$  0.048\pm0.034$&10   &$  0.001\pm0.031$&10&&\\
12-13 &$  0.038\pm0.022$&16   &$ -0.005\pm0.012$&16&&\\
13-14 &$  0.025\pm0.088$&17   &$ -0.031\pm0.137$&17&&\\
14-15 &$  0.042\pm0.018$&24   &$ -0.014\pm0.019$&24&&\\
15-16 &$  0.017\pm0.098$&50   &$ -0.011\pm0.022$&50&&\\
16-17 &$ -0.001\pm0.113$&92   &$ -0.040\pm0.084$&92&&\\
17-18 &$ -0.004\pm0.122$&108  &$ -0.048\pm0.089$&108&&\\
18-19 &$  0.029\pm0.151$&97   &$ -0.068\pm0.159$&91&&\\
Lynga \& Palous (1987, pe) & &&&&&\\
$<$10 &$  0.041\pm0.000   $&1&$$&$$&$$&$$\\  
10-11 &$  0.010\pm0.000   $&1&$$&$$&$$&$$\\ 
11-12 &$  0.014\pm0.025   $&2&$ -0.005\pm0.006$&2&&\\
12-13 &$ -0.005\pm0.000   $&1&$ -0.030\pm0.000$&1&&\\
13-14 &$ -0.022\pm0.000   $&1&$ -0.045\pm0.000$&1&&\\
 Oja (1976, Pe) &&&&&&\\
10-11 &$  0.049\pm0.013  $&4&$ -0.018\pm0.010  $&4&&\\
11-12 &$  0.036\pm0.000  $&1&$  0.001\pm0.000  $&1&&\\
12-13 &$  0.054\pm0.000  $&1&$ -0.021\pm0.000  $&1&&\\
 Smyth \& Nandy (1962, pg) &&&&&&\\ 
10-11 &$ -0.026\pm0.067  $&16 &$ -0.001\pm0.033  $&15&&\\
11-12 &$ -0.033\pm0.061  $&32 &$ -0.009\pm0.032  $&32&&\\
12-13 &$ -0.056\pm0.060  $&61 &$ -0.011\pm0.032  $&61&&\\
13-14 &$ -0.090\pm0.048  $&51 &$ -0.034\pm0.035  $&51&&\\
\hline           
NGC 2548 &&&&&&\\
Pesch (1961, pe)&&&&&&\\ 
$<$9  &$  0.047\pm0.023   $&15 &&&&\\  
9-10  &$  0.047\pm0.014   $&11 &&&&\\
10-11 &$  0.051\pm0.015   $&5  &&&&\\
11-12 &$  0.050\pm0.000   $&1  &&&&\\
Oja (1976, pe) & &&&&&\\ 
$<$9  &$  0.026\pm0.000  $&1&&&&\\
9-10  &$  0.040\pm0.016  $&2&&&&\\
10-11 &$  0.029\pm0.025  $&2&&&&\\
11-12 &$  0.066\pm0.058  $&4&&&&\\
12-13 &$  0.175\pm0.000  $&1&&&&\\
13-14 &$  0.069\pm0.000  $&1&&&&\\

\enddata
\end{deluxetable}

\begin{deluxetable}{ r r r r r r r }
\tablewidth{0pt}
\tabletypesize{\small}
\tablehead{
&&& \multicolumn{2}{c}{Optical} &\multicolumn{2}{c}{2MASS} \\ 
\colhead{Cluster} &\colhead{$\alpha_{2000}$} &\colhead{$\delta_{2000}$} &\colhead{Core radius} &\colhead{Cluster extent} &\colhead{Core radius}&\colhead{ Cluster extent} \\
 &\colhead{(h:m:s)}&\colhead{(d:m:s)}&\colhead{arcmin (pc)}&\colhead{arcmin (pc)}&\colhead{arcmin (pc)}&\colhead{arcmin (pc)}
}
\tablecaption{Structural parameters of the target open clusters estimated from the projected radial density profile of main sequence stars having $V\le18$ mag.\label{Table: 5}}

\startdata
Be 62& $   01:01:15.5$&$  63:56:17  $&2.2 (1.5)&10.0 (6.8)&2.5 (1.7)&12 (8.1)\\  
NGC 1528&$ 04:15:24.2$&$  51:15:23  $&8.3 (2.6)&15.0 (4.8)&18.5 (5.9)&24 (7.6)\\   
NGC 1960&$ 05:36:20.8$&$  34:08:31  $&3.2 (1.2)&14.0 (5.4)&3.8 (1.5)&21 (8.1)\\    
NGC 2287&$ 06:45:58.7$&$ -20:44:09  $&1.4 (0.3)&12.0 (2.5)&12.7 (2.6)&16 (3.3)\\   
NGC 2301&$ 06:51:46.4$&$  00:27:30  $&1.9 (0.5)&9.0 (2.3) &4.5 (1.1)&20 (5.1)\\     
NGC 2323&$ 07:02:47.4$&$ -08:20:43  $&6.5 (1.8)&17.0 (4.7)&6.7 (1.9)&22 (6.1)\\    
NGC 2420&$ 07:38:24.8$&$  21:34:30  $&1.4 (1.0)&10.0 (7.2)&1.3 (0.9)&9 (6.5)\\     
NGC 2437&$ 07:41:58.1$&$ -14:49:28  $&6.8 (3.0)&20.0 (8.8)&9.6 (4.2)&25 (11.0)\\   
NGC 2548&$ 08:13:42.9$&$ -05:46:37  $&1.5 (0.3)&8.0 (1.8) &2.4 (0.5)&8.0 (1.8)\\    
\enddata
\end{deluxetable}

\begin{deluxetable}{ r r c r c c c c c} 
\tablewidth{0pt}
\tablehead{
& \multicolumn{2}{c}{Kharchenko et al. (2005)}& \multicolumn{2}{c}{Nilakshi et al. (2002)}& \multicolumn{2}{c}{Chen et al. (2004)}& \multicolumn{2}{c}{Bonatto \& Bica (2005)} \\
\colhead{ Name }   &\colhead{ $r_c$}&\colhead{ $r_{cl}$} & \colhead{  $r_c$} &\colhead{ $r_{cl}$  } &\colhead{$r_{1/3}$} &\colhead{$r_{out}$}&\colhead{ $R_{core}$}&\colhead{ $R_{lim}$}
}

\tabletypesize{\scriptsize}
\tablecaption{Recent estimates of sizes (in arcmin) of target clusters available in the literature.\label{Table: 6}}
\startdata
NGC 1528 & 8.4& 26.4 &  $ $ &  $ $  & $ $ & $ $&$ $&$ $  \\
NGC 1960 & 5.4& 16.2 &   3.2&  15.4 &5.7  &7.6&$ $&$ $  \\
NGC 2287 & 9.6& 30.0 &  $ $ &  $ $  & $ $ & $ $&4.7  & 30.1   \\
NGC 2301 & 4.8& 15.0 &  1.9 & 12.0  & $ $ & $ $&&  \\
NGC 2323 & 6.0& 22.2 &  2.6 &  16.7 & $ $ & $ $&&  \\
NGC 2420 & $ $& $ $  &  1.5 &  13.2 &  3.7&11.6&&  \\
NGC 2437 & 7.2& 22.8 &  5.2 &  26.6 & $ $ & $ $&&  \\
NGC 2548 &15.0& 43.8 &  $ $ &  $ $  & $ $ & $ $&3.9  & 37.8  \\
\enddata
\end{deluxetable}

\begin{deluxetable}{ r c c r r r r }
\tablewidth{0pt}
\tablehead{
\colhead{Cluster} &\colhead{$E(B-V)$} &\colhead{ Log age } &\multicolumn{2}{c}{Distance} &\colhead{$R_G$} \\
        &\colhead{(mag)}&\colhead{(yr)}&\colhead{($V-M_V$)} &\colhead{ (kpc)}&\colhead{(kpc)}
}

\tablecaption{The estimated parameters of the target clusters obtained in the present work using color-color diagram and color-magnitude diagrams. To determine the galactocentric distances ($R_G$) to the clusters, a value of 8.5 kpc has been assumed for the galactocentric distance of the Sun.\label{Table: 8}}
\startdata

Be 62   &0.70-1.00&7.2 & 14.0 &2.32&9.98 \\                 
NGC 1528&0.26&8.6 & 11.0 &1.09&9.48 \\        
NGC 1960&0.22&7.4 & 11.3 &1.33&9.82 \\            
NGC 2287&0.01&8.4 &  9.3 &0.71&8.96 \\         
NGC 2301&0.03&8.2 &  9.8 &0.87&9.25 \\               
NGC 2323&0.20&8.0 & 10.5 &0.95&9.23 \\            
NGC 2420&0.04&9.3 & 12.1 &2.48&10.76\\                
NGC 2437&0.10&8.4 & 11.2 &1.51&9.51 \\                
NGC 2548&0.03&8.6 &  9.5 &0.77&9.02 \\           
\enddata
\end{deluxetable}

\begin{deluxetable}{ r r l l l l }
\tablewidth{0pt}
\tablehead{
\colhead{Cluster} &\colhead{ Radius (arcmin)  }   &\colhead{   ${(I-V)}\over{(B-V)}$}&\colhead{ ${(J-V)}\over{(B-V)}$} &\colhead{ ${(H-V)}\over{(B-V)}$} &\colhead{ ${(K-V)}\over{(B-V)}$}
}

\tablecaption{Slopes of Two Color Diagrams (TCDs) for cluster Be 62, NGC 2420 and NGC 1960.\label{Table: 7}}
\startdata

Be 62      &$<10     $& $-0.99\pm0.01$ & $-1.78\pm0.02$ &  $-2.29\pm0.03$ & $-2.49\pm0.03$\\
NGC 2420   &$<10     $& $-0.97\pm0.01$ & $-1.69\pm0.02$ &  $-2.32\pm0.03$ & $-2.40\pm0.05$\\
NGC 1960   &$<14     $&$ -1.13\pm0.01$&$ -1.97\pm0.01 $&$-2.44\pm0.01 $&$-2.56\pm0.01$\\
Normal value$^1$  &&$ -1.10$&$ -1.96 $&$-2.42 $&$-2.60$\\
\enddata
\tablenotetext{1} { See text }
\end{deluxetable}

\clearpage

\begin{deluxetable} { r r r r r r }
\tablewidth{0pt}
\tabletypesize{\footnotesize}
\tablehead{
\colhead{Cluster} &\colhead{Observations*}  &\colhead{E(B-V)}  &\colhead{ Distance}&\colhead{Age}  &\colhead{ Reference}\\
                  &                         &\colhead{(mag)}   & \colhead{(pc)}    &\colhead{(yrs)}&
}
\tablecaption{ Comparison of estimates available in the literature with the present estimates. \label{Table: 9}}
\startdata
Be 62&pe  &   0.86 & $2050\pm240 $  & $10^7$      & Forbes(1981)\\
&CCD &    0.82 &  2704  &   $10^7$  & Phelps \& Janes(1994)\\
&CCD & 0.70  & 2320 &  $1.6\times10^7$ & Present work\\       
\hline
NGC 1528 & pg & 0.28    & 603    & $4\times10^8$&Francic (1989)\\ 
&CCD & 0.26  & 1090  & $4\times10^8$& Present work\\
\hline
NGC 1960 & CCD & $0.25\pm0.02$   & $1318\pm120$ & $16^{+6}_{-5} \times 10^6$&Sanner et al. (2000)\\
&  CCD  & 0.22  &  1330 & $ 2.5\times10^7$&  Present work\\
\hline                 
NGC 2287 &      pg         &         0.00           & 661   &     $2\times10^8$             &Hoag et al. (1961)      \\
&       $ $          &        $ $             &   $ $       &  $1.7\times10^8$              &Barbaro (1967)   \\
&     pe    &       0.03             & 758  &  $ $                                   &Eggen (1974) \\
&     pe  &         $ $            &     $ $     &    $8\times10^7$                       &Eggen (1975) \\
&     pe  &     $ 0.01\pm0.03$ & 630 & $6.0\pm10^7$                          &Feinstein et al. (1978)\\
&     pe    &        $ $&740& $10^8$                      &Eggen (1981) \\
& DDO&$0.07\pm0.02$ & 752&$(1.1\pm0.4)\times10^8$                          &Miriani \& Ursula (1983) \\
&pe  & 0.00  & 675  &&  Ianna et al. (1987)\\
&   pe &$0.064\pm0.042$  &700  & $2\times10^8$                              &Harris et al. (1993)    \\
&             CCD    &        0.01&           710          &   $2.5\times10^8$       &Present work      \\
\hline
NGC 2301& pe&0.04 & 840      &    $ $  &       Nissen (1988)      \\
&pg&0.04  &794        &  $ $    &       Mohan \& Sagar(1988)\\
&$ $&0.05  &692        &  $2\times10^8$  &  Napiwotzki et al. (1991)\\
&DDO&0.04  &855       &   $ $   &       Twarog et al. (1997)\\
&CCD&0.05  &832        &  $2.5\times10^8$  &Kim et al. (2001)      \\
&CCD&0.03  &870      &   $1.6\times10^8$  & Present work            \\
\hline
NGC 2323& pg         &     $ $        &     780-860    &    $ $                          & Trumpler (1930)     \\
&pg         &     $ $        &     500-800     &    $ $                           & Shapley (1930)      \\
&pg         &     $ $        &     675        &    $ $                           & Collinder (1931)    \\
&pg         &     $ $        &      520       &    $ $                           & Rieke (1935)        \\
&pg         &     0.30       &     1210       &    $ $                           & Cuffey (1941)       \\
&pe        &  0.20-0.26     &     1170      &    $ $                            & Hoag et al. (1961)   \\
&pg         &   $ $          &     $ $        &    $6\times10^7$                            & Barbaro et al. (1969)\\
&pg         &    0.33        &     995       &   $ 1.4\times10^8$                       & Mostafa et al. (1983)      \\
&           & 0.04    & 750   &    $(1-1.4)\times10^8$                  & Lynga (1984)\\
&pe     &    0.25        &     931       &    $10^8$                               & Claria et al. (1998)\\
&CCD                 &     0.22       &     1000      &    $1.3\times10^8$                  & Kalirai et al. (2003)  \\
& CCD                &    0.20          &      950      &   $ 10^8$                & Present work         \\
\hline
NGC 2420 &pe\&pg     &       $ $  &  2900          &   $1.2\times10^9$               &Sarma \& Walker (1962)    \\
&pg    &       0.01         & 2400         &    $ $                          &West (1967)            \\
&pg    &       $ $          & $ $           &    $10^9$                       &Cannon \& Lloyd (1970)   \\
& pg\&pe &0.02& $1905\pm11$&$3.3\pm0.5\times10^9$                           &McClure et al. (1974)      \\
& pg         &      0.02    &1900            & $4\times10^9$            &McClure et al. (1978)      \\
& CCD         &     0.05      &2450       & $3.4\pm0.6\times10^9$      &Anthony-Twarog et al. (1990)   \\
&CCD    &    0.05    & 2450    &    $2\times10^9$     &                  Lee et al. (2002) \\
&  CCD        &      0.04     &  2480             &  $2\times10^9$           &Present work    \\
\hline
NGC 2548  & pe  & 0.04   &630 &              $ $          &          Pesch (1961)\\
          &CCD  & 0.03   & 770&              $4\times10^8$&           Present work \\
\tablenotetext{*} {pe: photoelectric; pg: photographic; CCD: Charged Couple Device}

\enddata

\end{deluxetable}

\begin{deluxetable}{ r c r r }
\tablewidth{0pt}
\tablehead{
\colhead{ Name} &\colhead{Core} &\colhead{Cluster } & \colhead{ $f_0$}      \\
         &\colhead{ $arcmin$ (pc)}  & \colhead{ $arcmin$ (pc)} &\colhead{$star\over{arcmin}^2$ ($star\over {pc}^2$)}
}
\tablecaption{ Parameters for target clusters for a mass limited sample having $M_V\le7$ mag. $f_0$ represents central star density of clusters. \label{Table: 10}}
\startdata
Be 62*   & 2.8 (1.9) &   14.0 (9.5)& 3.0  (6.6)\\
NGC 1528 & 8.3 (2.6) &   15.0 (4.8)& 1.7 (16.9) \\
NGC 1960 & 3.5 (1.4) &   14.0 (5.4)& 2.8 (18.7)\\
NGC 2287 & 1.0 (0.2) &   12.0 (2.5)& 2.2 (51.6)\\
NGC 2301 & 1.7 (0.4) &    9.0 (2.3)& 3.9 (60.9)\\
NGC 2323 & 7.2 (2.0) &   17.0 (4.7)& 2.5 (32.7)\\
NGC 2420 & 1.4 (1.0) &   10.0 (7.2)& 19.8 (38.1)\\
NGC 2437 & 8.3 (3.6) &   20.0 (8.8)& 3.5 (18.1)\\
NGC 2548 & 1.7 (0.4) &    8.0 (1.8)& 1.4 (27.9)\\
\enddata
\tablenotetext{*} { for $M_V\le 6$ mag}
\end{deluxetable}

\clearpage

\begin{deluxetable}{ r c c r c r c r}
\tablewidth{0pt}
\tabletypesize{\scriptsize}
\tablehead{
\colhead{Name} &\colhead{Log age} &\colhead{Distance}&\colhead{$R_G$} &\colhead{Core radius} &\colhead{Cluster radius} &\colhead{Limiting magnitude}&\colhead{Reference}\\
 &\colhead{(yr)} &\colhead{(kpc)} &\colhead{(kpc)}&\colhead{$arcmin$ (pc)}&\colhead{$arcmin$ (pc)}&\colhead{ $M_V\le$}&
}
\tablecaption{ Parameters of previously studied clusters (see text). \label{Table: 11}}
\startdata
NGC 663  & 7.2  & 2.4& 10.19 &3.9 (2.7)  & 17.5 (12.2)& 4.6& Pandey et al. (2005)\\ 
NGC 1912 & 8.5  & 1.4&  9.89 &4.5 (1.8)  & 14.0 (5.7)& 6.5& Pandey et al. (in preparation)\\
NGC 7654 & 8.2  & 1.4&  9.12 &3.8 (1.6)  & 11.7 (4.8)& 4.5& Pandey et al. (2001)\\
NGC 654  & 7.3  & 2.4& 10.18 &0.9 (0.6)  & 3.7 (2.6)& 5.3& Pandey et al. (2005)\\
NGC 1907 & 8.5  & 1.8& 10.33 &2.0 (1.1)  & 6.0 (3.1)& 5.0& Pandey et al. (in preparation)\\
King 5   & 9.0  & 1.9& 10.09 &0.8 (0.4) & 2.8 (1.6)& 5.0& Durgapal \& Pandey (2001)\\
King 7   & 8.8  & 2.2& 10.46 &1.1 (0.7)  & 2.9 (1.9)& 3.4& Durgapal \& Pandey (2001)\\
Be 20    & 9.7  & 9.0& 17.11 &0.6 (1.6) & 2.5 (6.6)& 3.9& Durgapal \& Pandey (2001)\\
\enddata
\end{deluxetable}

\end{document}